\documentclass[
  aps,
  prx,
  twocolumn,
  superscriptaddress
]{revtex4-2}

\usepackage{amsmath,amssymb,amsthm}
\usepackage{physics}
\newcommand{\imag}{\mathrm{i}}
\newcommand{\phicl}{\phi_\mathrm{cl}}

\DeclareMathOperator{\poly}{poly}

\newcommand{\E}{\mathbb E}

\usepackage{xcolor}

\usepackage{mathtools}

\usepackage{makecell}

\usepackage{hyperref}
\usepackage[capitalise,compress,nameinlink]{cleveref}

\newtheorem{theorem}{Theorem}
\newtheorem{problem}{Problem}

\newtheorem{lemma}{Lemma}
\newtheorem{definition}{Definition}

\usepackage[most]{tcolorbox}

\newcounter{protocol}
\renewcommand{\theprotocol}{\arabic{protocol}}

\crefname{protocol}{Protocol}{Protocols}
\Crefname{protocol}{Protocol}{Protocols}

\newlength{\protowidth}

\newcommand{\pprotocol}[5]{
\begin{figure*}[#4]
    \refstepcounter{protocol}
    \label{#3}

    \centering
    \setlength{\protowidth}{0.96\textwidth}

    \begin{minipage}{\protowidth}
        \hrule
        \vspace{0.7em}

        \small

        \begin{center}
            \textbf{#1}
        \end{center}

        #5

        \vspace{0.4em}
        \hrule
    \end{minipage}

    \vspace{0.6em}

    \begin{minipage}{\protowidth}
        \small
        \textbf{Protocol \theprotocol.} #2
    \end{minipage}

    \vspace{-2ex}
\end{figure*}
}

\newcommand{\protocol}[4]{
    \pprotocol{#1}{#2}{#3}{tbh!}{#4}
}

\usepackage{bbm}

\begin{document}
\title{Quantum-optimized phase closure for robust interferometric imaging}

\author{Zhenning~Liu}
\email{zhenning@umd.edu}
\affiliation{Joint Center for Quantum Information and Computer Science,
NIST/University of Maryland, College Park, MD 20742, USA}
\affiliation{Department of Computer Science,
University of Maryland, College Park, MD 20742, USA}

\author{Emil~T.~Khabiboulline}
\email{ekhabibo@umd.edu}
\affiliation{Joint Center for Quantum Information and Computer Science,
NIST/University of Maryland, College Park, MD 20742, USA}
\affiliation{Joint Quantum Institute,
NIST/University of Maryland, College Park, MD 20742, USA}

\author{Lorcan~O.~Conlon}
\affiliation{Joint Center for Quantum Information and Computer Science,
NIST/University of Maryland, College Park, MD 20742, USA}
\affiliation{Joint Quantum Institute,
NIST/University of Maryland, College Park, MD 20742, USA}

\author{Brittany~McClinton}
\affiliation{NSF NOIRLab, Tucson, AZ 85719, USA}

\author{Aqil~Sajjad}
\affiliation{Department of Electrical and Computer Engineering, University of Maryland, College Park, MD 20740, USA}

\author{Saikat~Guha}
\affiliation{Department of Electrical and Computer Engineering, University of Maryland, College Park, MD 20740, USA}

\author{Jayadev~Rajagopal}
\affiliation{NSF NOIRLab, Tucson, AZ 85719, USA}

\author{Alexey~V.~Gorshkov}
\affiliation{Joint Center for Quantum Information and Computer Science,
NIST/University of Maryland, College Park, MD 20742, USA}
\affiliation{Joint Quantum Institute,
NIST/University of Maryland, College Park, MD 20742, USA}

\author{Daniel~Gottesman}
\affiliation{Joint Center for Quantum Information and Computer Science,
NIST/University of Maryland, College Park, MD 20742, USA}
\affiliation{Department of Computer Science,
University of Maryland, College Park, MD 20742, USA}
\author{Stephen~T.~Ridgway}
\affiliation{NSF NOIRLab, Tucson, AZ 85719, USA}

\begin{abstract}

Closure phase enables imaging with ground-based interferometric arrays by canceling telescope-dependent phase errors due to atmospheric fluctuations. It can be obtained by measuring fringe phases on individual baselines forming a triangle and summing them. We show that closure phases can be measured more accurately with fewer photons by performing joint measurements on multiple photons. We first establish that at least three photons from the same atmospheric temporal phase-stability window are required for closure-phase information to survive the unknown telescope-dependent phase noise, and show that joint three-photon measurements already provide an advantage over single-photon measurements in the minimal three-telescope setting. We then turn to large arrays and consider estimating all bispectra---complex quantities whose phases are the
closure phases---under a global mean-square error criterion. Although an $m$-telescope array contains $\Theta(m^2)$ independent triangle bispectra, we show that physical constraints on the visibilities make the total information content grow only nearly linearly with $m$ at fixed accuracy. Exploiting this structure, we construct an efficient protocol using at most three-photon joint measurements with photon cost $\widetilde{O}(m)$, whereas any sequential single-photon protocol requires $\Omega(m^2)$ photons. When the number of detected photons per temporal stability window scales linearly with the array size, this allows the required exposure time to remain independent of $m$ up to logarithmic factors. We further analyze the complementary high-precision regime using multiparameter quantum metrology and find a similar advantage for collective measurements. Overall, our results show that quantum information processing techniques can substantially improve the sensitivity and robustness of optical interferometry.
\end{abstract}

\maketitle

\section{Introduction}

\begin{figure}[h]
    \centering
    \includegraphics[width=\linewidth]{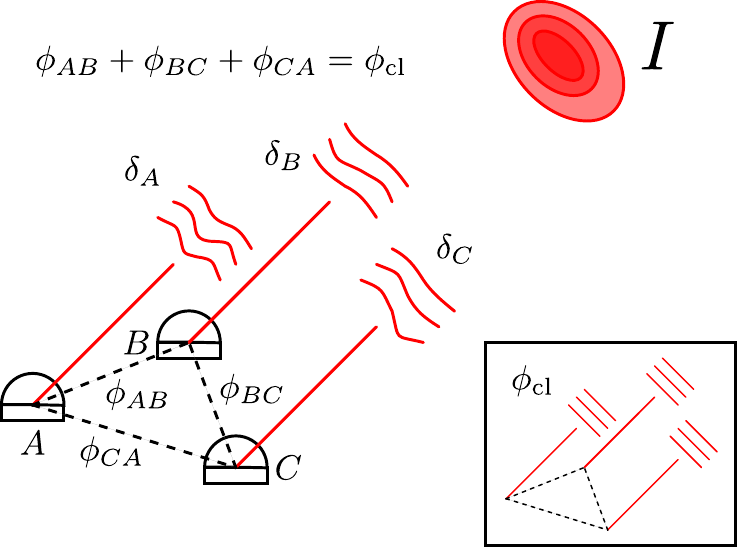}
    \caption{Telescope array relative to a source with intensity distribution $I$ in the far-field limit. Uncorrelated local noise $\delta_i$ distorts the wavefront at each telescope site $i$, corrupting the reconstructed image. Summing pairwise relative phases $\phi_{ij}$ in a cycle gives a closure phase $\phi_\text{cl}$, which is insensitive to this noise and corresponds to the noiseless image.}
    \label{fig:setup}
\end{figure}

In astronomy, interferometry operates with sparse telescope arrays, providing a small sampling of the complex coherence in the spatial frequency plane. 
Interferometric imaging typically proceeds by performing a Fourier transform on the coherences
~\cite{Goodman_2005,Thompson2017RadioAstronomy} to yield the intensity distribution of the source, as described by the Van Cittert-Zernike theorem~\cite{Zernike1938Theorem}. However, phase measurements are subject to time-varying and often uncalibrated fluctuations due to the atmosphere. Therefore, simple Fourier inversion of complex visibilities is generally not a robust imaging strategy for ground-based astronomical observations.
One of the most powerful and dependable observables that allows coherent accumulation of information across different atmospheric stability windows is the complex \emph{bispectrum}, and in particular its phase, known as the \emph{closure phase}, that the astronomical scene projects onto the interferometric array~\cite{Rogers1974,Hofmann1988ClosureSpeckle}. For any triangle of telescopes, the bispectrum is the product of the three complex visibilities around the triangle, and its phase—the closure phase—is invariant under telescope-dependent phase shifts. Classically, closure phase gives constraints to the imaging and is typically reconstructed from many photons either as the phase of the bispectrum averaged over many atmospheric stability windows or by measuring pairwise visibility phases between sites and summing them together.

Quantum technology enables longer-baseline optical interferometry by processing incoming photons nonlocally with quantum entanglement~\cite{Gottesman2012LongBaseline,Khabiboulline2019OpticalInterferometry,Khabiboulline2019TelescopeArrays, Rajagopal:2024suu, Padilla2026a, Padilla2026b, Liu2026LensingDelays, Stas_2026, chahine2026}, instead of physically interfering them at a classical beam combiner. Such quantum-enabled telescopes promise extremely high angular resolution (typically at \emph{microarcsecond} level) and a correspondingly tiny field of view. Usually, the amount of incoming light of such long-baseline arrays is severely limited. This makes the estimation of any observables, including closure phase, particularly challenging.

Beyond enabling long baselines, quantum information processing also enlarges the class of measurements that can be performed on the incoming light. For instance, coherent nonlocal measurements can outperform local detection for weak thermal light~\cite{Tsang2011WeakThermalLight}. Suitably chosen spatial-mode measurements can extract information inaccessible to conventional direct-imaging or pairwise-interferometric measurements~\cite{Cosmo2020,Sajjad2024}.
Joint measurements across multiple photons can further improve the information efficiency of conventional imaging tasks~\cite{deshler2026attaining}.

In this work, we investigate the fundamental limits and optimal strategies for measuring noise-invariant observables in the photon-starved regime for interferometric arrays. In particular, we study whether multi-photon collective measurements, typically enabled by storing multiple photons arriving at different times coherently in a quantum memory (or equivalently in long optical delay lines that can be rapidly and accurately tuned) followed by a quantum circuit and measurement (or linear optics plus photon counters for limited types of measurements~\cite{hong1987measurement}), can provide any advantage in terms of efficiency or robustness of closure phase or bispectrum estimation.

We first establish an information-theoretic lower bound on the number of photons required within a single atmospheric stability window to access closure-phase information. 
Then, we focus on the minimal setting of three telescopes forming a single triangle, where we characterize the fundamental limits of collective three-photon measurements and of strategies that measure the photons individually.
Collective measurements allow us to nondestructively measure the total photon number at each telescope, so that we can quickly identify the informative block that carries closure-phase information and avoid measurements that introduce noise.
Guided by this comparison, we construct a quantum-memory-assisted three-photon protocol, proving that it estimates closure phase more efficiently than any single-photon measurement strategy.

We then turn to larger interferometric arrays containing many telescope sites. We analyze the information content of the full set of complex bispectra and prove that, in astronomically relevant parameter regimes, they are highly compressible.
Surprisingly, at fixed target accuracy, the total amount of information grows only linearly with the number of telescopes, rather than with the number of independent bispectra.
This information-content analysis suggests that the number of photons sufficient to estimate all bispectra may be much lower than expected.
In this setting, occupation-number measurements enabled by collective processing provide an even stronger advantage: using joint measurements on only up to three photons, we construct a protocol that estimates all bispectra with asymptotically fewer photons than any sequential single-photon strategy. Practically speaking, telescope arrays can grow in size without having to increase observing time to collect all the relevant information. We also find the same trend in the regime where bispectra are estimated to high precision.

We organize the remainder of this paper as follows. In \cref{sec:prelim}, we introduce the physical model of photons from the scene how we do initial processing of them through nondemolition measurements. In \cref{sec:noise_invariant}, we study the mathematical structure of multi-photon noisy states and show the minimum photon number requirement for accessing the noise-invariant quantities, including closure phases. In \cref{sec:single}, we analyze the optimal measurement of a single closure phase in the minimal possible telescope array. In particular, we summarize a simple-to-implement measurement scheme in \cref{prot:three_photon_single_bispectrum}. 
Next, in \cref{sec:full_bispectra}, we generalize the study to larger telescope arrays and propose protocols to measure many noise-invariant quantities efficiently. We summarize our proposed measurement strategy in \cref{prot:three_photon_full_bisepctra}. Finally, in \cref{sec:discussion}, we discuss the implication, limitation, and outlook of our results.

We leave additional mathematical and astronomical details to the appendices.
In particular, we introduce the model of atmospheric turbulence in \cref{sec:kolmogorov}. In \cref{sec:bright}, we present a calculation of quantum Fisher information in the special case that the scene is sufficiently bright. \cref{appendix:single_photon_limit_for_m=3} shows the optimal approach of learning closure phase from single-photon measurements. 
In \cref{sec:details}, we present the details of the quantum Fisher information calculation for three-photon states. 
In \cref{sec:TorusQFT}, we generalize \cref{prot:three_photon_single_bispectrum} for more photons and sites.
\cref{sec:three_photon_information,sec:single_photon_large_array} contain additional proofs of theorems in \cref{sec:full_bispectra}.  Finally, in \cref{sec:multiparameterCramerRao}, we calculate Cramer-Rao bounds in the infinitesimal-precision regime, and demonstrate an incompatibility gap between different classes of collective measurements.

\section{Preliminaries}
\label{sec:prelim}
We begin by formulating the relevant optical concepts in terms of quantum states of photons. (See the SPIE conference version~\cite{Gorshkov2026SPIEClosure} of this paper for an exposition focused on astronomy.) This quantum description is not essential for conventional measurements that process photons individually, which can be described within the standard framework of linear optics. In this work, however, we focus on quantum-memory-assisted protocols and joint measurements of multiple photons, for which the quantum-state picture provides a more natural and unified framework.

We focus on the photon-starved regime and work in a fixed spectral channel. We describe the incoming light using orthogonal temporal modes, labeled by $t$, and assume that these modes are matched across the telescopes. Each temporal mode across the array is dominated by vacuum and has a small probability of containing one photon~\cite{Tsang2011WeakThermalLight}. We neglect events containing two or more photons in the same temporal mode.

For our memory-assisted protocols, we assume the incoming light is coherently stored in quantum memories, and the occupied temporal modes are identified without revealing which telescope received the photon~\cite{Khabiboulline2019OpticalInterferometry, Khabiboulline2019TelescopeArrays}. This nondestructive readout can be implemented by entanglement-assisted nonlocal parity checks. Within the vacuum--single-photon approximation, these checks distinguish empty from occupied modes while preserving the spatial coherence of the stored single-photon state. We condition on the single-photon outcomes and read out the temporal-mode labels as classical information.

We start with two telescopes labeled by $A$ and $B$.
Conditioned on one photon from a point source at angle $\theta$
occupying temporal mode $t$, its state is
\begin{equation}
    \ket{\psi_{\theta,t}}
    =
    \frac{1}{\sqrt{2}}
    \left(
        \ket{1}_{A,t}\ket{0}_{B,t}
        +
        e^{\frac{2\pi\imag d\sin\theta}{\lambda}}
        \ket{0}_{A,t}\ket{1}_{B,t}
    \right),
    \label{eq:photon_theta}
\end{equation}
where $\ket{n}_{j,t}$ denotes $n$ photons at site $j$ in temporal
mode $t$, $d$ is the baseline length, $\theta$ is measured from
the direction perpendicular to the baseline, and $\lambda$ is
the central wavelength.

In first-quantized notation, the same single-photon state is
\begin{equation}
    \begin{aligned}
        \ket{\psi_{\theta,t}}
        &=
        \frac{1}{\sqrt{2}}
        \left(
            \ket{A}
            +
            e^{\frac{2\pi\imag d\sin\theta}{\lambda}}\ket{B}
        \right)\otimes\ket{t} \\
        &\equiv
        \ket{\psi_\theta}\otimes\ket{t},
    \end{aligned}
    \label{eq:photon_theta_1st}
\end{equation}
where the spatial register encodes the telescope site and
$\ket{t}$ labels the temporal mode.
Since $t$ has been read out without revealing the site, we may
suppress the temporal information and work with the spatial
state $\ket{\psi_\theta}$. We label the photons by the known temporal modes in which they arrive, and use one register to represent the spatial state of each photon. For example, two independent photons from directions $\theta_1$ and $\theta_2$, occupying distinct modes $t_1$ and $t_2$, are represented by
\begin{equation}
    \ket{\Psi_{\theta_1,\theta_2}}
    =
    \ket{\psi_{\theta_1}}_{t_1}
    \otimes
    \ket{\psi_{\theta_2}}_{t_2}.
\end{equation}
Here the subscripts identify the registers by their temporal
modes.
More generally, conditioned on $n$ distinct occupied temporal modes in
an $m$-site array, the spatial state is represented on
$(\mathbb C^m)^{\otimes n}$, an $m^n$-dimensional Hilbert space.
We use this representation throughout the paper, suppressing
the temporal labels when no confusion can arise.

Now, we extend the imaging scenario from a point source \cref{eq:photon_theta,eq:photon_theta_1st} to an extended \emph{scene}, which is characterized by the intensity distribution $I(\theta)$.  Photons are emitted incoherently across the scene, and we do not know a priori what angle every photon comes from.
Therefore, the state of the single photon from the incoming light in the above two-telescope setting is described by a density matrix rather than a pure state:
\begin{equation}
    \rho^\mathrm{noiseless} = \int_\theta d\theta \, I(\theta) \ket{\psi_\theta}\bra{\psi_\theta} = \frac{1}{2}\begin{pmatrix}
       1 & V^\mathrm{obj}_{AB}\\
       V^\mathrm{obj*}_{AB}  & 1
    \end{pmatrix},
\end{equation}
where the off-diagonal term $V^\mathrm{obj}_{AB} = |V_{AB}^{\rm obj}|e^{\imag\phi_{AB}^{\rm obj}}$ is the scene's \emph{intrinsic} complex visibility between sites $A$ and $B$. However, due to atmospheric fluctuations, the \emph{measured} visibility $V_{AB} = |V_{AB}|e^{\imag \phi_{AB}}$ also includes telescope-dependent complex multiplicative contributions, denoted by $ g_i=|g_i|e^{\imag \delta_i}$,
so that
\begin{equation}
V_{AB} = g_A g_B^* V_{AB}^{\rm obj}=
|g_A||g_B||V_{AB}^{\rm obj}|
e^{\imag (\phi_{AB}^{\rm obj}+\delta_A-\delta_B)}.    
\end{equation}
where $\delta_i$ is the telescope-dependent phase error caused by atmospheric fluctuations (we also use $\vec{\delta}\coloneqq (\delta_A, \delta_B, \dots)$ to denote the collection of all piston phases). The $\delta_i$s are also called the local \emph{piston phases}. 
In this work, we mainly study the instability of the phase as it is most susceptible to errors; henceforth we assume $|g_i|=1$ and thus $|V_{AB}| = |V_{AB}^\mathrm{obj}|$. Without additional techniques (such as adaptive optics), the piston error is unknown, precluding the measurement of $V_{AB}^\mathrm{obj}$. Moreover, the piston error varies rapidly in time such that measuring $V_{AB}$ is also difficult for faint targets.
In astronomy, the time variation of $\vec{\delta}$ is typically characterized by the \emph{atmospheric coherence time} (ACT, denoted by $\tau_c$): one can assume if two time instances are separated by time shorter than $\tau_c$, then the piston errors are the same, otherwise they are independently uniformly random from $[0,2\pi)$, given that the phase fluctuation scales as path fluctuation divided by the extremely short wavelength in optical/infrared bands.
One caveat here is that photons within the same ACT but having significantly different frequencies experience different piston phases caused by atmospheric fluctuations. We therefore also need to restrict our discussions to be in a small spectral band. The bandwidth ensuring piston phases remain unchanged for light signals in the same ACT is typically $\sim 10^{11}\,\mathrm{Hz}$. We present a more detailed introduction to this error model in Appendix~\ref{sec:kolmogorov}.

Now, we consider arrays with $m=3$ sites, labeled by $A,B,C$ (Fig.~\ref{fig:setup}). Three-site arrays are the minimum setting to form a bispectrum. In one ACT, the single-photon density matrix is
\begin{equation}
\label{eq:dm_noisy}
\begin{aligned}
    \rho(\vec{\delta})  
&=\frac{1}{3} \begin{pmatrix}
1 & V_{AB} &  V_{AC}\\
 V_{AB}^*   & 1 &  V_{BC}  \\
V_{AC}^*   &  V^*_{BC}  & 1
\end{pmatrix} \,.
\end{aligned}
\end{equation}

\begin{definition}[Bispectrum and closure phase] \label{def:bispectrum}
We define the \emph{bispectrum} corresponding to three sites labeled by $i,j,k$ as
\begin{equation}
    B_{ijk}
\coloneqq
V_{ij}V_{jk}V_{ki} =
|V_{ij}V_{jk}V_{ki}| e^{i\phicl}
\end{equation} 
whose phase is the \emph{closure phase}, defined as
\begin{equation}
\begin{aligned}
    \phicl 
    &= \phi_{ij}^{\rm obj} + \phi_{jk}^{\rm obj} + \phi_{ki}^{\rm obj} \\
    &\quad +
    (\delta_i - \delta_j)+(\delta_j - \delta_k)+(\delta_k - \delta_i)\\
    &=\phi_{ij}^{\rm obj} + \phi_{jk}^{\rm obj} + \phi_{ki}^{\rm obj}.
\end{aligned}
\end{equation}
    
\end{definition}
Observe that, indeed, the closure phase $\phicl$ eliminates all station-based piston terms, leaving only quantities representative of the source \cite{Jennison_1958}.
Although two degrees of freedom in the intrinsic visibilities are lost when estimating $\phicl$ only, closure phase still provides a significant amount of information to infer various properties of the scene, such as its spatial extent and asymmetry~\cite{Monnier2007Altair,Lazareff2017Disks}.

Note that for any three distinct site labels $i,j,k$, cyclic permutations leave the bispectrum unchanged, whereas exchanging any two indices gives its complex conjugate:
\begin{equation}
\label{eq:equivalent_Bijk}
    \begin{aligned}
        B_{ijk}&=B_{jki}=B_{kij}, \\
        B_{ikj}&=B_{kji}=B_{jik}=B_{ijk}^*.
    \end{aligned}
\end{equation}
Thus, for any $\pi\in S_3$ permuting the three positions of $(i,j,k)$, $B_{\pi(ijk)}=B_{ijk}$ if $\pi$ is even, and $B_{\pi(ijk)}=B_{ijk}^*$ if $\pi$ is odd.

With the single-photon state well understood, we turn to the density matrix describing multiple photons received in the same ACT. Assuming exactly $n$ photons are received in every ACT, the joint state of photons in one specific ACT (corresponding to a specific local phase error vector $\vec{\delta}$) is $\rho(\vec{\delta})^{\otimes n}$. However, since in reality $\vec{\delta}$ is unknown, the density matrix of any such $n$-photon state is the classical mixture of $n$-photon density matrices characterized by all possible $\vec{\delta}$s, i.e.,
\begin{equation} \label{eq:dm_noisy_atmosphere}
\begin{aligned}
     \rho^\text{noisy}_n &\coloneqq \mathbb{E}_{\vec\delta}[\rho(\vec\delta)^{\otimes n}] \\
    &= \frac{1}{(2\pi)^m} \int_{0}^{2\pi} d\delta_1 \dots \int_{0}^{2\pi} d\delta_m \, \rho(\vec\delta)^{\otimes n} \,.
\end{aligned}
\end{equation}

Throughout this work, a collective $n$-photon strategy may coherently store photons arriving within the same ACT and jointly process them. In the photon-starved regime, the average number of photons per temporal-frequency mode is typically much lower than one ($\lesssim 10^{-7}$ for magnitude $10$ stars~\cite{Monnier2026QuantumTechnologies}), while the number of such modes in each ACT within the characteristic bandwidth is $\sim 10^{-2}\,\mathrm{s}\, \times 10^{12}\,\mathrm{Hz} \sim 10^{10}$.

Joint measurements therefore require storing photons in quantum memory for up to an ACT while waiting for the other photons to arrive. By contrast, a single-photon strategy measures each photon individually and destructively in any single-photon basis and maintains only classical information in the record. Each measurement may depend on the preceding classical outcomes and can typically be implemented with linear optics followed by single-photon detectors.

Later in this paper, we also design a protocol that only uses copies of $\rho_3^\mathrm{noisy}$ even if there are $n\gg 3$ in every ACT. It is useful to note that three-photon noisy states can be constructed from $\rho_n^\mathrm{nosy}$ artificially.
We can simply partition the photons into $\lfloor n/3\rfloor$ disjoint triples. For the $r$th triple, we can independently sample an artificial local phase vector $\vec{\delta}^{(r)}$ uniformly randomly and apply $D_{\vec{\delta}^{(r)}}^{\otimes 3}$ to the three-photon state, where $D_{\vec{\delta}} =\sum_j e^{i\delta_j}\ket{j}\!\bra{j}$, while discarding the classical record of $\vec{\delta}^{(r)}$. For any fixed $\vec{\delta}$, $\vec{\delta}+\vec{\delta}^{(r)}$ is again uniformly distributed, and hence each triple is transformed into
\begin{equation}
    \mathbb{E}_{\vec{\delta}^{(r)}}
    \left[
        \rho\!\left(
            \vec{\delta}+\vec{\delta}^{(r)}
        \right)^{\otimes 3}
    \right]
    =
    \rho_3^{\rm noisy}.
\end{equation}
Since the artificial phase vectors are chosen independently for different triples, the resulting triples are independent copies of $\rho_3^{\rm noisy}$.

We next ask, for each number of photons per ACT, what information about the scene survives the atmospheric phase averaging.

\section{Structure of multi-photon states and photon-number thresholds}
\label{sec:noise_invariant}
In this section, we conduct a closer examination to the mathematical structure of the $n$-photon state, $\rho_n^\mathrm{noisy}$, and use the results to prove the threshold of the photon number per ACT that allows for access to the closure phase information. 
In \cref{sec:occupation_sectors}, we first show that the $n$-photon state $\rho_n^\mathrm{noisy}$ is block-diagonal, where each block corresponds to a certain pattern of photon occupation among the $m$ telescopes. 
Using this information, in \cref{sec:photon_number_threshold}, we present 
a (quantum) information theoretic proof 
that at least $3$ photons are needed to access closure phases.

\subsection{Occupation sectors under atmospheric averaging and projections onto them}
\label{sec:occupation_sectors}

We start with analyzing what information survives atmospheric phase averaging for general $m$ and $n$ in the expression of $\rho_n^\mathrm{noisy}$ [see Eq.\ (\ref{eq:dm_noisy_atmosphere})].
For an $m$-site array, 
$\rho(\vec{\delta})$ has matrix elements 
\begin{equation}
\bra{i}\rho(\vec\delta)\ket{j}=V_{ij}^{\mathrm{obj}}e^{\imag(\delta_i-\delta_j)}/m    
\end{equation}
with $V_{ii}^{\mathrm{obj}}=1$. Let $\mathbf{i}=(i_1,\ldots,i_n)$ and $\mathbf{j}=(j_1,\ldots,j_n)$ be \emph{site strings} of the $n$ photons, i.e., $j_k$ is the label of the telescope that the $k$th photon is received.
Then we let $N(\mathbf{i})$ denote the \emph{occupation-number vector} of length $m$, whose $\ell$th element, denoted by $N_\ell(\mathbf{i})$, counts the photons at telescope $\ell$. For example, for three telescopes $A,B,C$ and four photons, $N(A,B,C,C)=N(C,B,C,A)=(1,1,2)$ and $N(B,A,C,B)=(1,2,1)$.

Now, we observe that the phase factor due to the atmospheric fluctuations $\vec{\delta}$ in the matrix element   $\bra{\mathbf{i}}\rho(\vec\delta)^{\otimes n}\ket{\mathbf{j}}$ is
\begin{equation}
    \exp\left[
        \imag\sum_{\ell=1}^m
        \bigl(N_\ell(\mathbf{i})-N_\ell(\mathbf{j})\bigr)\delta_\ell
    \right].
\end{equation}
In other words, every site in $\mathbf{i}$ (or $\mathbf{j}$) contributes its own piston phase positively (or negatively) to the overall phase error corresponding to the $\ket{\mathbf{i}}\bra{\mathbf{j}}$ element.

Now, observe that the only way to make the $\ket{\mathbf{i}}\bra{\mathbf{j}}$ element survives the averaging over the independent uniform piston phases is to ensure the error contribution of $\mathbf{i}$ and $\mathbf{j}$ exactly cancel, i.e.,
$N(\mathbf{i})=N(\mathbf{j})$. Therefore,
\begin{equation}
\label{eq:rho_n_general_noisy}
\begin{aligned}
    \bra{\mathbf{i}}\rho_n^{\mathrm{noisy}}\ket{\mathbf{j}}
    &=\frac{ \mathbbm{1}_{N(\mathbf{i})=N(\mathbf{j})} }{m^n}
    \prod_{a=1}^{n}V_{i_a j_a}^{\mathrm{obj}}.
\end{aligned}
\end{equation}
Now, we can see why the noisy $n$-photon state has the block diagonal structure: we first define the occupation projector for a given occupation-number vector $\boldsymbol{\nu}$:
\begin{equation}
    \Pi_{\boldsymbol\nu}
    =
    \sum_{\mathbf{i}:\,N(\mathbf{i})=\boldsymbol{\nu}}
    \ket{\mathbf{i}}\bra{\mathbf{i}},
\end{equation}
then $\rho_n^\mathrm{noisy}$ has the block-diagonal decomposition
\begin{equation}
\label{eq:occupation_general}
    \rho_n^{\mathrm{noisy}}
    =
    \bigoplus_{\boldsymbol\nu:\,\sum_j\nu_j=n}
    \Pi_{\boldsymbol\nu}
    \rho_n^{\mathrm{noisy}}
    \Pi_{\boldsymbol\nu}
\end{equation}
because every element outside of these blocks vanishes after averaging over piston errors.

\protocol{$n$-photon collective measurement for the occupation-number vector $\boldsymbol\nu$}{The spatially local collective nondemolition measurement of multiple photons that gives the occupation-number vector $\boldsymbol\nu$.}{prot:occupation_number}
{
Given an integer $n\geq1$ as the number of photons per ACT and a copy of $\rho_n^\mathrm{noisy}$, corresponding to $n$ photons received in $n$ distinct temporal modes by an array of $m$ telescopes.

\begin{enumerate}
    \item \textbf{Coherently count photons at each telescope.}
    At each telescope $j$, introduce an auxiliary counter initialized to zero. Coherently add to this counter the occupation of site $j$ across all $n$ stored temporal modes, without measuring which individual temporal modes contain photons at that site.

    \item \textbf{Read out the occupation vector.}
    After all $n$ photon registers have contributed, measure the
    local counters. The outcomes give the occupation vector
    $\boldsymbol\nu=(\nu_1,\ldots,\nu_m)$
    where $\nu_j$ is the total number of photons received by
    telescope $j$.
    The measurement projects the $n$-photon state onto
    $\Pi_{\boldsymbol\nu}
    \rho_n^{\mathrm{noisy}}
    \Pi_{\boldsymbol\nu}$
    while preserving coherence between different
    photon-to-site assignments with the same
    $\boldsymbol\nu$.
    One may then keep or discard the state according to the
    desired occupation pattern.
\end{enumerate}
}

As the minimum example, we can consider the $3$-photon state on $m=3$ sites labeled by $A,B,C$. The $3$-photon noisy state can be written as the direct sum of blocks labeled by the occupation-number vector, where $\boldsymbol\nu = (1,2,0)$ means $1$ photon at site $A$ and $2$ photons at site $B$, for instance. This gives
\begin{equation}
\begin{aligned}
    \rho_3^{\mathrm{noisy}} &= M_{(3,0,0)} \oplus  M_{(0,3,0)} \oplus M_{(0,0,3)} \oplus M_{(2,1,0)} \\
    &\quad \oplus M_{(2,0,1)} \oplus M_{(1,2,0)} \oplus M_{(1,0,2)} \oplus M_{(0,2,1)}\\
    &\quad \oplus M_{(0,1,2)} \oplus M_{(1,1,1)}.
\end{aligned}
\end{equation}
Since $M_{(1,1,1)}$ is the block that $A,B,C$ each receives one photon, we also give it an alias: $M_{ABC}\coloneqq M_{(1,1,1)}$.

The block-diagonal decomposition suggests that one can always measure the occupation vector first to project $\rho_n^\mathrm{noisy}$ onto one of the blocks, without disturbing any coherence that survives the atmospheric averaging. The occupation outcome $\boldsymbol\nu$ occurs with probability
\begin{equation}
    \Pr[\boldsymbol\nu] = \frac{n!}{m^n\prod_j \nu_j!}.
\end{equation}

Operationally, the occupation-number vector $\boldsymbol\nu$ can be measured using only local operations at each telescope. Each site nondestructively measures its total photon number across the $n$ stored temporal modes, without resolving the occupation of individual modes. This can be implemented by coherently accumulating the local occupations in an auxiliary counter and measuring the counter after all $n$ photons have arrived. Combining the local counts gives $\boldsymbol\nu$, while preserving coherence between different photon-to-site assignments within the same occupation sector.

Although the readout is local across telescope sites, this is a \emph{collective measurement} across the $n$ photons because their contributions are combined coherently before any occupation information is revealed, rather than measured separately and summed classically. We summarize this quantum-memory-based implementation in \cref{prot:occupation_number}.

\subsection{Photon-number threshold and closure-phase-sensitive blocks }
\label{sec:photon_number_threshold}

The naive strategy to estimate $\phicl$ is measuring all visibilities $V_{AB},V_{BC},V_{CA}$ in each ACT window and computing the closure phase classically. However, to estimate the closure phase within a reasonable precision (for instance, $\sim 0.1$), this method requires at least $\sim 10^2$ photons per ACT window to measure all visibilities within $\sim 0.1$ precision. Moreover, the number of photons received during one ACT window, $n$, follows a Poisson distribution with a mean characterized by the length of ACT $\tau_c$, scene brightness, light collection area, etc., and is typically small for longer-baseline optical interferometry. To achieve optimal robustness to atmospheric fluctuations (or equivalently maximum sensitivity for a given $\tau_c$), we wish to use the minimum possible $n$ to measure closure phases and ideally all complex bispectra.
Although Refs.~\cite{wirnitzer1985bispectral,nakajima1988signal,nakajima1989diffraction} hinted that $n=3$ is the minimum for closure phase measurements in single-aperture telescopes, an information-theoretic proof of the lower bound remains unexplored.

Here, using the block-diagonal decomposition derived in \cref{sec:occupation_sectors}, we formally prove the lower bound of $3$ photons per ACT.

\begin{theorem}[Photon-number thresholds for noise-robust parameters]
\label{thm:photonumber2}
For an arbitrary $m$-site array with $m\geq3$, $\rho_1^{\mathrm{noisy}}$ contains no source-dependent coherence.
However, given access to sufficiently many copies of $\rho_2^{\mathrm{noisy}}$, one can unambiguously measure all pairwise visibility amplitudes $|V_{ij}|$ but not the closure phases.
Given access to sufficiently many copies of $\rho_3^{\mathrm{noisy}}$, one can unambiguously measure all complex bispectra $B_{ijk}$.
\end{theorem}

\begin{proof}
For a single photon, \cref{eq:rho_n_general_noisy} removes all off-diagonal coherences. The diagonal elements remain equal to $1/m$, so $\rho_1^{\mathrm{noisy}}=\mathbbm{1}/m$, containing no information about the scene.

For two photons, the only surviving off-diagonal elements exchange two distinct occupied sites. For $i\neq j$,
\begin{equation}
    \bra{ij}\rho_2^{\mathrm{noisy}}\ket{ji}
    =\frac{1}{m^2}V_{ij}^{\mathrm{obj}}V_{ji}^{\mathrm{obj}}
    =\frac{1}{m^2}|V_{ij}|^2.
\end{equation}
All diagonal elements are $1/m^2$. Thus, $\rho_2^{\mathrm{noisy}}$ depends on the visibilities only through their amplitudes, and contains no closure-phase information.

For three photons, consider any three distinct sites $i,j,k$ each receiving exactly one photon, \cref{eq:rho_n_general_noisy} gives
\begin{equation}
\label{eq:Bijk_oexample_in_3photon}
\begin{aligned}
    \bra{ijk}\rho_3^{\mathrm{noisy}}\ket{jki}
    &=\frac{1}{m^3}V_{ij}^{\mathrm{obj}}V_{jk}^{\mathrm{obj}}V_{ki}^{\mathrm{obj}}\\
    &=\frac{1}{m^3}B_{ijk}.
\end{aligned}
\end{equation}
Hence every triangle bispectrum appears as a coherence in the off-diagonal term of two site strings connected via a cyclic shift in the three-photon state. Therefore, three photons are the first to give access to closure-phase information. Naively, with sufficiently many copies, one can conduct state tomography on $\rho_3^\mathrm{noisy}$ to learn all bispectra.

The remaining off-diagonal elements within this one-photon-per-site sector connect basis states that differ by a transposition and depend only on pairwise visibility amplitudes.
\end{proof}

More formally, for $n=3$, the proof of \cref{thm:photonumber2} indicates that closure phase information is only accessible when three photons are received by three distinct sites $i,j,k$. (For the simplest $m=3$ example, this is the $M_{(1,1,1)}$ block.)
The corresponding occupation sector is the following six-dimensional subspace of $\mathbb{C}^{m^3}$:
\begin{equation}
    \mathcal H_{ijk}
    =\operatorname{span}\{\ket{\pi(ijk)}:\pi\in S_3\}.
\end{equation}

Equipped with the knowledge that closure phases are only accessible in these one-photon-per-site blocks of $3$-photon states, we start considering two estimation problems. We first investigate closure-phase estimation in an array of only $3$ telescopes, and then study how to estimate all complex bispectra in a larger array of $m\gg 3$ telescopes.

\section{Single closure phase estimation} \label{sec:single}
Since $n=3$ is the minimum number of photons to have access to closure phase, and $m=3$ is the minimum number of sites to form a triangular closure, we consider the minimal $m=n=3$ setting as the first step.
In this section, we first describe the setting in \cref{sec:mn3setting}; then we compute the fundamental quantum limit in this scenario and how to attain it in \cref{sec:3photon_qfi}; finally, we present an easy-to-implement measurement protocol for the complex bispectrum in \cref{sec:direct_bispectrum}.

\subsection{Minimal $m=n=3$ setting}
\label{sec:mn3setting}
Although there is only one closure phase when $m=3$, it is already useful in a variety of astronomical imaging problems, such as estimating the relative brightness of two binary stars~\cite{Monnier2007Altair,Lazareff2017Disks}. Therefore, our goal in this case is to measure the only closure phase with high precision to provide high-quality data for these applications.

Recall from \cref{sec:photon_number_threshold} that all closure-phase dependence of $\rho_3^{\mathrm{noisy}}$ is confined to the occupation sector with one photon at each site, and the subspace containing all such site strings is
\begin{equation}
\label{eq:HABC}
    \mathcal H_{ABC}
    =\operatorname{span}\{\ket{\pi(ABC)}:\pi\in S_3\}.
\end{equation}
Let $\Pi_{ABC}$ be the projector onto $\mathcal H_{ABC}$, and let $M_{ABC}=\Pi_{ABC}\rho_3^{\mathrm{noisy}}\Pi_{ABC}$ denote the corresponding unnormalized block.
Its trace is $\Tr(M_{ABC})=2/9$. All dependence on $\phi_{\mathrm{cl}}$ lies in $M_{ABC}$; the other occupation blocks are independent of $\phi_{\mathrm{cl}}$.
The two-site blocks, along with two-photon events, can be used to learn visibility amplitudes. In fact, if we are in the photon-starved regime where the average photon number per ACT is not greater than $1$, then there are many more ACTs with $2$ photons than with $3$ photons. Therefore, we assume that all three visibility amplitudes are already measured by $2$-photon states, and focus on using the $3$-photons states to learn the closure phase with high efficiency.

\subsection{Fundamental limits of single-photon versus collective measurements}
\label{sec:3photon_qfi}

Note that a single-photon strategy measures the three photons individually and may adapt each measurement to the preceding classical outcomes. A collective strategy may jointly measure many photons.
One can apply any valid quantum measurement, i.e., positive operator-valued measure (POVM), on $\rho_3^\mathrm{noisy}$ [see \cref{eq:dm_noisy_atmosphere}] or even multiple copies of $\rho_3^\mathrm{noisy}$, since as described earlier, they are stored in quantum memory after initial processing~\cite{Khabiboulline2019OpticalInterferometry,Khabiboulline2019TelescopeArrays}.

\begin{problem}
    What is the closure phase estimation strategy for $m=n=3$ that requires the minimum number of photons to achieve high precision? Is there any advantage in using multi-photon collective measurements?
\end{problem}

The main result of this section 
is a provable constant-factor separation in the efficiency of closure phase measurements between the optimal single-photon schemes and general collective-measurement schemes.

Before stating the theorem formally, we first give a simple intuition
for the collective-measurement advantage. As shown in
\cref{sec:photon_number_threshold}, all closure-phase information in
$\rho_3^{\mathrm{noisy}}$ is confined to the $(1,1,1)$ occupation
block. If every copy were in this block, one could directly perform
the desired closure-phase measurement. In reality, this block occurs
with probability $p=2/9$. A nondestructive occupation-number
measurement [see \cref{prot:occupation_number}] identifies these
informative events without disturbing their internal coherence.
Thus, collective processing incurs only a $1/p=9/2$ overhead in the
number of input copies needed to obtain a given number of informative
samples.

By contrast, photon-by-photon measurements cannot herald the $(1,1,1)$ block. The closure-phase-sensitive contribution from this block is therefore mixed with closure-phase-insensitive background from all other occupation sectors, reducing the mean of the estimator by a factor $p$. Recovering the original signal requires rescaling the estimator by $1/p$, which amplifies the statistical error by the same factor. Consequently, the sample cost scales as $1/p^2$ rather than $1/p$, giving an additional factor
$1/p=9/2$. Intuitively, collective processing gains this factor by
discarding uninformative events before they contribute statistical
noise.

Next, we present the formal result to show this constant factor advantage in the low-visibility limit, which typically happens when the three baselines are sufficiently long such that the angular size of the scene is comparable with the field of view.

Our result is based on the analysis of classical and quantum Fisher information (CFI and QFI), which characterize the precision of single-parameter estimation~\cite{liu2020quantum}. Let $\kappa$ be the parameter of interest. For a fixed measurement on a state $\rho(\kappa)$, let $\mathcal I_\kappa$ denote the CFI per copy. With $N$ independent repetitions, a locally unbiased estimator $\tilde\kappa$ satisfies the Cram\'er--Rao bound
$\mathrm{Var}(\tilde\kappa)\geq1/(N\mathcal I_\kappa)$.
The QFI, $F_\kappa$, is the supremum of the CFI over all measurements and gives the quantum bound
$\mathrm{Var}(\tilde\kappa)\geq1/(NF_\kappa)$.
It can be computed as $F_\kappa=\Tr[\rho L_\kappa^2]$, where the symmetric logarithmic derivative (SLD) denoted by $L_\kappa$ is defined by
$\partial_\kappa\rho=(L_\kappa\rho+\rho L_\kappa)/2$.
Locally, measuring in the eigenbasis of the SLD attains the QFI. These quantities therefore allow us to compare the precision of single-photon and collective measurements in the limit of many repetitions.

\begin{theorem}[Single-photon and collective-measurement Fisher information in the weak-visibility limit]
\label{thm:single_joint_fisher}
For $m=n=3$, take known positive visibility amplitudes $|V_{AB}|,|V_{BC}|,|V_{AC}|$ and let $s_2=|V_{AB}|^2+|V_{BC}|^2+|V_{AC}|^2$. In the weak-visibility limit $s_2\to0$, the optimal classical Fisher information per ACT from sequential single-photon measurements and the joint three-photon QFI satisfy
\begin{equation}
 \begin{aligned}
 \mathcal I_{\mathrm{single}}^{\max}
 &=\frac8{81}|V_{AB}|^2 |V_{BC}|^2 |V_{AC}|^2[1+O(s_2)],\\
 F_{\mathrm{cl}}&=\frac{4}{9}|V_{AB}|^2 |V_{BC}|^2 |V_{AC}|^2 [1+O(s_2)].
 \end{aligned}
\end{equation}
Each single-photon measurement may use all preceding classical outcomes. The ratio satisfies
$F_{\mathrm{cl}}/\mathcal I_{\mathrm{single}}^{\max}= 9/2 +O(s_2)$.
\end{theorem}

\protocol{Three-photon collective measurements for a single bispectrum}{Our three-photon collective measurements for a single bispectrum whose argument is the closure phase.}{prot:three_photon_single_bispectrum}
{
Choose an integer $M_B\geq1$ as the target accepted-sample count
and let $\omega=e^{2\pi\imag/3}$.

\begin{enumerate}
    \item \textbf{Collect three-photon states and select the informative block.}
    Collect three photons from the same ACT and store them in quantum memory, obtaining a copy of $\rho_3^{\mathrm{noisy}}$. Then, perform \cref{prot:occupation_number}: at each telescope, nondestructively measure the total photon number across the three stored temporal modes,
    without resolving individual modes.
    Keep the state only if the occupation outcome is
    $\boldsymbol\nu=(1,1,1)$, i.e., one photon at each site.

    \item \textbf{Read out the cyclic eigenvalues.}
    For each accepted three-photon state, measure the eigenspaces
    of the cyclic site permutation $C_{ABC}$ by measuring
    each photon in the three-site Fourier basis
   $\ket{u_s}
        =
        \frac{1}{\sqrt3}
        \left(
            \ket A+\omega^{-s}\ket B+\omega^{-2s}\ket C
        \right)$ 
    for $s=0,1,2$ using a three-port Fourier interferometer followed by
    photodetection.
    From the three output labels $s_1,s_2,s_3$, record $ q=(s_1+s_2+s_3)\bmod3$ and compute $ z=\omega^q$.
    Repeat these two steps using independent ACTs until $M_B$
    states have been accepted, yielding $M_B$ records
    $z_1,\ldots,z_{M_B}$.

    \item \textbf{Reconstruct the bispectrum and closure phase.}
    From the $M_B$ records, compute
    \begin{equation}
        \widetilde B_{ABC}
        =
        \frac{1}{M_B}\sum_{\ell=1}^{M_B}z_\ell,
        \qquad
        \widetilde\phi_{\mathrm{cl}}
        =
        \arg\widetilde B_{ABC}.
    \end{equation}
    If $\widetilde B_{ABC}=0$, report the phase as undetermined.
\end{enumerate}
}

In \cref{appendix:single_photon_limit_for_m=3}, we prove an upper bound for single-photon measurements' CFI and provide a scheme attaining it. 
In the remainder of this subsection, we present the main techniques to compute the QFI ($F_\mathrm{cl}$) and the SLD ($L_\mathrm{cl}$) corresponding to $\phicl$.

The fact that all blocks other than $M_{ABC}$ in $\rho_3^\mathrm{noisy}$ are independent of $\phi_\text{cl}$ implies that the overall QFI is simply $2/9$ times the QFI of $\rho_{ABC}$, so that we only need to compute the QFI and SLD corresponding to $\rho_{ABC}$.
Now, recall that the basis states of $\mathcal H_{ABC}$ [see \cref{eq:HABC}] are permutations of $\ket{ABC}$, they form the \emph{regular representation} of $S_3$.
By \cref{eq:dm_noisy_atmosphere}, $\rho_3^{\mathrm{noisy}}$ is an average of $\rho(\vec{\delta})^{\otimes 3}$ and is therefore invariant under permutations of the three photon registers. 
Now, letting $P_\pi$ be the $6\times6$ permutation matrix of $\pi\in S_3$ in $\mathcal{H}_{ABC}$, we have
\begin{equation}
    [P_\pi, M_{ABC}]= 0, \quad \forall \pi\in S_3.
\end{equation}
Note that the regular representation of $S_n$ can be decomposed into the direct sum of all irreducible representations (irreps). This implies that $P_\pi \in S_3$ is the direct sum of its operations in the trivial irrep, $1$, the sign irrep, $\mathrm{sgn}(\pi)$, and the standard irrep, $O_\mathrm{std}(\pi)\in \mathbb{C}^{2\times 2}$, i.e.,
\begin{equation}
    P_\pi = 1 \oplus \mathrm{sgn}(\pi) \oplus (I^{2\times 2}\otimes O_\mathrm{std}(\pi)).
\end{equation}
In other words, in the (nonabelian) Fourier basis of $S_3$, $P_\pi$ is block-diagonal for all $\pi\in S_3$.
Using the commutation relation between $M_{ABC}$ and $P_\pi$ and Schur's lemma, we have
\begin{equation}
\label{eq:MABC_block}
    M_{ABC} = p_\mathrm{trivial} \oplus p_\mathrm{sign} \oplus (M_\mathrm{std} \otimes I^{2\times 2}),
\end{equation}
where $p_\mathrm{trivial},p_\mathrm{sign} \in \mathbb{C}$, and $M_\mathrm{std}\in\mathbb{C}^{2\times 2}$ acts in the multiplicity space of the standard irrep. Therefore, in the Fourier basis of $S_3$, $M_{ABC}$ is a block-diagonal matrix containing two complex numbers and two identical $2\times 2$ Hermitian matrices (which can be seen as an unnormalized single-qubit density matrix).

Since the QFI (SLD) for block-diagonal density matrices is simply the sum (direct sum) of all blocks' individual QFIs (SLDs), the computation is significantly simplified---the only nonclassical step is computing the QFI and SLD of a single-qubit state $M_\mathrm{std}$.
We leave the remaining technical details of the computation to \cref{sec:details}, which computes $F_\mathrm{cl}$ in \cref{thm:single_joint_fisher}.

The optimal protocol to attain the QFI first uses nondestructive occupation-number measurement [see \cref{prot:occupation_number}] to postselect the outcome $\boldsymbol\nu=(1,1,1)$.
The resulting normalized state is $ \rho_{ABC} = \frac{9}{2}M_{ABC}$. Then, it measures $\rho_{ABC}$ in the eigenbasis of the SLD
$L_{\mathrm{cl}}$. Such a measurement requires a unitary to transform $\rho_{ABC}$ to its block-diagonal basis [see \cref{eq:MABC_block}], which is the Fourier basis of the $S_3$ group. This can be done using the finite-group quantum Fourier transform. We leave the exact form of the measurement basis to \cref{sec:details}. 

Note that the above SLD measurement saturates QFI universally for all possible visibility amplitudes and $\phicl$ values, and generally requires low-precision premeasurements to provide prior knowledge. In the low-visibility limit, it could be an overkill.
In the next subsection, we  instead present a conceptually and practically simpler measurement scheme to measure the complex bispectrum. It only saturates the QFI in the low-visibillity limit when it is used to only measure $\phicl$.

\subsection{Direct bispectrum measurement}
\label{sec:direct_bispectrum}
In this subsection, we propose a protocol to directly estimate the complex bispectrum $B_{ABC}=V_{AB}V_{BC}V_{CA} = |V_{AB}V_{BC}V_{CA}| \,e^{i\phi_{\rm cl}}$ from $\rho_3^\mathrm{noisy}$ in an $m=3$ telescope array. 
Also, it is simple to implement and will serve as the basic primitive for the general $m>3$ protocol in
\cref{sec:full_bispectra}. A generalization of this protocol to arbitrary $m,n$ is presented in \cref{sec:TorusQFT}.

This direct bispectrum estimation protocol uses the intuition that bispectrum generally appears in the off-diagonal element between a site string $\mathbf{i} $ and its \emph{cyclic permutation} [see \cref{eq:Bijk_oexample_in_3photon} for an example]. 
We start with the nondestructive multi-photon occupation-number measurement in \cref{prot:occupation_number}. Conditioned on the outcome $\boldsymbol\nu=(1,1,1)$, we apply the following operations to the $\rho_{ABC}$.
We define the \emph{cyclic site permutation} $\sigma_{ABC}$ 
as the following relabeling of the sites: $A\mapsto B\mapsto C\mapsto A$. We let $C_{ABC}$ denote the operation of $\sigma_{ABC}$ on the site strings, such that in the $\mathcal H_{ABC}$ subspace, $C_{ABC}$ acts as
\begin{equation}
\begin{aligned}
    \ket{ABC} & \mapsto\ket{BCA} \mapsto \ket{CAB}
     \mapsto \ket{ABC},\\
    \ket{ACB} & \mapsto \ket{BAC} \mapsto \ket{CBA}
    \mapsto \ket{ACB}.
\end{aligned}
\end{equation}
A direct calculation gives
\begin{equation}
    \Tr(\rho_{ABC} C_{ABC})
    = V_{AB}V_{BC}V_{CA} = B_{ABC}.
    \label{eq:bispectrum-cycle}
\end{equation}
Thus the complex bispectrum is the expectation value of the unitary $C_{ABC}$.

We estimate $B_{ABC}$ by measuring the projectors corresponding to the eigenspaces of $C_{ABC}$.
Since $C_{ABC}^3=I$, the outcome $q\in\{0,1,2\}$ labels an
eigenvalue $\omega^q$, where $\omega=e^{2\pi i/3}$, with projector
\begin{equation}
    P_q
    =
    \frac{1}{3}
    \left(
        I+\omega^{-q}C_{ABC}
        +\omega^{-2q}\left(C_{ABC}\right)^2
    \right).
    \label{eq:cyclicP}
\end{equation}

The standard quantum-computing style approach to perform this measurement is through the quantum phase estimation algorithm~\cite{Cleve1998QuantumAlgorithms}.
We prepare an auxiliary qutrit in
$(\ket{0}+\ket{1}+\ket{2})/\sqrt3$,
apply $\sum_{r=0}^2\ket{r}\bra{r}\otimes \left(C_{ABC}\right)^r$,
and then perform the inverse three-dimensional Fourier transform
on the qutrit before measuring it.
The measurement operator for outcome $q$ is precisely $P_q$.
The circuit resolves the fixed eigenphases of $C_{ABC}$; and the unknown closure phase is inferred from the outcome probabilities.

A fault-tolerant quantum algorithm might be an overkill because we need only measure according to $C_{ABC}$ rather than a general unitary. Indeed, using only single-photon Fourier-basis measurements after the occupation-sector postselection, one can also measure $q$.
Let $X$ be the operator that cyclically relabels the sites as $A\mapsto B\mapsto C \mapsto A$ for a single photon, we observe that $C_{ABC}$ is simply the restriction of $X^{\otimes3}$ to $\mathcal H_{ABC}$.
We can simply measure each photon individually in the eigenbasis of $X$,
\begin{equation}
    \ket{u_s}
    =
    \frac{1}{\sqrt3}
    \left(
        \ket A+\omega^{-s}\ket B+\omega^{-2s}\ket C
    \right),
    \quad s=0,1,2,
\end{equation}
using a three-port Fourier interferometer followed by
photodetection.
If the three outcomes are $s_1,s_2,s_3$, record
$q=(s_1+s_2+s_3)\bmod3$.
Since $X\ket{u_s}=\omega^s\ket{u_s}$, this coarse-grained
readout has exactly the outcome probabilities of $\{P_q\}$
on $\mathcal H_{ABC}$. This measurement requires only linear optics and photon counters. However, as mentioned before, one must first perform the collective measurement \cref{prot:occupation_number} to project the state to $\rho_{ABC}$, which makes the whole protocol a multi-photon scheme.

Next, if the outcome $q$ is recorded as the complex number $z=\omega^q$, then
\begin{equation}
    \mathbb{E}[z|ABC]
    =
    \Tr(\rho_{ABC} C_{ABC})
    =
    B_{ABC}.
    \label{eq:bispectrum-unbiased}
\end{equation}
Hence repeated measurements give an unbiased estimator of the full
complex bispectrum simply by averaging the observed values of $z$. More explicitly, the conditional outcome probabilities are
\begin{equation}
\begin{aligned}
    &\quad \,\, \Pr(q | ABC)\\
    &=
    \frac{1}{3}
    \left[
        1
        +2|V_{AB} V_{BC} V_{AC}|\cos\left(
            \phi_{\rm cl}-\frac{2\pi q}{3}
        \right)
    \right].
\end{aligned}
    \label{eq:cyclic-prob}
\end{equation}
Therefore, $B_{ABC}=|V_{AB} V_{BC} V_{AC}|e^{\imag \phi_{\rm cl}}$ can be inferred from the measured $q$ values conditioned on the $M_{ABC}$ block. Since the $M_{ABC}$ block occurs with probability $2/9$, the corresponding unconditional probabilities per three-photon ACT are $(2/9)\Pr(q|ABC)$.

We summarize this measurement scheme in \cref{prot:three_photon_single_bispectrum}.
Although this measurement is not exactly QFI-optimal for arbitrary $\phi_{\rm cl}$ at finite visibility, the gap vanishes rapidly in the weak-visibility regime. 
One can compute the classical Fisher information according to $(2/9)\Pr(q|ABC)$ and take
the weak-visibility limit:
\begin{equation}
\mathcal{I}_{\rm Protocol2}
=
\frac{4\Gamma^2}{9}
\left[
1+\Gamma\cos(3\phi_{\rm cl})
+O(\Gamma^2)
\right],
\end{equation}
where we define $\Gamma \coloneqq |V_{AB} V_{BC} V_{AC}|$. Indeed, it agrees with the QFI to leading order.

\section{Full bispectrum estimation in large arrays}
\label{sec:full_bispectra}

In this section, we turn to measuring all complex bispectra from an interferometric array with $m\geq 3$ telescopes. In \cref{sec:many_site_setup}, we introduce the setup of an astronomically relevant finite-precision bispectra estimation problem. Then in \cref{sec:measurement_primitives}, we present some basic measurement primitives using only two or three photons. In \cref{sec:compressibility}, we show the surprising fact that  all $O(m^2)$ bispectra can be compressed to $\tilde{O}(m)$ bits in the context of the finite-precision estimation problem.
Next, in \cref{sec:3_photon_m_sites}, we present an asymptotic separation (i.e.\ different asymptotic scaling with $m$) between $3$-photon collective measurements and single-photon measurements by providing a specific $3$-photon measurement plus classical data processing strategy to measure all bispectra with $\tilde{O}(m)$ photons.
Next, in \cref{sec:shadow}, we briefly review the shadow tomography approach which can in principle attain the information-theoretic lower bound. Finally, in \cref{sec:high_precision}, we investigate the asymptotically high precision regime in which Cramer-Rao bound type analysis is performed.

For the clarity of presentation, in this section, we use numbers ($1,2,3,\dots,m$) rather than letters ($A,B,C,\dots$) to label the telescope sites.

\subsection{Motivation and problem setup}
\label{sec:many_site_setup}
Most existing optical/infrared interferometers contain only a handful of telescopes: for example, the Very Large Telescope Interferometer (VLTI) combines $4$ telescopes at a time, while the Center for High Angular Resolution Astronomy (CHARA) array consists of $6$ telescopes~\cite{eisenhauer2023advances}. In contrast, radio interferometry is already moving toward much larger arrays; the Deep Synoptic Array, currently under construction, is designed to contain $1650$ dishes~\cite{otto2026electromagnetic,cohen2026optimal}. This motivates us to ask whether optical interferometry can, in principle, scale toward array sizes comparable to those being pursued in radio astronomy. Large arrays of smaller apertures are also of independent interest as an alternative architecture to relying on a few very large optical telescopes, whose construction becomes increasingly challenging and costly with aperture size~\cite{petrov2016cophasing}.

Increasing the number of telescopes is valuable for two complementary reasons. First, for similarly sized apertures, the total collecting area grows linearly with $m$. If each telescope receives on average $a$ photons per ACT window, then the total photon number per ACT scales as $n \simeq am$.
Second, an $m$-site array contains $\binom{m}{2} = \Theta(m^2)$ baselines, providing increasingly dense sampling of the spatial frequencies of the scene. 
Interferometric imaging in the noiseless case makes use of the full set of pairwise complex visibilities~\cite{Rogers1974,Monnier_2007}.
Even though the telescope array is subject to piston phases, the number of independent closure phases also grows quadratically with $m$~\cite{Monnier2003OpticalInterferometry,Blackburn2020ClosureStatistics}, as shown below. 

Observe that telescope-dependent piston phases create only $m-1$ independent nuisance degrees of freedom (it is not $m$ because a global phase is irrelevant). Consequently, the number of independent gauge-invariant
phase combinations is
\begin{equation}
\label{eq:number_of_closures}
    \binom{m}{2}-(m-1)
    =
    \binom{m-1}{2} = \Theta(m^2).
\end{equation}
Indeed, one can construct ${m-1 \choose 2}$ independent closure phases through counting independent triangles:
fixing site $1$, one convenient independent set is
\begin{equation}
\label{eq:cycleBasis}
    \{ \phi_{1i} + \phi_{ij} + \phi_{j1} \}_{2\leq i < j \leq m}.
\end{equation}

We further observe that since $ \binom{m-1}{2} / \binom{m}{2} = 1-\frac{2}{m} \rightarrow 1$,
almost all pairwise phase degrees of freedom become gauge-invariant in
the large-array limit.

Let $T=\{(i,j,k):1\leq i<j<k\leq m\}$ denote the set of telescope triangles, with $|T|=\binom{m}{3}=\Theta(m^3)$.
Note that the complete collection of complex bispectra contains all  $|T|$ triangle bispectra
$B_{ijk}=V_{ij}V_{jk}V_{ki}$. 
However, these $\binom{m}{3}$ complex numbers are highly redundant: their phases have only $\binom{m-1}{2}$ independent gauge-invariant combinations, as shown in \cref{eq:number_of_closures}.

Unlike the $m=3$ case where high-precision measurement of the only closure phase is a reasonable goal, for large $m$, estimating every closure phase and visibility amplitude individually to very high precision is not necessarily the most relevant operational goal, because different noise-robust quantities may play different roles in imaging and have different difficulties of measurements. One such example is that measuring the closure phase of $B_{ijk}$ corresponding to extremely small $|B_{ijk}|=|V_{ij}V_{jk}V_{ki}|$ is extremely difficult, as suggested by our QFI computation in \cref{thm:single_joint_fisher}.

Therefore, one natural questions one would ask is: what is the astronomically relevant task when doing robust interferometry with $m\gg 3$; or more specifically, how would astronomers use the closure phase or bispectrum data from an $m$-site interferometry array? The answer to this question is typically a model-fitting type task, in which closure phases (or bispectra) constrain various parameters of the scene, such as stellar shapes, companions, or circumstellar
structure~\cite{Monnier2007Altair,LeBouquinAbsil2012,Lazareff2017Disks}. More specifically, astronomers aim to compute certain parameters via fitting the full complex bispectra, denoted by $B^\mathrm{fitted} = \{ B^\mathrm{fitted}_t\}_{t\in T}$ such that its weighted mean-square error (WMSE) to the actual bispectra $B = \{ B\}_{t\in T}$ is smaller than some prescribed $\epsilon^2$. In other words, the goal is to ensure
\begin{equation}
\label{eq:goal2}
    L(B,B^\mathrm{fitted}) \leq \epsilon^2,
\end{equation}
where $L$ is the WMSE between two sets of bispectra:
\begin{equation}
\label{eq:bispectrum_loss}
    L(B^{(1)}, B^{(2)})
    \coloneqq
    \sum_{t\in T}
    w_t |B^{(1)}_t- B^{(2)}_t|^2
\end{equation}
with $w_t\geq0$ and $\sum_{t\in T}w_t=1$. 

In practical analyses, the weights $w_t$ may depend on the measurement covariance, array geometry, and the particular source or scientific target~\cite{blackburn2020closure,le2012sensitivity}. However, for simplicity, we focus on uniform weights ($w_t=1/|T|$ for all $t$) in this work. In other words, instead of WMSE, we work with mean-square error (MSE). 
We claim that this assumption is not essential when the weights are moderately nonuniform: if $w_t\leq \zeta /|T|$ for all $t$, with $\zeta =O(1)$
independent of $m$, then the weighted loss is at most $\zeta$ times
the uniform loss. Hence such bounded variations of the weights do not change any of our asymptotic error or sample complexity scalings with $m$. Strongly nonuniform weights, which may arise from a particular
array geometry, noise model, or scientific target, can be treated
as a task-specific extension.

In practice, we let $\widetilde B=\{\tilde B_t\}_{t\in T}$ denote the estimated bispectra.
We can measure the quality of  $\tilde B$ via $L(B,\tilde{B})$. Since
$\sqrt{L}$ is a Euclidean distance, if
\begin{equation}
\label{eq:astro_need}
    L(B,\widetilde B)\leq \frac{\epsilon^2}{4},
    \qquad
    L(\widetilde B,B^{\rm fitted})\leq \frac{\epsilon^2}{4},
\end{equation}
then the triangle inequality automatically gives
$L(B,B^{\rm fitted})\leq \epsilon^2$.
Therefore, to ensure \cref{eq:goal2} is satisfied, it suffices to estimate the bispectra to the prescribed
MSE before performing source-model fitting. 

With the goal now well established, the remaining task is to determine the minimum number of photons required to achieve \cref{eq:goal2} and the optimal measurement strategy for attaining this limit.

\begin{problem}
What is the number of photons required for estimating all bispectra of an $m$-telescope array to a fixed weighted mean-square accuracy $L(B,\widetilde B)\leq\epsilon^2$? Moreover, how does the photon number scale with $m$ when using single-photon measurements and multi-photon measurements? Do collective measurements offer any scaling advantage?
\end{problem}

A key distinction from the previous high-precision single-parameter analysis is that here $\epsilon^2$ is regarded as a fixed target accuracy rather than an asymptotically small parameter. Local asymptotic tools such as the Cram\'er--Rao bound therefore do not by themselves characterize the finite-accuracy photon complexity of this problem. We instead study the problem directly from information-theoretic and finite-sample perspectives. Before doing so, we first introduce several measurement techniques as primitives.

\subsection{Measurement primitives}
\label{sec:measurement_primitives}

Our large-array measurement protocol uses the following two kinds of records: two-photon pair records and three-photon triangle records, generated by the following mesurement primitives.

\paragraph{Two-photon pair record.}
Consider $\rho_2^\mathrm{noisy}$.
We first measure its occupation numbers using \cref{prot:occupation_number}. Conditioned on two photons occupying distinct sites $i$ and $j$, their normalized state is supported on $\{\ket{ij},\ket{ji}\}$. We measure the SWAP operator
\begin{equation}
    S_{ij}
    =
    \ket{ij}\!\bra{ji}
    +
    \ket{ji}\!\bra{ij},
\end{equation}
whose outcome satisfies
\begin{equation}
    s\in\{-1,1\},
    \qquad \E[s|\{i,j\}]=|V_{ij}|^2.
    \label{eq:pair_record}
\end{equation}
This gives the pair record $(s,\{i,j\})$. Conditioned on two distinct occupied sites, the pair is uniformly sampled from all unordered pairs.

\paragraph{Three-photon triangle record.}
Consider $\rho^\mathrm{noisy}_3$.
The cyclic measurement of \cref{sec:direct_bispectrum} [see also \cref{prot:three_photon_single_bispectrum}] can be generalized to larger arrays. For any triangle $t=(i,j,k)$, $i<j<k$, we can define $C_{ijk}$  ($=C_{t}$) acting in its postselected 
one-photon-per-site block as the following operation:
\begin{equation}
\begin{aligned}
    \ket{ijk}\mapsto\ket{jki} \mapsto \ket{kij}
     \mapsto \ket{ijk},\\
    \ket{ikj} \mapsto \ket{jik} \mapsto \ket{kji}
    \mapsto \ket{ikj}.
\end{aligned}
\end{equation}
Measuring in the eigenbasis of $C_t$ gives an outcome [see \cref{eq:bispectrum-cycle}]
\begin{equation}
    z\in\{1,\omega,\omega^2\},
    \qquad \mathbb E[z | t]=B_t.
    \label{eq:triangle_record}
\end{equation}
Thus each accepted event gives a classical record $(t,z)$ for estimating the bispectrum of the observed triangle. Conditioned on three distinct occupied sites, $t$ is uniformly distributed over all $\binom m3$ triangles.

\subsection{Compressibility of full bispectra}
\label{sec:compressibility}
Since $\epsilon$ is typically set by the constant target accuracy, we consider the asymptotic regime of large $m$, with the goal of determining the optimal photon-number scaling with $m$.

Before designing a measurement protocol, we ask how much classical information is required to specify the full collection of bispectra within $\epsilon^2$ MSE, as this suggests an information-theoretically possible number of photons to measure the complete set of all bispectra.

Let $N_{B,\epsilon}$ denote the minimum number of classical bits needed, in the worst case, to represent a physically allowed collection of bispectra $B$ within MSE $\epsilon^2$.
Naively, one might expect $N_{B,\epsilon}$ to scale as
$O(m^2\log(1/\epsilon))$, since a general visibility matrix contains
$\Theta(m^2)$ real degrees of freedom after removing the telescope
phase gauge and redundant closure phases.  This estimate, however, ignores the positive-semidefinite
constraint on physical visibility matrices.  Surprisingly, this constraint makes the set of physically realizable bispectra much more compressible.

\begin{theorem}[Compression of the full bispectra]
\label{thm:number_of_bits}
The collection of all physically allowed bispectra can
be represented to MSE $\epsilon^2$ using $\widetilde O(m/\epsilon^2)$
classical bits.
\end{theorem}

\begin{proof}[Proof]
Without loss of generality, we choose telescope site $1$ as a reference and gauge-transform the visibility
matrix so that all baselines from telescope $1$ have zero phase.  In other words, one can always find a suitable diagonal unitary $D$ implementing the gauge transform such that
\begin{equation}
    R \coloneqq D V D^\dagger
\end{equation}
satisfies that $R_{1i}=|V_{1i}|$ for all $i\geq2$.
$R/m$ is still a valid density matrix, so $R\succeq0$, $R_{ii}=1$, and the gauge transformation leaves every
bispectrum invariant:
\begin{equation}
    B_{ijk}
    =
    V_{ij}V_{jk}V_{ki}
    =
    R_{ij}R_{jk}R_{ki}.
\end{equation}
Equivalently, the phase of $R_{ij}$ is the closure phase on the
triangle $(1,i,j)$.

Since $R$ is a positive semidefinite matrix with unit diagonal, it is also a Gram matrix: there exist unit vectors $v_1,\ldots,v_m\in\mathbb C^m$ such that $R_{ij}=\langle v_i,v_j\rangle$. The Johnson--Lindenstrauss lemma~\cite{johnson1984extensions,dasgupta2003elementary} allows these $m$ vectors to be compressed to dimension $O(\log m/\epsilon^2)$ while preserving all pairwise inner products to additive error $O(\epsilon)$.
Because each bispectrum is the product of three such inner products and all of them have magnitude at most one, this also preserves every bispectrum to additive error $O(\epsilon)$, which is a sufficient condition to ensure the whole collection of bispectra has $O(\epsilon^2)$ MSE.
Representing the $m$ compressed vectors to finite precision therefore requires only $\widetilde O(m/\epsilon^2)$ bits. 
This proves the upper bound on the number of bits describing the whole collection of bispectra.
\end{proof}

The significance of \cref{thm:number_of_bits} is that the apparent $\Theta(m^2)$ number of independent visibility parameters does not translate into $\Theta(m^2)$ information at fixed global accuracy.
Under the MSE criterion, the physically allowed collections of bispectra occupy a much smaller effective set whose description length grows only nearly linearly with $m$.

The theorem above is only a compression statement: it shows that a short description of the bispectra exists but does not by itself give a measurement that recovers this description from photons.  We next show that a closely related low-dimensional representation of all bispectra can be learned using $O(m/\epsilon^4 + 1/\epsilon^8)$ photons via collective measurements on at most three photons, while single-photon measurements cost at least $\Omega(m^2/\epsilon^2)$ photons in the worst case.

\protocol{Three-photon collective measurements for full bispectra}{Our three-photon collective measurements for all bispectra.}{prot:three_photon_full_bisepctra}
{
Choose $\tau,\lambda=\Theta(\epsilon^2)$ and target accepted-sample
counts $M_H=\widetilde O(m/\epsilon^4)$ and $M_B=O(\epsilon^{-8})$.

\begin{enumerate}
    \item \textbf{Collect and postselect two-photon states.}
    Collect two photons from the same ACT
    and store them in quantum memory, obtaining a copy of
    $\rho_2^{\mathrm{noisy}}$. Then, perform \cref{prot:occupation_number}: at each telescope, nondestructively measure the total photon number across the two stored temporal modes, without resolving individual modes. Keep the state only if the photons occupy two distinct sites, and record the occupied pair as $\{i,j\}$ with $i<j$.

    \item \textbf{Estimate the squared-visibility matrix.}
    For each accepted two-photon state on sites $i,j$, measure
    $ S_{ij} = \ket{ij}\!\bra{ji} + \ket{ji}\!\bra{ij},$ and record the outcome $s\in\{-1,1\}$. Repeat until $M_H$ states have been accepted. From the $M_H$ records $(\{i_\ell,j_\ell\},s_\ell)$, compute
    \begin{equation}
        \widehat H
        =
        I_m+\frac{\binom m2}{M_H}
        \sum_{\ell=1}^{M_H}s_\ell
        \left(
            \ket{i_\ell}\bra{j_\ell}
            +
            \ket{j_\ell}\bra{i_\ell}
        \right).
    \end{equation}

    \item \textbf{Construct the shared features.}
    Diagonalize $\widehat H$ and collect its orthonormal
    eigenvectors with eigenvalues at least $2\tau m$ as the
    columns of $W \in\mathbb R^{m\times r}$.
    For every triangle $t=(i,j,k)$ with $i<j<k$, define
    \begin{equation}
        f_t[\alpha,\beta,\gamma]
        =
        m^{3/2}W_{i\alpha}W_{j\beta}W_{k\gamma},
        \qquad \alpha,\beta,\gamma=1,\ldots,r.
    \end{equation}

    \item \textbf{Collect and postselect three-photon states.}
    Collect
    $\rho_3^{\mathrm{noisy}}$, each consisting of three stored
    photons from the same ACT.
    Then, for each copy, perform \cref{prot:occupation_number}: nondestructively measure the total photon
    number at each telescope across the three stored temporal
    modes, without resolving individual time stamps.
    Keep the state only if the photons occupy three distinct
    sites, and record the occupied triangle as $t=(i,j,k)$ with $i<j<k$.

    \item \textbf{Read out the cyclic eigenvalues.}
    For each accepted three-photon state on triangle $t=(i,j,k)$,
    measure the eigenspaces of the cyclic site permutation
    $C_{ijk}$ and record its eigenvalue
    $z\in\{1,\omega,\omega^2\}$, where
    $\omega=e^{2\pi\imag/3}$.
    Repeat until $M_B$ states have been accepted, yielding $M_B$ triangle records $(t_\ell,z_\ell)$.

    \item \textbf{Reconstruct all bispectra.}
    Compute
    \begin{equation}
        G=\frac{1}{|T|}\sum_{t\in T}f_tf_t^T,
        \qquad
        \widehat c=\frac{1}{M_B}
        \sum_{\ell=1}^{M_B}f_{t_\ell}z_\ell,
    \end{equation}
    solve $(G+\lambda I)\widehat\theta=\widehat c$ for
    $\widehat\theta$, and output
    \begin{equation}
        \widetilde B_t
        =
        \frac{f_t^T\widehat\theta}
        {\max\{1,|f_t^T\widehat\theta|\}},
        \qquad t\in T.
    \end{equation}
\end{enumerate}
}

\subsection{Asymptotic advantage of three-photon collective measurements}
\label{sec:3_photon_m_sites}

In this subsection, we focus on using up to three-photon collective measurements to estimate the whole bispectra within MSE $\epsilon^2$. Our main result is the following theorem that separates photon-by-photon measurements from collective measurements, and a concrete collective measurement plus classical data processing scheme (\cref{prot:three_photon_full_bisepctra}) that uses only $\tilde{O}(m)$ photons.
\begin{theorem}[Asymptotic advantage of three-photon joint measurements
at fixed bispectrum precision]
\label{thm:joint_3_m_scaling}
Suppose each ACT contains exactly $n=3$ photons. 
For $0<\epsilon\le 0.1$, 
Consider estimating all bispectra with the expected MSE, i.e., $\E[L(B,\tilde{B})]\leq \epsilon^2$.
Joint measurements on at most three photons achieve this with $\widetilde O(m/\epsilon^4+\epsilon^{-8})$ input photons and $\poly(m,1/\epsilon)$-time classical computation. Any sequential single-photon protocol that achieves the same MSE requires $\Omega(m^2/\epsilon^2)$ photons in the worst case.
\end{theorem}
We prove the $\tilde{O}(m)$ upper bound for collective measurements by constructing a concrete protocol to achieve it. A key idea is that the $\binom m3$ triangular bispectra (only $\Theta(m^2)$ of them are independent) should not be learned independently. Instead, we use the same intuition as in \cref{thm:number_of_bits} that the full collection of bispectra may be compressible. Indeed, we use the \emph{squared-visibility matrix} $H\in\mathbb{R}^{m\times m}$ (defined as $H_{ij}=|V_{ij}|^2$) as a proxy to show that:
once the dominant low-dimensional eigenspace of $H$ is known, all complex bispectra $B_{ijk}$ admit a common low-dimensional representation corresponding to the dominant eigenspace of $H$. 
Therefore, to learn the full collection of bispectra (compressed to $\epsilon^2$ MSE), we first learn the large eigenvectors of $H$ through measuring $\tilde{O}(m/\epsilon^4)$ copies of $\rho_2^\mathrm{noisy}$; then, we collect the measurement outcomes of $O(1/\epsilon^8)$ copies of $\rho_3^\mathrm{noisy}$ and find the low-dimensional parameters generating a collection of bispectra that fit the outcomes best through regression. Note that the data fitting algorithm is classical, but only two- and three-photon collective measurements provide the raw data with the requried quality and efficiency.

We provide a roadmap of the construction of \cref{prot:three_photon_full_bisepctra} and the proof of \cref{thm:joint_3_m_scaling}: we first show the dominant eigenspace of $H$ is low-dimensional in \cref{lem:dominant_eigenspace_dimension}; we then construct the explicit low-dimensional representation of $B$ through \cref{lem:bispectrum_low_rank}; then, we show how to learn the dominant eigenspace of $H$ via \cref{lem:learn_H}; next, we give the algorithm to learn the low-dimensional representation of $B$ via \cref{lem:learn_theta}; finally, combing these results together, we can show the number of two- and three-photon samples required to achieve the $\epsilon^2$ MSE requirement is $\tilde{O}(m/\epsilon^4 + 1/\epsilon^8)$, proving \cref{thm:joint_3_m_scaling}.

\begin{lemma}[Dimension of the dominant eigenspace]
\label{lem:dominant_eigenspace_dimension}
Fix $0<\tau\leq 1$, and let
$u_1,\ldots,u_r \in\mathbb R^{m}$ be the orthonormal eigenvectors of $H$ whose eigenvalues are at least $\tau m$. Then
$r\leq 1/\tau$.     
\end{lemma}
The short proof, including why $H\succeq0$ and $\Tr H=m$, is given in
\cref{app:dominant_eigenspace_dimension}.

We let $\mathrm{span}(u_1,\dots,u_r)$ be the \emph{dominant eigenspace} and define $W=(u_1,\dots,u_r) \in \mathbb{R}^{m\times r}$.

\begin{lemma}[Low-dimensional representation of the bispectra]
\label{lem:bispectrum_low_rank}
There exist a vector
$\theta\in\mathbb C^{r^3}$ with $\|\theta\|_2\leq 1$ and vectors
$f_{ijk}\in\mathbb R^{r^3}$ defined as \begin{equation}
    f_{ijk}[\alpha,\beta,\gamma]
    =
    m^{3/2}W_{i\alpha}W_{j\beta}W_{k\gamma}
    \label{eq:bispectrum_feature}
\end{equation}
for $\alpha,\beta,\gamma\in\{1,\ldots,r\}$, such that
\begin{equation}
    \frac{1}{\binom m3}
    \sum_{i<j<k}
    \left|B_{ijk}-f_{ijk}^T\theta\right|^2
    =O(\tau).
    \label{eq:low_rank_bispectrum_error}
\end{equation}
\end{lemma}

\begin{proof}
To obtain the low-dimensional representation, write the visibility
matrix as a Gram matrix, $V_{ij}=\langle g_i,g_j\rangle$, and define
$Q_i=\ket{g_i}\!\bra{g_i}$. Then
\begin{equation}
    H_{ij}=\Tr(Q_iQ_j),
    \qquad
    B_{ijk}=\Tr(Q_iQ_jQ_k).
    \label{eq:H_B_Q}
\end{equation}
For convenience, we extend $B_{uvw}=\Tr(Q_uQ_vQ_w)$ to all ordered triples, including repeated indices.

The main proof idea is to project every $Q_j$ onto the dominant eigenspace of $H$, and show that the projected version can reconstruct $B_{ijk}$ with small error. Let $P=WW^T$ be the projector onto the dominant eigenspace. 
We project
the dependence of the operators $Q_i$ on the telescope index by defining
\begin{equation}
    Q_i^P
    \coloneqq
    \sum_{u=1}^m P_{iu}Q_u
    =
    \sum_{\alpha=1}^r\sum_{u=1}^m
    W_{i\alpha}W_{u\alpha}Q_u .
    \label{eq:projected_Q}
\end{equation}
Expanding the trace gives
\begin{align}
    &\quad \, \Tr(Q_i^P Q_j^P Q_k^P)\\
    &=
    \sum_{\alpha,\beta,\gamma=1}^r
    W_{i\alpha} W_{j\beta} W_{k\gamma}
    \sum_{u,v,w=1}^m
    W_{u\alpha} W_{v\beta} W_{w\gamma}
    B_{uvw}.
\end{align}
Now define
\begin{equation}
    \theta[\alpha,\beta,\gamma]
    \coloneqq
    \frac{1}{m^{3/2}}
    \sum_{u,v,w=1}^m
    W_{u\alpha} W_{v\beta} W_{w\gamma}B_{uvw}.
    \label{eq:theta_def}
\end{equation}
Together with \cref{eq:bispectrum_feature}, this gives
\begin{equation}
\begin{aligned}
\Tr(Q_i^P Q_j^P Q_k^P) &= f_{ijk}^T\theta \\
&\equiv \sum_{\alpha,\beta,\gamma=1}^r f_{ijk}[\alpha,\beta,\gamma]\, \theta[\alpha\beta\gamma]    
\end{aligned}
\label{eq:projected_bispectrum_feature}
\end{equation}
as the low-rank approximation of $B_{ijk}$.

The normalization in \cref{eq:theta_def} ensures
$\|\theta\|_2\leq1$. Indeed, applying $W^T$ to each of the three
indices cannot increase the Frobenius norm because the columns of $W$
are orthonormal. Hence
\begin{equation}
    \|\theta\|_2^2
    \leq
    \frac{1}{m^3}\sum_{u,v,w=1}^m |B_{uvw}|^2
    \leq 1,
\end{equation}
where we used $|B_{uvw}|\leq1$.

It remains to bound the error introduced by the projection. We first
show that it suffices to control the error from projecting one tensor
index at a time.

Regard $B$ as a tensor over all ordered triples, and let
$\mathcal P_\ell$ denote the orthogonal projection that applies $P$
to its $\ell$th index. For example,
\begin{equation}
    (\mathcal P_1 B)_{ijk}
    =
    \sum_u P_{iu}B_{ujk}.
\end{equation}
The fully projected tensor satisfies
$(\mathcal P_1\mathcal P_2\mathcal P_3B)_{ijk}
=\Tr(Q_i^P Q_j^P Q_k^P)$.
Adding and subtracting the intermediate projections gives
\begin{equation}
    \begin{aligned}
        B-\mathcal P_1\mathcal P_2\mathcal P_3B
        &=
        (I-\mathcal P_1)B \\
        &\quad+
        \mathcal P_1(I-\mathcal P_2)B \\
        &\quad+
        \mathcal P_1\mathcal P_2(I-\mathcal P_3)B.
    \end{aligned}
\end{equation}
These three terms are the errors incurred when projecting the first,
second, and third indices successively. Applying
$\|X+Y+Z\|_F^2
\leq 3(\|X\|_F^2+\|Y\|_F^2+\|Z\|_F^2)$
and using that orthogonal projections cannot increase the Frobenius
norm, we obtain
\begin{equation}
    \left\|
        B-\mathcal P_1\mathcal P_2\mathcal P_3B
    \right\|_F^2
    \leq
    3\sum_{\ell=1}^3
    \left\|(I-\mathcal P_\ell)B\right\|_F^2.
    \label{eq:projection_error_reduction}
\end{equation}
Thus the total error is controlled by the three errors obtained by
projecting each index of the original tensor separately.

We now bound the first-index error. Define
$R_i\coloneqq Q_i-Q_i^P$.
Since $R_i=\sum_u(I-P)_{iu}Q_u$ and $P$ is real and symmetric,
the $R_i$ are Hermitian and their Hilbert--Schmidt Gram matrix is
\begin{equation}
    \Tr(R_i^\dagger R_j)
    =
    \big[(I-P)H(I-P)\big]_{ij},
\end{equation}
where we used $\Tr(Q_uQ_v)=H_{uv}$.
Since $P$ contains all eigenvectors of $H$ with eigenvalues at least
$\tau m$,
\begin{equation}
    0\preceq(I-P)H(I-P)\preceq\tau m I.
    \label{eq:H_residual_exact}
\end{equation}

The first-index error has entries
$[(I-\mathcal P_1)B]_{ijk}=\Tr(R_iQ_jQ_k)$.
For fixed $j,k$, \cref{lem:gram_trace_bound}, proved in
\cref{app:gram_trace_bound}, converts the Gram-matrix bound
\cref{eq:H_residual_exact} into
\begin{equation}
    \begin{aligned}
        \sum_i
        \left|
            [(I-\mathcal P_1)B]_{ijk}
        \right|^2
        &=
        \sum_i |\Tr(R_iQ_jQ_k)|^2 \\
        &\leq
        \tau m\|Q_jQ_k\|_F^2 \\
        &=
        \tau m H_{jk}
        \leq \tau m.
    \end{aligned}
\end{equation}
Here we use $\|Q_jQ_k\|_F^2=H_{jk}\leq1$, since the $Q_i$s are rank-one projectors onto unit vectors. Summing over $j,k$ gives $\|(I-\mathcal P_1)B\|_F^2\leq\tau m^3$. By cyclicity of the trace, the same argument applies to the other two indices. Substituting these three bounds into \cref{eq:projection_error_reduction} therefore yields
\begin{equation}
    \sum_{i,j,k}
    \left|
        B_{ijk}-\Tr(Q_i^P Q_j^P Q_k^P)
    \right|^2
    \leq9\tau m^3.
    \label{eq:bispectrum_projection_error}
\end{equation}

For distinct $i,j,k$, cyclic permutations leave both quantities
unchanged and reversing the order complex conjugates them. Hence the
six orderings of each triangle have the same squared error. Since
$p_{\rm dist}\coloneqq6\binom m3/m^3$, we obtain
\begin{equation}
    \frac{1}{\binom m3}
    \sum_{i<j<k}
    |B_{ijk}-f_{ijk}^T\theta|^2
    \leq
    \frac{9\tau}{p_{\rm dist}}
    =O(\tau).
\end{equation}
\end{proof}

The proof above uses only two properties of $P$: its rank is
$O(1/\tau)$, and the residual matrix $(I-P)H(I-P)$ has operator norm $O(\tau m)$. We next show that the dominant eigenspace can be
learned efficiently from two-photon collective measurements.

\begin{lemma}[Learning the dominant eigenspace]
\label{lem:learn_H}
For $0<\tau\leq1$ and 
$0 < \eta < 1$, 
$M_H=O\!\left(m\tau^{-2}\log(m/\eta)\right)$ two-photon pair records suffice to construct an estimate $\widehat H$ satisfying
\begin{equation}
    \|\widehat H-H\|_{\rm op}\leq\tau m
\end{equation}
with probability at least $1-\eta$. If $U$ contains the eigenvectors of
$\widehat H$ with eigenvalues at least $2\tau m$ and $P=UU^T$, then
\begin{equation}
    r\leq\frac1\tau,
    \qquad
    0\preceq(I-P)H(I-P)\preceq3\tau m I.
    \label{eq:estimated_H_residual}
\end{equation}
\end{lemma}

\begin{proof}
We use the pair records of \cref{sec:measurement_primitives}, which satisfy $\E[s|\{i,j\}]=H_{ij}$ and sample unordered pairs uniformly.

Let $q=\binom m2$ and $A_{ij}=\ket{i}\!\bra{j}+\ket{j}\!\bra{i}$. The random matrix $Y=I_m+qsA_{ij}$ satisfies $\E Y=H$. Set $\widehat H = \left(\sum_{\ell=1}^{M_H}Y_\ell\right)/M_H$. Applying matrix Bernstein inequality to independent copies of $Y$, we conclude that $M_H=O(m\tau^{-2}\log(m/\eta))$ records suffice for $\|\widehat H-H\|_{\rm op}\leq\tau m$. We present more detailed second-moment calculation and concentration bound calculation in \cref{app:matrix_bernstein}.

It remains to check that the estimated dominant eigenspace obtained from $\widehat H$ has the properties required above. For every eigenvector $u_\alpha$ in this dominant eigenspace,
\begin{equation}
    u_\alpha^T H u_\alpha
    \geq
    u_\alpha^T\widehat H u_\alpha-\tau m
    \geq\tau m.    
\end{equation}
Summing over its $r$ orthonormal eigenvectors gives
$r\tau m\leq\Tr(PH)\leq\Tr H=m$, and hence $r\leq1/\tau$.

On the other hand, for every unit vector $v$ orthogonal to the dominant
eigenspace,
$v^T\widehat Hv<2\tau m$, and therefore
$v^THv\leq3\tau m$. Since $H\succeq0$, this proves
\cref{eq:estimated_H_residual}.
\end{proof}

The proof of Lemma~\ref{lem:bispectrum_low_rank} depends on $P$ only
through the two properties stated above. Hence Lemma~\ref{lem:learn_H}
implies that, conditioned on successful estimation of $H$, there are
known vectors $f_t$ and a vector
$\theta\in\mathbb C^{r^3}$ with $\|\theta\|_2\leq1$ satisfying
\begin{equation}
    L(B,f^T \theta) = \frac{1}{\binom m3}\sum_t|B_t-f_t^T\theta|^2=O(\tau).
    \label{eq:estimated_feature_approx}
\end{equation}
We now only need to learn the $r^3$ coefficients of $\theta$.

\begin{lemma}[Learning the shared coefficients]
\label{lem:learn_theta}
Suppose the vectors $f_t\in\mathbb R^{r^3}$ for all $t\in T$ are given, and we use \cref{sec:measurement_primitives} to produce three-photon measurement records $(t,z)$ where $t$ is a uniformly sampled triangle and $z$ is a complex outcome satisfying $|z|=1$ and $\E[z|t]=B_t$.
For any $\lambda>0$ and integer $M_B\geq1$, there exists an
estimator $\widetilde B$, based on the given $f_t$ vectors and $M_B$
independent accepted records, satisfying
\begin{equation}
    \E L(B,\widetilde B)
    \leq
    L(B,f^T\theta)
    +
    \lambda\|\theta\|_2^2
    +
    \frac{r^3}{M_B}
    \label{eq:regression_summary}
\end{equation}
for every $\theta\in\mathbb C^{r^3}$.
The expectation is over the measurement records.
\end{lemma}
\begin{proof}
We notice that each measurement record $(t_\ell,z_\ell)$ gives a noisy observation of $B_{t_\ell}$, since $\E[z_\ell | t_\ell]=B_{t_\ell}$.
Rather than estimate each bispectrum separately, we use these
records to fit a common coefficient vector $\theta$, so that
$f_t^T\theta$ predicts the bispectrum of every triangle.

A natural starting point is least-squares fitting: choose $\theta$
to minimize the squared differences between the predictions and
the observed values,
\begin{equation}
    \frac{1}{M_B}\sum_{\ell=1}^{M_B}
    |z_\ell-f_{t_\ell}^T\theta|^2.
\end{equation}
The minimizing coefficients satisfy the linear equations
\begin{equation}
    \left(
        \frac{1}{M_B}\sum_{\ell=1}^{M_B}
        f_{t_\ell}f_{t_\ell}^T
    \right)\theta
    =
    \frac{1}{M_B}\sum_{\ell=1}^{M_B}f_{t_\ell}z_\ell.
\end{equation}
The matrix on the left depends only on the sampled triangles,
not on the measurement outcomes.
Since all vectors $f_t$ are known and triangles are sampled
uniformly, we replace this sample average by its exactly known
mean. Define
\begin{equation}
    G=\frac{1}{|T|}\sum_{t\in T}f_tf_t^T,
    \qquad
    \widehat c
    =
    \frac{1}{M_B}\sum_{\ell=1}^{M_B}f_{t_\ell}z_\ell.
\end{equation}
This suggests that one can estimate the coefficients $\theta$ by solving
$G\theta=\widehat c$.

Directly inverting $G$ may be problematic: it can be singular,
and small eigenvalues can make the coefficients sensitive to
errors in $\widehat c$.
To stabilize the solution, we add a penalty
$\lambda\|\theta\|_2^2$ to the modified fitting objective and
minimize
\begin{equation}
    \theta^\dagger G\theta
    -2\operatorname{Re}(\theta^\dagger\widehat c)
    +\lambda\|\theta\|_2^2.
\end{equation}
The penalty discourages large coefficients and replaces $G$
in the fitting equation by $G+\lambda I$.
For $\lambda>0$, the unique solution is therefore
\begin{equation}
    \widehat\theta=(G+\lambda I)^{-1}\widehat c.
\end{equation}
This is a variant of ridge regression~\cite{hoerl1970ridge}
using the exactly known matrix $G$.
We predict each bispectrum by $f_t^T\widehat\theta$,
with projection onto the unit disk at the end.

For the error analysis, note that $\widehat c$ is an unbiased
estimate of
\begin{equation}
    c\coloneqq\E[\widehat c]
    =
    \frac{1}{|T|}\sum_{t\in T}f_tB_t.
\end{equation}
We compare $\widehat\theta$ with its mean,
\begin{equation}
    \theta_\lambda
    \coloneqq
    \E[\widehat\theta]
    =
    (G+\lambda I)^{-1}c,
\end{equation}
and write $e=\widehat\theta-\theta_\lambda$ for its random
fluctuation. Since $\E[e]=0$, expanding the squared prediction
error gives
\begin{equation}
    \E L(B,f^T\widehat\theta)
    =
    L(B,f^T\theta_\lambda)
    +
    \E[e^\dagger G e].
    \label{eq:regression_error_decomposition}
\end{equation}
The first term is the prediction error of the mean coefficient
vector; the second is the additional error caused by finite
sampling. We bound these two terms separately, as in standard
ridge-regression analyses~\cite{hsu2014random}.

For the first term, expanding the squared loss shows that
$\theta_\lambda$ minimizes
$L(B,f^T\theta)+\lambda\|\theta\|_2^2$.
Consequently, for every $\theta\in\mathbb C^{r^3}$,
\begin{equation}
    L(B,f^T\theta_\lambda)
    +
    \lambda\|\theta_\lambda\|_2^2
    \leq
    L(B,f^T\theta)
    +
    \lambda\|\theta\|_2^2.
\end{equation}
Dropping the nonnegative term
$\lambda\|\theta_\lambda\|_2^2$ on the left gives
\begin{equation}
    L(B,f^T\theta_\lambda)
    \leq
    L(B,f^T\theta)
    +
    \lambda\|\theta\|_2^2.
\end{equation}
Thus the mean prediction performs nearly as well as any
coefficient vector $\theta$, with an additional error allowance
of $\lambda\|\theta\|_2^2$.

For the second term, independence of the measurement records
and $|z_\ell|=1$ give
\begin{equation}
    \E\!\left[
        (\widehat c-c)(\widehat c-c)^\dagger
    \right]
    =
    \frac{G-cc^\dagger}{M_B}
    \preceq
    \frac{G}{M_B}.
\end{equation}
Since $e=(G+\lambda I)^{-1}(\widehat c-c)$, we obtain
\begin{equation}
    \begin{aligned}
        \E[e^\dagger G e]
        &\leq
        \frac{1}{M_B}
        \Tr\!\left[
            \bigl(G(G+\lambda I)^{-1}\bigr)^2
        \right] \\
        &=
        \frac{1}{M_B}
        \sum_{a=1}^{r^3}
        \left(
            \frac{\mu_a}{\mu_a+\lambda}
        \right)^2 \\
        &\leq
        \frac{r^3}{M_B},
    \end{aligned}
\end{equation}
where $\mu_1,\ldots,\mu_{r^3}\geq0$ are the eigenvalues of $G$.
Each eigenvector direction therefore contributes at most
$1/M_B$ to the additional mean-square prediction error.
Combining the two bounds with
\cref{eq:regression_error_decomposition} yields
\begin{equation}
    \E L(B,f^T\widehat\theta)
    \leq
    L(B,f^T\theta)
    +
    \lambda\|\theta\|_2^2
    +
    \frac{r^3}{M_B}.
\end{equation}

Finally, define the estimator by projecting each prediction
onto the unit disk:
\begin{equation}
    \widetilde B_t
    =
    \frac{f_t^T\widehat\theta}
    {\max\{1,|f_t^T\widehat\theta|\}}.
\end{equation}
This leaves predictions of modulus at most one unchanged.
Since $|B_t|\leq1$, the projection cannot increase the error,
which proves \cref{eq:regression_summary}.
\end{proof}

\begin{proof}[Proof of \cref{thm:joint_3_m_scaling}]
We now combine the two measurements. Choose
$\tau=\Theta(\epsilon^2)$ and use Lemma~\ref{lem:learn_H} to learn the
dominant eigenspace of $H$. 
The rank safeguard is never triggered when the
spectral estimate succeeds. With probability at least $1-\eta$, we therefore
have $r=O(\epsilon^{-2})$, and
\cref{eq:estimated_feature_approx} gives a representation satisfying
$L(B,f^T\theta)=O(\epsilon^2)$ with $\|\theta\|_2\leq1$.

We then apply Lemma~\ref{lem:learn_theta} with
$\lambda=\Theta(\epsilon^2)$ and
$M_B=O(r^3/\epsilon^2)$. Since $r=O(\epsilon^{-2})$, the two stages use
\begin{equation}
    M_H=\widetilde O(m/\epsilon^4),
    \qquad
    M_B=O(\epsilon^{-8})
\end{equation}
accepted records, respectively.

Finally, the probabilities that two photons occupy distinct sites and
that three photons occupy three distinct sites are
$1-1/m$ and $(m-1)(m-2)/m^2$, respectively. Both are bounded below by
constants for $m\geq6$. We use a fixed number of attempts for each
stage and output zero if too few records are accepted. As proved in
\cref{app:failure_accounting}, this changes the photon cost only by a
constant factor and adds only $O(\eta)$ to the failure probability.
All final estimates are projected onto the unit disk. Hence the loss
is at most $4$ on every failure event, and choosing
$\eta=\Theta(\epsilon^2)$ makes its contribution to the expected loss
$O(\epsilon^2)$. Adjusting constant factors in $\tau$, $\lambda$, and
$\eta$ therefore gives
\begin{equation}
    \E L(B,\widetilde B)\leq\epsilon^2
\end{equation}
using
\begin{equation}
    \widetilde O(m/\epsilon^4+\epsilon^{-8})
\end{equation}
input photons. All classical postprocessing consists of matrix
diagonalization and polynomial-dimensional linear algebra, and is
therefore polynomial in $m$ and $1/\epsilon$; details are given in
\cref{app:classical_computation}. The single-photon lower bound in the
statement is proved in \cref{sec:single_photon_large_array}.
\end{proof}

Note that the key intuition why the three-photon collective measurement has scaling advantage
is that the occupation-number vector measurements enable every three-photon copy to provide one sample of $(t,z)$ with constant probability, while single-photon sequential measurements cannot directly provide such information. Naively, if we divide $m$ sites into $m/3$ triangles, and for each photon, measure which triangle it belongs to, then the probability for three photons in the same ACT to be received by the same triangle is only $O(1/m^2)$, leading to a quadratic penalty compared with joint measurements. Although the optimal sequential measurement strategy may have better overall photon number scaling, we manage to prove an $\Omega(m^2/\epsilon^2)$ lower bound by constructing a counterexample, as detailed in \cref{sec:single_photon_large_array}. 

We briefly sketch the proof idea in \cref{sec:single_photon_large_array} as follows.
The main idea is to reduce a hypothesis testing problem to the parameter estimation problem. In particular, we construct two visibility matrices $V$ and $V'$ with identical amplitudes and different bispectra on a constant fraction of all triangles. One can show that single-photon measurement of each three-photon ACT produces classical records whose distributions differ by only $O(\epsilon^2/m^2)$ in relative entropy.
Distinguishing the two candidates well enough to estimate their bispectra therefore requires $\Omega(m^2/\epsilon^2)$ photons.

\subsection{Shadow tomography}
\label{sec:shadow}
The compressibility of bispectra [see \cref{thm:number_of_bits}] suggests that nearly linear photon complexity may be achievable, but does not by itself establish a photon-complexity lower bound.
Fortunately, there is a protocol that can in principle achieve the $\tilde{O}(m/\epsilon^2)$ photon number cost for finite $\epsilon$: 
\emph{shadow tomography}. Originally proposed in Ref.~\cite{aaronson2020shadow} and later improved in Refs.~\cite{badescu2021improved,watts2024quantum,chen2024optimal,sinha2025dimension,pelecanos2025debiased,chen2026online,jeronimo2026dimension}, shadow tomography is a set of techniques that allows for measuring multiple parameters (which are not necessarily compatible with each other) with much improved sample complexity than measuring them sequentially. Typical shadow tomography approaches utilize joint measurements of multiple copies of the state, thus generally requiring quantum computational power.

Here the sample complexity counts copies of the input state rather than individual photons. Therefore, if the sample complexity has $\log D$-dependence, then it also introduce additional powers of $m$ when each copy contains $n=\Theta(m)$ photons, since $ D=m^n$.
In contrast, the dimension-independent result of
\textcite{jeronimo2026dimension} has only polylogarithmic dependence on the
number of observables. Therefore, one can potentially apply this approach to achieve linear-in-$m$ sample complexity.

To apply this result to bispectrum estimation, we must account for the normalization that relates observable expectations to bispectra.
Consider the $n$-photon noisy state $\rho_n^{\rm noisy}$ from each ACT.
For any fixed triangle $t=(i,j,k)$, the probability that all three sites
receive at least one photon is
\begin{equation}
\begin{aligned}
    p(m,n) &=
    1-3\left(1-\frac{1}{m}\right)^n
    \\
    &\quad \, +3\left(1-\frac{2}{m}\right)^n
    -\left(1-\frac{3}{m}\right)^n.
\end{aligned}
\end{equation}
For $n=am$ with fixed $a>0$, this becomes
$p(m,n)=(1-e^{-a})^3+O(1/m)$, a positive constant independent of $m$.
For each triangle, define an operator $A_t$ as follows.
Within an occupation sector that covers $t$, average the cyclic relabeling
$i\to j\to k\to i$ over all $N_i N_j N_k$ choices of one photon register
at each site where $N_i,N_j,N_k$ are the occupation numbers of site $i,j,k$, respectively [see~\cref{sec:occupation_sectors}]. On sectors that do not cover $t$, set $A_t=0$. Observe that
\begin{equation}
    \|A_t\|_{\rm op}\leq 1,
    \qquad
    \operatorname{Tr}\!\left(\rho_n^{\rm noisy}A_t\right)
    =
    p(m,n)B_t.
\end{equation}
We can now apply shadow tomography to the real and imaginary parts of these operators, shifted and rescaled to have spectra in $[0,1]$.
Estimating their expectations to additive accuracy $O(p(m,n)\epsilon)$
therefore estimates all bispectra to additive accuracy $O(\epsilon)$,
using
\begin{equation}
    K
    =
    \tilde{O}\!\left(
        \frac{1}{p(m,n)^2\epsilon^2}
    \right)
    \end{equation} 
copies, corresponding to
$nK= \tilde{O}\!\left(\frac{m}{\epsilon^2}\right)$ photons.

By contrast, using $\rho_3^{\rm noisy}$ as the input gives
$p(m,3)=6/m^3$.
The same direct application would then require observable expectations
to be estimated to accuracy $O(\epsilon/m^3)$, yielding only a $\tilde{O}(m^6/\epsilon^2)$ photon upper bound. Thus, the favorable normalization here relies on using the full $n=\Theta(m)$-photon ACT state rather than first reducing it to independently randomized triples.

However, a critical drawback is that the \textcite{jeronimo2026dimension} approach is not guaranteed to be implementable using $\poly(m)$ quantum gates. For the specific bispectra estimation problem of our interest, it is in principle possible that there exists an efficient ($\poly(m,1/\epsilon)$-gate) circuit implementing their protocol. We leave it as an open problem.

\subsection{High-precision regime}
\label{sec:high_precision}
Next, we consider the limit of a constant number of sites $m$ and estimating the bispectra to high precision $\epsilon \to 0$. We describe the behavior of the Holevo Cramer-Rao bound $C_H$, since it most generally constrains  collective processing (see Appendix~\ref{sec:multiparameterCramerRao} for related multiparameter Cramer-Rao bounds), lower bounding the normalized variance of the parameters estimated:
\begin{equation}
    C_H \leq\frac{1}{\binom{m-1}{2}}\sum_{2\leq a < b \leq m} \text{Var}[B_{1ab}] \,.
\end{equation}
In particular, we prove that $C_H = \Theta(1)$ for $m\sim n$. The proof is in two parts. First, we show the upper bound  (Theorem~\ref{thm:HolevoAttain}) for any scene. Second, we provide a lower bound (Theorem~\ref{thm:holevo}) using a specific visibility matrix.

The Holevo bound can be achieved in the limit of a large number of photons using a variety of techniques~\cite{tsang2026approaching,zhou2026quantum}. We provide a construction showing attainability for our specific problem. 
\begin{theorem} [Holevo attainability for bispectra] \label{thm:HolevoAttain}
    For any visibility matrix $V$ with unit diagonal,
    \begin{equation}
        C_H \leq \frac{2}{p(m,n)} \sum_t w_t (1-\lvert B_t^2 \rvert) \,,
    \end{equation}
    where
    \begin{equation}
        p(m,n)= 1 - 3\left(1-\frac{1}{m}\right)^n + 3\left(1-\frac{2}{m}\right)^n - \left(1-\frac{3}{m}\right)^n
    \end{equation}
    is the probability of any given triangle to receive a photon per site.

    For large arrays
\begin{equation}
    p(m,n) \xrightarrow[]{m\to \infty} (1-e^{-n/m})^3.
\end{equation}
As an implication, if $n\sim m$, the Holevo bound is constant.
\end{theorem}
\begin{proof}
    For each triangle, define the operators
    \begin{align}
        X_{t,R} &= \frac{\sqrt{w_t}}{p} \left(\frac{A_t + A_t^\dagger}{2} - \text{Re}(B_t P_t) \right) \,,\\
        X_{t,I} &= \frac{\sqrt{w_t}}{p} \left(\frac{A_t - A_t^\dagger}{2i} - \text{Im}(B_t P_t) \right) \,,
    \end{align}
    where $A_t$ is $C_{ABC}$ for $(A,B,C)\in t$, $P_t$ is the projector onto $t$, and $B_t$ is the associated bispectrum. By construction, these operators have mean 0.

    Observe that
    \begin{align}
        &\Tr[\rho (X_{t,R}^2+X_{t,I}^2)] \\
        &= \frac{w_t}{p^2} \left(\frac{1}{2} \Tr[\rho(A_t^\dagger A_t + A_t A_t^\dagger)]-p \lvert B_t \rvert^2\right)  \\
        &\leq \frac{w_t}{p}(1-\lvert B_t \rvert^2) \,,
    \end{align}
    where we used $A_t = P_t A_t P_t $, $\langle A_t \rangle = p B_t$, and $\langle P_t \rangle = p$ in the first line and $A_t^\dagger A_t \leq P_t$ in the second.

    Write the covariance matrix
    \begin{equation}
        Z_{\mu \nu} = \Tr[ \rho X_\mu X_\nu] \,.
    \end{equation}
    It is positive semidefinite and can be expressed as $Z=S+iD$ where $S=\text{Re} Z$ and $D=\text{Im} Z$. We have
    \begin{equation}
        \Tr[S] = \sum_t \Tr[\rho(X_{t,R}^2+X_{t,I}^2)]
    \end{equation}
    and
    \begin{align}
        \lVert D \rVert_1 &= \frac{1}{2} \lVert Z - \bar{Z} \rVert_1 \\
        &\leq \frac{1}{2}(\lVert Z \rVert_1 +\lVert \bar{Z} \rVert_1) \\
        &= \Tr[S] \,.
    \end{align}

    The Holevo bound is the infimum over $\Tr[S] + \lVert D \rVert_1$, so
    \begin{align}
        C_H &\leq 2 \Tr[S] \\
        &\leq \frac{2}{p} \sum_t w_t(1 - \lvert B_t \rvert^2) \,.
    \end{align}
\end{proof}

\begin{theorem}[Holevo lower bound for bispectra] \label{thm:holevo}
    Assume $V_{ij}=v$ for $0<v<1$. For estimation of all bispectra $\{B_{1ab}\}_{a,b}$, the Holevo bound is
    \begin{equation}
        C_H \geq \frac{3m (1-v)v^4}{2n} \,.
    \end{equation}
\end{theorem}
\begin{proof}
    
    We calculate (see Appendix~\ref{sec:bright} for $m=3$) that the single-photon QFI is 
    \begin{equation}
        F_t^{(1)} = \frac{2v^2}{3m(1-v)} \,. 
    \end{equation}
    The QFI for $n$ photons, without averaging over the piston term, is 
    \begin{equation} \label{eq:nPhotonQFI}
        F_t^{(n)} = n F_t^{(1)} \,, 
    \end{equation}
    since the SLD is the sum of the $n$ single-photon SLDs, given that each has zero mean. Next, observe that averaging over the noise cannot increase the Fisher information, and the Holevo bound is an infimum, so we work with the unrandomized state $\rho$.

    Consider a perturbation of the form
    \begin{equation}
        V_{ij} = v e^{ix/3} \,,
    \end{equation}
    for all edges $(i,j)$ of triangle $t$. Then $B_t=v^3 e^{ix}$. For this perturbation,
    \begin{equation}
        v^3=\Tr[\frac{d \rho}{dx}\lvert_{x=0} X_{t,I}] \,,
    \end{equation}
    also for an observable saturating the Holevo bound, $X_{t,I}$. 
    Using the SLD equation for SLD $L$,
    \begin{equation}
        v^3 = \text{Re} \Tr[\rho L X_{t,I}] \,.
    \end{equation}
    By the Cauchy-Schwarz inequality,
    \begin{align}
        v^6 &\leq \Tr[\rho L^2] \Tr[\rho X_{t,I}^2] \\
        &= F_t^{(n)} \Tr[\rho X_{t,I}^2] \,.
    \end{align}
    Substituting \cref{eq:nPhotonQFI},
    \begin{equation}
        \Tr[\rho X_{t,I}^2] \geq \frac{3m(1-v)v^4}{2n} \,.
    \end{equation}

    The covariance matrix is 
    \begin{equation}
        Z_{\mu \nu } = \Tr[\rho X_\mu X_\nu]
    \end{equation}
    and
    \begin{equation}
        C_H = \inf \Tr[\text{Re} Z] + \lVert \text{im} Z \rVert_1 \,.
    \end{equation}
    By the non-negativity of the trace norm and variances, 
    \begin{align}
        C_H &\geq \Tr[\rho X_{t,I}^2] \\
        &\geq \frac{3m(1-v)v^4}{2n} \,.
    \end{align}
\end{proof}

\section{Discussions and outlook}
\label{sec:discussion}
In conclusion, we have studied how to extract noise-robust information about the imaged scene in telescope arrays when general quantum operations are possible. We focus on the case where few photons are available, which corresponds to the most relevant limiting scenario in high-resolution imaging, and construct optimal measurement protocols using few-photon collective measurements in this setting. Generally, when more photons are available, such as when the size of the array grows, we find that collective processing allows more efficient estimation of all bispectra.

In this work, we treat the closure phase or the complete collection of bispectra as being the scientific target. In practice, particular telescope geometries and source distributions influence which bispectra are most useful; an end-to-end optimization would be interesting but depends on the particular scientific use case~\cite{eldredge2018optimal, proctor2018multiparameter, zhuang2018distributed, allen2025quantum, khan2025quantum, sarovar2023quantum}. Furthermore, while we have considered closure phase in the context of general imaging, it is useful in other settings. For characteristic scenes, closure phases often take discrete values $\{0,\pi\}$, and detecting change in the binary value can reveal the size of the imaged object. In addition, closure phase diagnoses asymmetries in the scene. For instance, it is $0$ for a single point source but can be $\pi$ for a binary. As such, constructing noise-robust algorithms for classification tasks using the closure phase is a promising future direction.

\section*{AI disclosure}
The authors came up with the main ideas and algorithms. ChatGPT 5.6 Sol and 6 Astra assisted in calculations and preparation of the manuscript, and the authors checked all the mathematical expressions. These tools also helped with with the proofs for general bispectra estimation. 

The research problem and the main conceptual ideas of this work were developed manually by the authors.
In particular, the authors derived the block structure and photon-number lower bound in \cref{sec:noise_invariant}, proposed the nondemolition occupation-number measurements in \cref{prot:occupation_number}, computed the quantum Fisher information of $3$-photon noisy states in \cref{sec:single}, proposed the bispectrum measurement protocol in \cref{prot:three_photon_single_bispectrum}, and introduced the general framework of using high-quality data generated by collective measurements to enable efficient classical regression in \cref{sec:full_bispectra} and \cref{prot:three_photon_full_bisepctra}.
 
Generative AI, including ChatGPT 5.5, 5.6 Sol, and 6 Astra, derived a substantial portion of the technical details used to propose \cref{prot:three_photon_full_bisepctra} in \cref{sec:full_bispectra} and prove its efficiency, including providing the concrete low-rank representation of the full bispectra, proving the efficiency of building the visibility-square matrix, and constructing the classical ridge regression procedure to reconstruct all bispectra with a rigoursly proved sample complexity. Furthermore, AI provided insights to prove the compressibility of bispectra in \cref{sec:full_bispectra} and helped prove the limitation of single-photon measurements in \cref{appendix:single_photon_limit_for_m=3,sec:single_photon_large_array}.

AI also assisted with literature searches and the writing and editing of the manuscript. All AI-generated arguments, calculations, results, references, and text were independently checked and verified by the authors, who take full responsibility for the content and correctness of the paper.

\section*{Acknowledgments.}
We thank Andrew Childs, Parth Darekar, Nico Deshler, and Yu-Jie Liu for helpful discussions. E.T.K.~and A.V.G.~were supported in part by ONR MURI, ARL (W911NF-24-2-0107), DoE ASCR Quantum Testbed Pathfinder program (award No.~DE-SC0024220), NSF QLCI (awards No.~OMA-2120757 and No.~OSI-2553574), NSF STAQ program, AFOSR MURI, and NQVL:QSTD:Design:FTL. E.T.K.~and A.V.G.~also acknowledge support from the U.S.~Department of Energy, Office of Science, National Quantum Information Science Research Centers, Quantum Systems Accelerator (award No.~DE-SCL0000121) and from the U.S.~Department of Energy, Office of Science, Accelerated Research in Quantum Computing, Fundamental Algorithmic Research toward Quantum Utility (FAR-Qu). D.G. and Z.L.\ acknowledge support from the National Science Foundation (QLCI grant OMA-2120757). The work of B.M.M., J.R.,  and S.T.R. is supported by NOIRLab, which is managed by the Association of Universities for Research in Astronomy (AURA) under a cooperative agreement with the U.S. National Science Foundation.

\bibliography{main}

@article{Khabiboulline2019OpticalInterferometry,
  author        = {Khabiboulline, E. T. and Borregaard, J. and De Greve, K. and Lukin, M. D.},
  title         = {{Optical Interferometry with Quantum Networks}},
  journal       = {Physical Review Letters},
  year          = {2019},
  volume        = {123},
  number        = {7},
  pages         = {070504},
  numpages      = {7},
  month         = aug,
  publisher     = {American Physical Society},
  doi           = {10.1103/PhysRevLett.123.070504}
}

@article{Khabiboulline2019TelescopeArrays,
  author        = {Khabiboulline, E. T. and Borregaard, J. and De Greve, K. and Lukin, M. D.},
  title         = {{Quantum-assisted telescope arrays}},
  journal       = {Physical Review A},
  year          = {2019},
  volume        = {100},
  number        = {2},
  pages         = {022316},
  numpages      = {9},
  month         = aug,
  publisher     = {American Physical Society},
  doi           = {10.1103/PhysRevA.100.022316}
}

@article{conlon2021efficient,
  author        = {Conlon, Lorc{\'a}n O and Suzuki, Jun and Lam, Ping Koy and Assad, Syed M},
  title         = {{Efficient computation of the {N}agaoka--{H}ayashi bound for multiparameter estimation with separable measurements}},
  journal       = {npj Quantum Information},
  year          = {2021},
  volume        = {7},
  number        = {1},
  pages         = {110},
  publisher     = {Nature Publishing Group UK London},
  doi           = {10.1038/s41534-021-00414-1}
}

@article{suzuki2020quantum,
  author        = {Suzuki, Jun and Yang, Yuxiang and Hayashi, Masahito},
  title         = {{Quantum state estimation with nuisance parameters}},
  journal       = {Journal of Physics A: Mathematical and Theoretical},
  year          = {2020},
  volume        = {53},
  number        = {45},
  pages         = {453001},
  publisher     = {IOP Publishing},
  doi           = {10.1088/1751-8121/ab8b78}
}

@book{Goodman_2005,
  author    = {Goodman, Joseph W.},
  title     = {{Introduction to Fourier Optics}},
  year      = {2005},
  edition   = {3},
  publisher = {Roberts \& Company Publishers},
  address   = {Englewood, CO},
  isbn      = {9780974707723},
  url       = {https://books.google.com/books?id=ow5xs_Rtt9AC}
}

@article{Monnier_2007,
  author        = {Monnier, J. D.},
  title         = {{Phases in interferometry}},
  journal       = {New Astronomy Reviews},
  year          = {2007},
  volume        = {51},
  number        = {8--9},
  pages         = {604--616},
  publisher     = {Elsevier},
  doi           = {10.1016/j.newar.2007.06.006}
}

@article{Rogers1974,
  author        = {Rogers, A. E. E. and Hinteregger, H. F. and Whitney, A. R. and Counselman, C. C. and Shapiro, I. I. and Wittels, J. J. and Klemperer, W. K. and Warnock, W. W. and Clark, T. A. and Hutton, L. K.},
  title         = {{The structure of radio sources 3C 273B and 3C 84 deduced from the `closure' phases and visibility amplitudes observed with three-element interferometers}},
  journal       = {The Astrophysical Journal},
  year          = {1974},
  volume        = {193},
  pages         = {293},
  doi           = {10.1086/153162}
}

@article{Jennison_1958,
  author        = {Jennison, R. C.},
  title         = {{A Phase Sensitive Interferometer Technique for the Measurement of the {F}ourier Transforms of Spatial Brightness Distributions of Small Angular Extent}},
  journal       = {Monthly Notices of the Royal Astronomical Society},
  year          = {1958},
  volume        = {118},
  number        = {3},
  pages         = {276--284},
  month         = jun,
  doi           = {10.1093/mnras/118.3.276},
  url           = {https://doi.org/10.1093/mnras/118.3.276}
}

@article{liu2020quantum,
  author        = {Liu, Jing and Yuan, Haidong and Lu, Xiao-Ming and Wang, Xiaoguang},
  title         = {{Quantum {F}isher information matrix and multiparameter estimation}},
  journal       = {Journal of Physics A: Mathematical and Theoretical},
  year          = {2020},
  volume        = {53},
  number        = {2},
  pages         = {023001},
  publisher     = {IOP Publishing},
  doi           = {10.1088/1751-8121/ab5d4d}
}

@article{Gottesman2012LongBaseline,
  author        = {Gottesman, Daniel and Jennewein, Thomas and Croke, Sarah},
  title         = {{Longer-Baseline Telescopes Using Quantum Repeaters}},
  journal       = {Physical Review Letters},
  year          = {2012},
  volume        = {109},
  number        = {7},
  pages         = {070503},
  numpages      = {5},
  month         = aug,
  publisher     = {American Physical Society},
  doi           = {10.1103/PhysRevLett.109.070503},
  url           = {https://link.aps.org/doi/10.1103/PhysRevLett.109.070503}
}

@book{Thompson2017RadioAstronomy,
  author        = {Thompson, A. Richard and Moran, James M. and Swenson, Jr., George W.},
  title         = {{Interferometry and Synthesis in Radio Astronomy}},
  year          = {2017},
  edition       = {3},
  publisher     = {Springer Cham},
  doi           = {10.1007/978-3-319-44431-4}
}

@inproceedings{chahine2026,
  author        = {Yousef K. Chahine and Chaohan Cui and William DeRocco and Daniel Gottesman and Saikat Guha and Emil T. Khabiboulline and Zhenning Liu and Brittany McClinton and Jayadev Rajagopal and Fredrik Rantakyro and J. Gabriel Richardson and Stephen Ridgway and Aqil Sajjad and Joohyung Song},
  editor        = {Jens Kammerer and Gioia Rau and Joel Sanchez-Bermudez},
  title         = {{Eliminating photon transport in long-baseline optical interferometry using quantum memories}},
  booktitle     = {{Optical and Infrared Interferometry and Imaging X}},
  year          = {2026},
  volume        = {14148},
  pages         = {1414819},
  publisher     = {SPIE},
  organization  = {International Society for Optics and Photonics},
  doi           = {10.1117/12.3105495},
  url           = {https://doi.org/10.1117/12.3105495}
}

@article{Tsang2011WeakThermalLight,
  author        = {Tsang, Mankei},
  title         = {{Quantum Nonlocality in Weak-Thermal-Light Interferometry}},
  journal       = {Physical Review Letters},
  year          = {2011},
  volume        = {107},
  number        = {27},
  pages         = {270402},
  numpages      = {5},
  month         = dec,
  publisher     = {American Physical Society},
  doi           = {10.1103/PhysRevLett.107.270402}
}

@article{Stas_2026,
  author        = {Stas, P.-J. and Wei, Y.-C. and Sirotin, M. and Huan, Y. Q. and Yazlar, U. and Abdo Arias, F. and Knyazev, E. and Baranes, G. and Machielse, B. and Grandi, S. and Riedel, D. and Borregaard, J. and Park, H. and Lončar, M. and Suleymanzade, A. and Lukin, M. D.},
  title         = {{Entanglement-assisted non-local optical interferometry in a quantum network}},
  journal       = {Nature},
  year          = {2026},
  volume        = {651},
  number        = {8105},
  pages         = {326--332},
  doi           = {10.1038/s41586-026-10171-w},
  issn          = {1476-4687}
}

@article{Zernike1938Theorem,
  author        = {F. Zernike},
  title         = {{The concept of degree of coherence and its application to optical problems}},
  journal       = {Physica},
  year          = {1938},
  volume        = {5},
  number        = {8},
  pages         = {785--795},
  doi           = {10.1016/S0031-8914(38)80203-2},
  issn          = {0031-8914}
}

@misc{Liu2026LensingDelays,
  author        = {Zhenning Liu and William DeRocco and Shiming Gu and Emil T. Khabiboulline and Soonwon Choi and Andrew M. Childs and Anson Hook and Alexey V. Gorshkov and Daniel Gottesman},
  title         = {{Measuring gravitational lensing time delays with quantum information processing}},
  year          = {2025},
  doi           = {10.48550/arXiv.2510.07898},
  url           = {https://arxiv.org/abs/2510.07898},
  eprint        = {2510.07898},
  archiveprefix = {arXiv},
  primaryclass  = {quant-ph}
}

@inproceedings{Gorshkov2026SPIEClosure,
  author        = {Alexey V. Gorshkov and Daniel Gottesman and Emil T. Khabiboulline and Zhenning Liu and Brittany McClinton and Jayadev Rajagopal and Stephen T. Ridgway},
  editor        = {Jens Kammerer and Gioia Rau and Joel Sanchez-Bermudez},
  title         = {{Quantum-optimized phase closure for robust imaging}},
  booktitle     = {{Optical and Infrared Interferometry and Imaging X}},
  year          = {2026},
  volume        = {14148},
  pages         = {141482S},
  publisher     = {SPIE},
  organization  = {International Society for Optics and Photonics},
  doi           = {10.1117/12.3104954},
  url           = {https://doi.org/10.1117/12.3104954}
}

@article{tsang2026approaching,
  author        = {Tsang, Mankei},
  title         = {{Approaching the ultimate limit of quantum multiparameter estimation by many-body physics}},
  journal       = {Physical Review Letters},
  year          = {2026},
  volume        = {137},
  number        = {7},
  pages         = {070804},
  publisher     = {APS},
  doi           = {10.1103/cghv-j44h}
}

@article{aaronson2020shadow,
  author        = {Aaronson, Scott},
  title         = {{Shadow tomography of quantum states}},
  journal       = {SIAM Journal on Computing},
  year          = {2020},
  volume        = {49},
  number        = {5},
  pages         = {STOC18--368--STOC18--394},
  publisher     = {SIAM},
  doi           = {10.1137/18M120275X}
}

@inproceedings{badescu2021improved,
  author        = {B{\u{a}}descu, Costin and O'Donnell, Ryan},
  title         = {{Improved quantum data analysis}},
  booktitle     = {{Proceedings of the 53rd Annual ACM SIGACT Symposium on Theory of Computing}},
  year          = {2021},
  pages         = {1398--1411},
  doi           = {10.1145/3406325.3451109}
}

@inproceedings{watts2024quantum,
  author        = {Watts, Adam Bene and Bostanci, John},
  title         = {{Quantum event learning and gentle random measurements}},
  booktitle     = {{15th Innovations in Theoretical Computer Science Conference (ITCS 2024)}},
  year          = {2024},
  volume        = {287},
  pages         = {97:1--97:22},
  publisher     = {Schloss Dagstuhl--Leibniz-Zentrum f{\"u}r Informatik},
  doi           = {10.4230/LIPIcs.ITCS.2024.97}
}

@article{chen2024optimal,
  author        = {Chen, Sitan and Li, Jerry and Liu, Allen},
  title         = {{Optimal high-precision shadow estimation}},
  journal       = {arXiv preprint arXiv:2407.13874},
  year          = {2024},
  doi           = {10.48550/arXiv.2407.13874}
}

@inproceedings{sinha2025dimension,
  author        = {Sinha, Pulkit},
  title         = {{Dimension independent and computationally efficient shadow tomography}},
  booktitle     = {{Proceedings of the 57th Annual ACM Symposium on Theory of Computing}},
  year          = {2025},
  pages         = {1242--1253},
  doi           = {10.1145/3717823.3718253}
}

@article{pelecanos2025debiased,
  author        = {Pelecanos, Angelos and Spilecki, Jack and Wright, John},
  title         = {{The debiased {K}eyl's algorithm: A new unbiased estimator for full state tomography}},
  journal       = {arXiv preprint arXiv:2510.07788},
  year          = {2025},
  doi           = {10.48550/arXiv.2510.07788}
}

@article{chen2026online,
  author        = {Chen, Sitan and O'Donnell, Ryan and Pelecanos, Angelos and Wright, John},
  title         = {{Online shadow tomography matching the classical bounds}},
  journal       = {arXiv preprint arXiv:2607.29686},
  year          = {2026},
  doi           = {10.48550/arXiv.2607.29686}
}

@article{jeronimo2026dimension,
  author        = {Jeronimo, Fernando Granha and Huang, Qizhao and Liu, Lenny},
  title         = {{Dimension-free polylogarithmic quantum shadow tomography from sequential pretty-good measurements}},
  journal       = {arXiv preprint arXiv:2608.06345},
  year          = {2026},
  doi           = {10.48550/arXiv.2608.06345}
}

@article{Monnier2007Altair,
  author        = {Monnier, J. D. and Zhao, M. and Pedretti, E. and Thureau, N. and Ireland, M. and Muirhead, P. and Berger, J.-P. and Millan-Gabet, R. and van Belle, G. and ten Brummelaar, T. and McAlister, H. and Ridgway, S. and Turner, N. and Sturmann, L. and Sturmann, J. and Berger, D.},
  title         = {{Imaging the Surface of {Altair}}},
  journal       = {Science},
  year          = {2007},
  volume        = {317},
  number        = {5836},
  pages         = {342--345},
  doi           = {10.1126/science.1143205}
}

@article{LeBouquinAbsil2012,
  author        = {Le Bouquin, J.-B. and Absil, O.},
  title         = {{On the sensitivity of closure phases to faint companions in optical long baseline interferometry}},
  journal       = {Astronomy \& Astrophysics},
  year          = {2012},
  volume        = {541},
  pages         = {A89},
  doi           = {10.1051/0004-6361/201117891}
}

@article{Lazareff2017Disks,
  author        = {Lazareff, B. and Berger, J.-P. and Kluska, J. and Le Bouquin, J.-B. and Benisty, M. and Malbet, F. and Koen, C. and Pinte, C. and Thi, W.-F. and Absil, O. and Baron, F. and Delboulb{\'e}, A. and Duvert, G. and Isella, A. and Jocou, L. and Juhasz, A. and Kraus, S. and Lachaume, R. and M{\'e}nard, F. and Millan-Gabet, R. and Monnier, J. D. and Moulin, T. and Perraut, K. and Rochat, S. and Soulez, F. and Tallon, M. and Thi{\'e}baut, E. and Traub, W. and Zins, G.},
  title         = {{Structure of {Herbig AeBe} disks at the milliarcsecond scale: A statistical survey in the {H} band using {PIONIER-VLTI}}},
  journal       = {Astronomy \& Astrophysics},
  year          = {2017},
  volume        = {599},
  pages         = {A85},
  doi           = {10.1051/0004-6361/201629305}
}

@misc{Monnier2026QuantumTechnologies,
  author        = {Monnier, John D.},
  title         = {{Development of quantum technologies for optical/infrared interferometry}},
  year          = {2026},
  doi           = {10.48550/arXiv.2609.04311},
  url           = {https://arxiv.org/abs/2609.04311},
  eprint        = {2609.04311},
  archiveprefix = {arXiv},
  primaryclass  = {astro-ph.IM}
}

@phdthesis{nakajima1989diffraction,
  author = {Nakajima, Tadashi},
  title  = {{Diffraction-limited imaging on the 200-inch telescope}},
  school = {California Institute of Technology},
  year   = {1989},
  doi    = {10.7907/hgnx-x451}
}

@article{nakajima1988signal,
  author        = {Nakajima, Tadashi},
  title         = {{Signal-to-noise ratio of the bispectral analysis of speckle interferometry}},
  journal       = {Journal of the Optical Society of America A},
  year          = {1988},
  volume        = {5},
  number        = {9},
  pages         = {1477--1491},
  publisher     = {Optical Society of America},
  doi           = {10.1364/JOSAA.5.001477}
}

@article{wirnitzer1985bispectral,
  author        = {Wirnitzer, B},
  title         = {{Bispectral analysis at low light levels and astronomical speckle masking}},
  journal       = {Journal of the Optical Society of America A},
  year          = {1985},
  volume        = {2},
  number        = {1},
  pages         = {14--21},
  publisher     = {Optical Society of America},
  doi           = {10.1364/JOSAA.2.000014}
}

@article{zhou2026quantum,
  author        = {Zhou, Sisi},
  title         = {{Quantum metrology of mixed states via purification}},
  journal       = {arXiv preprint arXiv:2605.03975},
  year          = {2026},
  doi           = {10.48550/arXiv.2605.03975}
}

@incollection{johnson1984extensions,
  author        = {Johnson, William B. and Lindenstrauss, Joram},
  title         = {{Extensions of {L}ipschitz mappings into a {H}ilbert space}},
  booktitle     = {{Conference in Modern Analysis and Probability}},
  series        = {Contemporary Mathematics},
  year          = {1984},
  volume        = {26},
  pages         = {189--206},
  publisher     = {American Mathematical Society},
  doi           = {10.1090/conm/026/737400}
}

@article{dasgupta2003elementary,
  author        = {Dasgupta, Sanjoy and Gupta, Anupam},
  title         = {{An elementary proof of a theorem of {J}ohnson and {L}indenstrauss}},
  journal       = {Random Structures \& Algorithms},
  year          = {2003},
  volume        = {22},
  number        = {1},
  pages         = {60--65},
  doi           = {10.1002/rsa.10073}
}

@article{Hofmann1988ClosureSpeckle,
  author        = {K.-H. Hofmann},
  title         = {{Optical long-baseline interferometry: signal-to-noise ratio of image reconstruction by optical phase-closure imaging and by speckle masking}},
  journal       = {Applied Optics},
  year          = {1988},
  volume        = {27},
  number        = {10},
  pages         = {1943--1951},
  month         = may,
  publisher     = {Optica Publishing Group},
  doi           = {10.1364/AO.27.001943},
  url           = {https://opg.optica.org/ao/abstract.cfm?URI=ao-27-10-1943}
}

@article{Monnier2003OpticalInterferometry,
  author        = {Monnier, John D.},
  title         = {{Optical interferometry in astronomy}},
  journal       = {Reports on Progress in Physics},
  year          = {2003},
  volume        = {66},
  number        = {5},
  pages         = {789--857},
  doi           = {10.1088/0034-4885/66/5/203},
  url           = {https://doi.org/10.1088/0034-4885/66/5/203},
  eprint        = {astro-ph/0307036},
  archiveprefix = {arXiv}
}

@article{Blackburn2020ClosureStatistics,
  author        = {Blackburn, Lindy and Pesce, Dominic W. and Johnson, Michael D. and Wielgus, Maciek and Chael, Andrew A. and Christian, Pierre and Doeleman, Sheperd S.},
  title         = {{Closure Statistics in Interferometric Data}},
  journal       = {The Astrophysical Journal},
  year          = {2020},
  volume        = {894},
  number        = {1},
  pages         = {31},
  doi           = {10.3847/1538-4357/ab8469},
  url           = {https://doi.org/10.3847/1538-4357/ab8469},
  eprint        = {1910.02062},
  archiveprefix = {arXiv},
  primaryclass  = {astro-ph.IM}
}

@article{Cleve1998QuantumAlgorithms,
  author        = {Cleve, Richard and Ekert, Artur and Macchiavello, Chiara and Mosca, Michele},
  title         = {{Quantum algorithms revisited}},
  journal       = {Proceedings of the Royal Society of London. Series A: Mathematical, Physical and Engineering Sciences},
  year          = {1998},
  volume        = {454},
  number        = {1969},
  pages         = {339--354},
  doi           = {10.1098/rspa.1998.0164},
  url           = {https://doi.org/10.1098/rspa.1998.0164},
  eprint        = {quant-ph/9708016},
  archiveprefix = {arXiv}
}

@article{hoerl1970ridge,
  author        = {Hoerl, Arthur E. and Kennard, Robert W.},
  title         = {{Ridge Regression: Biased Estimation for Nonorthogonal Problems}},
  journal       = {Technometrics},
  year          = {1970},
  volume        = {12},
  number        = {1},
  pages         = {55--67},
  doi           = {10.1080/00401706.1970.10488634}
}

@article{hsu2014random,
  author        = {Hsu, Daniel and Kakade, Sham M. and Zhang, Tong},
  title         = {{Random Design Analysis of Ridge Regression}},
  journal       = {Foundations of Computational Mathematics},
  year          = {2014},
  volume        = {14},
  number        = {3},
  pages         = {569--600},
  doi           = {10.1007/s10208-014-9192-1}
}

@article{hong1987measurement,
  author        = {Hong, Chong-Ki and Ou, Zhe-Yu and Mandel, Leonard},
  title         = {{Measurement of subpicosecond time intervals between two photons by interference}},
  journal       = {Physical Review Letters},
  year          = {1987},
  volume        = {59},
  number        = {18},
  pages         = {2044--2046},
  publisher     = {APS},
  doi           = {10.1103/PhysRevLett.59.2044}
}

@article{eisenhauer2023advances,
  author        = {Eisenhauer, Frank and Monnier, John D. and Pfuhl, Oliver},
  title         = {{Advances in Optical/Infrared Interferometry}},
  journal       = {Annual Review of Astronomy and Astrophysics},
  year          = {2023},
  volume        = {61},
  pages         = {237--285},
  doi           = {10.1146/annurev-astro-121622-045019}
}

@inproceedings{petrov2016cophasing,
  author        = {Petrov, Romain G. and Boskri, Abdelkarim and Elhalkouj, Thami and Monnier, John and Ireland, Michael and Kraus, Stefan},
  title         = {{Cophasing the Planet Formation Imager}},
  booktitle     = {{Optical and Infrared Interferometry and Imaging V}},
  series        = {Proceedings of SPIE},
  year          = {2016},
  volume        = {9907},
  pages         = {99073W},
  publisher     = {SPIE},
  doi           = {10.1117/12.2231081}
}

@inproceedings{otto2026electromagnetic,
  author        = {Otto, Abraham J. and Kapp, Fran{\c{c}}ois B. and Hellbourg, Gregory},
  title         = {{Electromagnetic Compatibility Control Plan for the {DSA-2000} Radio Telescope}},
  booktitle     = {{2026 United States National Committee of URSI National Radio Science Meeting (USNC-URSI NRSM)}},
  year          = {2026},
  pages         = {437},
  doi           = {10.23919/NRSM68586.2026.11550929}
}

@misc{cohen2026optimal,
  author        = {Cohen, Tyler and Demorest, Paul B. and Baier, Jeremy G. and Lynch, Ryan S. and Ransom, Scott M. and Lam, Michael T. and Agazie, Gabriella and Anumarlapudi, Akash and Archibald, Anne M. and Arzoumanian, Zaven and Baker, Paul T. and Battaglia, Ava L. and Brook, Paul R. and Brueggemann, Anabelle L. and Cromartie, H. Thankful and Crowter, Kathryn and DeCesar, Megan E. and Dolch, Timothy and Ferrara, Elizabeth C. and Fiore, William and Fonseca, Emmanuel and Freedman, Gabriel E. and Garver-Daniels, Nate and Gentile, Peter A. and Glaser, Joseph and Good, Deborah C. and Hazboun, Jeffrey S. and Jennings, Ross J. and Jones, Megan L. and Kaplan, David L. and Kerr, Matthew and Lorimer, Duncan R. and Luo, Jing and Marquis, Hugo E. and McEwen, Alexander and McLaughlin, Maura A. and McMann, Natasha and Meyers, Bradley W. and Ng, Cherry and Nice, David J. and Pennucci, Timothy T. and Perera, Benetge B. P. and Pol, Nihan S. and Radovan, Henri A. and Ray, Paul S. and Schmiedekamp, Ann and Schmiedekamp, Carl and Shapiro-Albert, Brent J. and Stairs, Ingrid H. and Stovall, Kevin and Susobhanan, Abhimanyu and Swiggum, Joseph K. and Wahl, Haley M.},
  title         = {{Optimal Pulsar Timing Array Strategies with the Deep Synoptic Array}},
  year          = {2026},
  doi           = {10.48550/arXiv.2610.00836},
  eprint        = {2610.00836},
  archiveprefix = {arXiv},
  primaryclass  = {astro-ph.HE}
}

@article{eldredge2018optimal,
  author        = {Eldredge, Zachary and Foss-Feig, Michael and Gross, Jonathan A and Rolston, Steven L and Gorshkov, Alexey V},
  title         = {{Optimal and secure measurement protocols for quantum sensor networks}},
  journal       = {Physical Review A},
  year          = {2018},
  volume        = {97},
  number        = {4},
  pages         = {042337},
  publisher     = {APS},
  doi           = {10.1103/PhysRevA.97.042337}
}

@article{proctor2018multiparameter,
  author        = {Proctor, Timothy J and Knott, Paul A and Dunningham, Jacob A},
  title         = {{Multiparameter estimation in networked quantum sensors}},
  journal       = {Physical Review Letters},
  year          = {2018},
  volume        = {120},
  number        = {8},
  pages         = {080501},
  publisher     = {APS},
  doi           = {10.1103/PhysRevLett.120.080501}
}

@article{zhuang2018distributed,
  author        = {Zhuang, Quntao and Zhang, Zheshen and Shapiro, Jeffrey H},
  title         = {{Distributed quantum sensing using continuous-variable multipartite entanglement}},
  journal       = {Physical Review A},
  year          = {2018},
  volume        = {97},
  number        = {3},
  pages         = {032329},
  publisher     = {APS},
  doi           = {10.1103/PhysRevA.97.032329}
}

@article{allen2025quantum,
  author        = {Allen, Richard R and Machado, Francisco and Chuang, Isaac L and Huang, Hsin-Yuan and Choi, Soonwon},
  title         = {{Quantum computing enhanced sensing}},
  journal       = {arXiv preprint arXiv:2501.07625},
  year          = {2025},
  doi           = {10.48550/arXiv.2501.07625}
}

@article{khan2025quantum,
  author        = {Khan, Saeed A and Prabhu, Sridhar and Wright, Logan G and McMahon, Peter L},
  title         = {{Quantum computational-sensing advantage}},
  journal       = {arXiv preprint arXiv:2507.16918},
  year          = {2025},
  doi           = {10.48550/arXiv.2507.16918}
}

@inproceedings{sarovar2023quantum,
  author        = {Sarovar, Mohan},
  title         = {{Quantum computational imaging and sensing}},
  booktitle     = {{Quantum Nanophotonic Materials, Devices, and Systems 2023}},
  year          = {2023},
  volume        = {12657},
  pages         = {1265703},
  organization  = {SPIE},
  doi           = {10.1117/12.2680837}
}

@article{deshler2026attaining,
  author        = {Deshler, Nico and Warke, Aakash and Grace, Michael R and Ashok, Amit and Guha, Saikat},
  title         = {{Attaining Fundamental Limits of Multiparameter Incoherent Optical Imaging Using Joint-Detection Quantum Measurements}},
  journal       = {arXiv preprint arXiv:2608.19524},
  year          = {2026},
  doi           = {10.48550/arXiv.2608.19524}
}

@article{Padilla2026a,
  author        = {Padilla, Isack and Sajjad, Aqil and Saif, Babak N. and Guha, Saikat},
  title         = {{Superresolution Imaging with Entanglement-Enhanced Telescopy}},
  journal       = {Physical Review Letters},
  year          = {2026},
  volume        = {136},
  pages         = {010803},
  publisher     = {American Physical Society},
  doi           = {10.1103/354q-ch63},
  url           = {https://doi.org/10.1103/354q-ch63}
}

@article{Padilla2026b,
  author        = {Padilla, Isack and Sajjad, Aqil and Saif, Babak N. and Guha, Saikat},
  title         = {{Quantum resolution limit of long-baseline imaging using distributed entanglement}},
  journal       = {Physical Review A},
  year          = {2026},
  volume        = {113},
  pages         = {012608},
  publisher     = {American Physical Society},
  doi           = {10.1103/4npc-gpvh},
  url           = {https://doi.org/10.1103/4npc-gpvh}
}

@article{Cosmo2020,
  author        = {Lupo, Cosmo and Huang, Zixin and Kok, Pieter},
  title         = {{Quantum Limits to Incoherent Imaging are Achieved by Linear Interferometry}},
  journal       = {Physical Review Letters},
  year          = {2020},
  volume        = {124},
  number        = {8},
  pages         = {080503},
  numpages      = {6},
  month         = feb,
  publisher     = {American Physical Society},
  doi           = {10.1103/PhysRevLett.124.080503},
  url           = {https://link.aps.org/doi/10.1103/PhysRevLett.124.080503}
}

@article{Sajjad2024,
  author        = {Sajjad, Aqil and Grace, Michael R. and Guha, Saikat},
  title         = {{Quantum limits of parameter estimation in long-baseline imaging}},
  journal       = {Physical Review Research},
  year          = {2024},
  volume        = {6},
  pages         = {013212},
  publisher     = {American Physical Society},
  doi           = {10.1103/PhysRevResearch.6.013212},
  url           = {https://doi.org/10.1103/PhysRevResearch.6.013212}
}

@article{Rajagopal:2024suu,
  author        = {Rajagopal, Jayadev K. and others},
  title         = {{Towards quantum-enhanced long-baseline optical/near-IR interferometry}},
  journal       = {Proceedings of SPIE},
  year          = {2024},
  volume        = {13095},
  pages         = {130951N},
  doi           = {10.1117/12.3019518},
  eprint        = {2407.06302},
  archiveprefix = {arXiv},
  primaryclass  = {astro-ph.IM},
  reportnumber  = {Paper No. 13095-58}
}

@article{blackburn2020closure,
  author        = {Blackburn, Lindy and Pesce, Dominic W and Johnson, Michael D and Wielgus, Maciek and Chael, Andrew A and Christian, Pierre and Doeleman, Sheperd S},
  title         = {{Closure statistics in interferometric data}},
  journal       = {The Astrophysical Journal},
  year          = {2020},
  volume        = {894},
  number        = {1},
  pages         = {31},
  publisher     = {The American Astronomical Society},
  doi           = {10.3847/1538-4357/ab8469}
}

@article{le2012sensitivity,
  author        = {Le Bouquin, J-B and Absil, Olivier},
  title         = {{On the sensitivity of closure phases to faint companions in optical long baseline interferometry}},
  journal       = {Astronomy \& Astrophysics},
  year          = {2012},
  volume        = {541},
  pages         = {A89},
  publisher     = {EDP Sciences},
  doi           = {10.1051/0004-6361/201117891}
}

@article{absil2006performance,
  author        = {Absil, Olivier and Den Hartog, R and Gondoin, P and Fabry, P and Wilhelm, R and Gitton, P and Puech, F},
  title         = {{Performance study of ground-based infrared Bracewell interferometers: Application to the detection of exozodiacal dust disks with GENIE}},
  journal       = {Astronomy \& Astrophysics},
  year          = {2006},
  volume        = {448},
  number        = {2},
  pages         = {787--800},
  publisher     = {EDP Sciences},
  doi           = {10.1051/0004-6361:20053516}
}

\clearpage
\twocolumngrid
\appendix

\crefalias{section}{appendix}
\crefalias{subsection}{appendix}

\section*{Appendices}

Appendix~\ref{sec:kolmogorov} provides a detailed physical noise model that motivates our description of incident light. We begin by calculating the Fisher information of the noiseless model (Appendix~\ref{sec:bright}). Then, introducing noise, Appendix~\ref{appendix:single_photon_limit_for_m=3} describes the optimal approach of learning closure from sequential photons. Next, we calculate the Fisher information for the collective state (Appendix~\ref{sec:details}) and generalize the collective measurement (Appendix~\ref{sec:TorusQFT}). Appendices~\ref{sec:three_photon_information} and~\ref{sec:single_photon_large_array} contain the proofs for bispectra estimation with finite precision. In Appendix~\ref{sec:multiparameterCramerRao}, we provide calculations of bounds in the infinitesimal-precision regime, and demonstrate an incompatibility gap between different classes of collective measurements.

\section{Atmospheric phase noise model}
\label{sec:kolmogorov}

\subsection{The Kolmogorov atmospheric fluctuation model}
The key challenge given by atmospheric turbulence is that $\vec{\delta}$ changes with time. Here we elaborate on the Kolmogorov model, which is one the most common models used in the astronomy community. Note that for each telescope site, $\delta_j$ can be written as
\begin{equation}
\label{eq:delta_j}
    \delta_j(t) = \frac{2\pi}{\lambda}\int n(\vec{x}_j,z,t)dz  
\end{equation}
where $\vec{x}_j$ is the location of the $j$th telescope, and $n(\vec{x}_j,z,t)$ is the refractive index at height $z$ above the $j$th telescope at time $t$. The \emph{Taylor frozen-flow hypothesis} says that the atmosphere can be modeled as multiple static phase screens driven by winds. Each phase screen has its own height and is driven by different wind velocities. Therefore, we can write
\begin{equation}
    \delta_j(t) = \sum_l \phi_l(\vec{x_j}-\vec{v}_l t)
\end{equation}
where $\phi_l$ is the $l$th phase screen, and $\vec{v}_l$ is its wind velocity. Therefore, in this model, the temporal fluctuation at a certain telescope site can be described by the spatial fluctuations of phase screens multiplied by the wind velocities. For each phase screen, its spatial fluctuation is determined by a \emph{Gaussian} distribution with a second-order structure function
\begin{equation}
    D_\mathrm{\phi_l}(r) := \mathbb{E}_{\vec{x}}\left[\left(\phi_l(\vec{x}) - \phi_l(\vec{x}+r)\right)^2\right].
\end{equation}
The second-order structure function is given by
\begin{equation}
    D_{\phi_l}(r) = 6.88\left(\frac{r}{r_{0,l}}\right)^{\frac{5}{3}}
\end{equation}
where $r_{0,l}$ is the \emph{Fried parameter} of the $l$th layer, determined by its strength of turbulence. One can derive the temporal fluctuation from the spatial fluctuation through $r=\tau v_l$:
\begin{equation}
    D_{\phi_l}(\tau) := \mathbb{E}_{\vec{x}}\left[ (\phi_l(\vec{x}) - \phi_l(\vec{x}+v_l \tau))^2 \right] = 6.88\left(\frac{\tau v_
    l}{r_{0,l}} \right)^{\frac{5}{3}}.
\end{equation}
Now, assuming different phase screens are independent to each other, we observe that the the multiple-layer model can be seen as a single effective phase screen:
\begin{equation}
\begin{aligned}
 &D_{\delta}(\tau)\\
 &\quad=\mathbb{E}_{\vec{x}}\left[\left(\sum_l\phi_l(\vec{x})
 -\sum_l\phi_l(\vec{x}+\vec v_l\tau)\right)^2\right]\\
 &\quad=\sum_lD_{\phi_l}(\tau)\\
 &\quad=6.88\tau^{\frac53}\sum_l\left(\frac{v_l}{r_{0,l}}\right)^{\frac53}
 =\left(\frac{\tau}{\tau_c}\right)^{\frac53}.
\end{aligned}
\end{equation}
where
\begin{equation}
    \tau_c = \left[6.88 \sum_l \left(\frac{ v_
    l}{r_{0,l}} \right)^{\frac{5}{3}}\right]^{-\frac{3}{5}}
\end{equation}
is the \emph{atmospheric coherence time} (ACT). In optical/near-IR, typically we have $\tau_c \lesssim 10\,\mathrm{ms}$. With this, it is easy to see that even if $\delta_j(0) = 0 $, after any $t \gg \tau_c$, $\delta_j(t)$ has uncertainty much greater than $2\pi$ such that one can safely assume $\delta_j(t)$ is uniformly random between $0$ and $2\pi$ for a uniformly randomly chosen $t$.

We can now write down the probability distribution of $\delta_j$ using the second-order structure function. For some reference time $t$, the phase error at time $t+\tau$ follows the Gaussian distribution
\begin{equation}
    \delta_j(t+\tau) \sim \mathcal{N}(\delta_j(t),D_\delta(\tau)).
\end{equation}
In other words,
\begin{equation}
\begin{aligned}
 &p(\delta_j(t+\tau))\\
 &\quad=\frac{1}{\sqrt{2\pi D_\delta(\tau)}}
 \exp\left[-\frac{(\delta_j(t+\tau)-\delta_j(t))^2}{2D_\delta(\tau)}\right]\\
 &\quad=\frac{1}{\sqrt{2\pi(\tau/\tau_c)^{5/3}}}
 \exp\left[-\frac{(\delta_j(t+\tau)-\delta_j(t))^2}{2(\tau/\tau_c)^{5/3}}\right].
\end{aligned}
\end{equation}

Since we are in the long-baseline scenario where the baseline is much longer than the Fried parameter (or the coherence length) $r_0 \sim \tau_c v_\mathrm{wind} \sim 0.1\,\mathrm{m}$, we can also assume the phase errors at different telescope sites are independent to each other, i.e.,
\begin{equation}
    p(\vec{\delta}(t)) = \prod_{j=1}^3 p(\delta_j(t)).
\end{equation}

\subsection{Bandwidth restriction}
So far, we have treated all photons received within the same ACT as experiencing the same telescope-dependent piston phases. We now estimate the spectral bandwidth over which this approximation remains valid. In the nondispersive regime, let $S_j(t)$ denote the atmospheric piston optical-path fluctuation at telescope $j$, which is approximately independent of optical frequency across a sufficiently narrow spectral channel. According to \cref{eq:delta_j}, we have
\begin{equation}
    S_j =  \int n(\vec{x}_j,z,t)dz.
\end{equation}
At frequency $\nu$, the corresponding phase is
\begin{equation}
    \delta_j(\nu,t)=\frac{2\pi\nu}{c} S_j(t).
\end{equation}
Therefore, for a differential piston $S_{ij}=S_i-S_j$, the phase variation across a spectral channel of width $\Delta\nu$ is
\begin{equation}
    \Delta\delta_{ij}
    \simeq
    \frac{2\pi |S_{ij}|}{c}\Delta\nu.
\end{equation}
We require this variation to be smaller than some tolerance $\epsilon_\phi$, therefore
\begin{equation}
    \Delta\nu
    \lesssim
    \frac{\epsilon_\phi c}{2\pi |S_{ij}|}.
    \label{eq:piston_bandwidth}
\end{equation}
Note that this relation uses the assumption the air is nondispersive at the carrier frequency of our interest.

As a representative estimate, atmospheric differential piston fluctuations can be of order $p_{\rm rms}\sim 35\,\mu{\rm m}$ for a $47\,{\rm m}$ baseline under $1''$ seeing conditions~\cite{absil2006performance}. Taking $\epsilon_\phi=0.1$ rad, \cref{eq:piston_bandwidth} gives
\begin{equation}
    \Delta\nu
    \lesssim
    1.4\times10^{11}\ {\rm Hz}
    \simeq
    140\ {\rm GHz}.
\end{equation}
Thus, spectral channels with bandwidth of order $10^{11}\,{\rm Hz}$ are sufficiently narrow for the piston phase to be approximately common across the channel at the $0.1$-rad level.

\section{Quantum-optimal bright-scene closure phase estimation}
\label{sec:bright}

In this section, we investigate the fundamental quantum limit of closure phase estimation when there are sufficiently many photons from each ACT. We study this problem via computing the QFI in this scenario.
The caveat in this case is that even in the bright-scene imaging scenario, the density matrix is determined by not only the closure phase, but also the pairwise corrupted visibility phases. Since the piston phase error is unknown, these phases are \emph{nuisance parameters}. In this case, it might be impossible to saturate the QFI of the closure phase. Instead, we must compute the quantum Fisher information matrix (QFIM) that involves both the closure phase and the nuisance phases to analyze the quantum limit. Also, the optimal measurement scheme in this case will be the eigenbasis of the effective SLD which takes the nuisance parameters into account, meaning that some rough pre-measurements of the visibility phases might be necessary. For more details about parameter estimation in the presence of nuisance parameters, we refer the reader to Suzuki et al. \cite{suzuki2020quantum}.

Consider the 3-site density matrix as in Eq.~\eqref{eq:dm_noisy}:
\begin{widetext}
    \begin{equation}
 \rho= \frac{1}{3}\left(\begin{array}{@{}ccc@{}}
1 & |V_{AB}| e^{\imag\phi_{AB}} & |V_{CA}| e^{\imag(\phi_{AB}+\phi_{BC}-\phi_\mathrm{cl})} \\
|V_{AB}| e^{-\imag\phi_{AB}} & 1 & |V_{BC}| e^{\imag \phi_{BC}} \\
|V_{CA}| e^{-\imag(\phi_{AB}+\phi_{BC}-\phi_\mathrm{cl})} & |V_{BC}| e^{-\imag\phi_{BC}} & 1
\end{array}\right).
\end{equation}
\end{widetext}
We use a gauge transformation to simplify this density matrix. Letting the transformation unitary $U_\mathrm{gauge}$ be
\begin{equation}
    U_\mathrm{gauge} = \begin{pmatrix}
        1 & 0 & 0\\
        0 & e^{\imag \phi_{AB}} & 0\\
        0 & 0 & e^{\imag (\phi_{AB}+\phi_{BC})}
    \end{pmatrix},
\end{equation}
then
\begin{equation}
    \rho = U_\mathrm{gauge}^\dagger \tilde{\rho}  U_\mathrm{gauge},
\end{equation}
where
\begin{equation}
    \tilde{\rho} = \frac{1}{3} \begin{pmatrix}
1 & |V_{AB}| & |V_{CA}|e^{-\imag \phi_\mathrm{cl}} \\
|V_{AB}| & 1 & |V_{BC}| \\
|V_{CA}|e^{\imag \phi_\mathrm{cl}} & |V_{BC}| & 1
\end{pmatrix}.
\end{equation}

Now, we can write down
\begin{equation}
    \frac{\partial \rho}{\partial \phi_\mathrm{cl}} = \frac{1}{3}  U^\dagger_\mathrm{gauge} \begin{pmatrix}
0 & 0 & -\imag |V_{CA}| e^{-\imag \phi_\mathrm{cl}} \\
0 & 0 & 0 \\
\imag |V_{CA}| e^{\imag \phi_\mathrm{cl}} & 0 & 0
\end{pmatrix} U_\mathrm{gauge},
\end{equation}

\begingroup
\setlength{\arraycolsep}{2pt}
\begin{equation}
\begin{aligned}
&
    \frac{\partial \rho}{\partial \phi_{AB}}\\ &\quad= \frac{1}{3}U^\dagger_\mathrm{gauge} \left(\begin{array}{@{}ccc@{}}
0 & \imag |V_{AB}| & \imag |V_{CA}| e^{-\imag \phi_\mathrm{cl}} \\
-\imag |V_{AB}| & 0 & 0 \\
-\imag |V_{CA}| e^{\imag \phi_\mathrm{cl}} & 0 & 0
\end{array}\right) U_\mathrm{gauge},
\end{aligned}
\end{equation}
\endgroup

\begingroup
\setlength{\arraycolsep}{2pt}
\begin{equation}
\begin{aligned}
&
    \frac{\partial \rho}{\partial \phi_{BC}}\\ &\quad= \frac{1}{3} U^\dagger_\mathrm{gauge} \left(\begin{array}{@{}ccc@{}}
0 & 0 & \imag |V_{CA}| e^{-i\phi_\mathrm{cl}} \\
0 & 0 & \imag |V_{BC}| \\
-\imag |V_{CA}| e^{\imag \phi_\mathrm{cl}} & -\imag |V_{BC}| & 0
\end{array}\right) U_\mathrm{gauge}.
\end{aligned}
\end{equation}
\endgroup

The SLD corresponding to the closure phase, $L_\mathrm{cl}$, is defined via $\frac{\partial \rho}{\partial \phi_\mathrm{cl}} = \frac{1}{2}(L_\mathrm{cl}\rho + \rho L_\mathrm{cl})$. Using this definition, we can also derive the gauge transformed symmetric logarithmic derivative $\tilde{L}_\mathrm{cl} = U_\mathrm{gauge} L_\mathrm{cl} U_\mathrm{gauge}^\dagger$ as follows:
\begin{equation}
\label{eq:sldequation}
\begin{aligned}
&\frac{\partial\rho}{\partial\phi_\mathrm{cl}}
 =\frac12(L_\mathrm{cl}\rho+\rho L_\mathrm{cl})\\
 &\implies\frac13\begin{pmatrix}
 0&0&-\imag|V_{CA}|e^{-i\phi_\mathrm{cl}}\\
 0&0&0\\
 \imag|V_{CA}|e^{i\phi_\mathrm{cl}}&0&0
 \end{pmatrix}\\
 &\quad=\frac12U_\mathrm{gauge}(L_\mathrm{cl}\rho+\rho L_\mathrm{cl})U^\dagger_\mathrm{gauge}\\
 &\quad=\frac12U_\mathrm{gauge}L_\mathrm{cl}U_\mathrm{gauge}^\dagger
 U_\mathrm{gauge}\rho U_\mathrm{gauge}^\dagger\\
 &\qquad+\frac12U_\mathrm{gauge}\rho U_\mathrm{gauge}^\dagger
 U_\mathrm{gauge}L_\mathrm{cl}U^\dagger_\mathrm{gauge}\\
 &\quad=\frac12(\tilde L_\mathrm{cl}\tilde\rho+\tilde\rho\tilde L_\mathrm{cl}).
\end{aligned}
\end{equation}
The original SLD can also be easily computed from it. Observe that the gauge transformed SLD shares the same eigenvalues with the original SLD, and the eigenvectors are related by the gauge transformation: given $\tilde{L}_\mathrm{cl}\ket{\tilde{\phi}} = \lambda \ket{\tilde{\phi}}$, then
\begin{equation}
    L_\mathrm{cl} U^\dagger_\mathrm{gauge} \ket{\tilde{\phi}} = \lambda U^\dagger_\mathrm{gauge}\ket{\tilde{\phi}} =: \lambda \ket{\phi}.
\end{equation}
Therefore, the optimal measurement basis (when the nuisance phases are known) is simply the gauge transformed eigenstates of $\tilde{L}_\mathrm{cl}$. The same derivation also applies to $\tilde{L}_\mathrm{AB}$ and $\tilde{L}_{BC}$.

With Equation~\eqref{eq:sldequation}, we can solve all three gauge transformed SLDs, $\tilde{L}_\mathrm{cl},\tilde{L}_{AB},\tilde{L}_{BC}$. For simplicity, we let $|V_{AB}|=|V_{BC}|=|V_{CA}|=v$ and assume that we know $\phi_\mathrm{cl}$ is close to $0$ a priori. Now, we can compute the quantum Fisher information matrix $J$ defined as
\begin{equation}
         J = \begin{pmatrix}
        \mathcal{F}_{\phi_{AB},\phi_{AB}} & \mathcal{F}_{\phi_{AB},\phi_{BC}} & \mathcal{F}_{\phi_{AB},\phi_{c}}\\
\mathcal{F}_{\phi_{BC},\phi_{AB}} & \mathcal{F}_{\phi_{BC},\phi_{BC}} & \mathcal{F}_{\phi_{BC},\phicl}\\
\mathcal{F}_{\phicl,\phi_{AB}} & \mathcal{F}_{\phicl,\phi_{BC}} & \mathcal{F}_{\phicl,\phicl}
    \end{pmatrix}, 
\end{equation}
where each
\begin{equation}
    \mathcal{F}_{i,j} = \mathrm{Re}\Tr[\rho L_i L_j] = \mathrm{Re} \Tr[\tilde{\rho} \tilde{L}_i \tilde{L}_j]
\end{equation}
is gauge-invariant, which means that we can use the gauge transformed SLDs and density matrix to compute the QFIM. The effective QFI of $\phi_\mathrm{cl}$ with two nuisance parameters $\phi_{AB},\phi_{BC}$ is
\begin{equation}
\label{eq:qficlosure}
    \mathcal{F}^\mathrm{eff}_{\phi_\mathrm{cl},\phi_\mathrm{cl}} =\frac{1}{(J^{-1})_{\phi_\mathrm{cl},\phi_\mathrm{cl}}} = \frac{2v^2}{9(1-v)} \geq \frac{2v^2}{9} =  O(v^2).
\end{equation}
This effective QFI is the fundamental quantum limit of any measurement scheme. This means that the standard deviation of closure phase estimation cannot be lower than $\sim \frac{1}{v\sqrt{n}}$ when using $n$ photons in this bright-scene setting, for $v$ bounded away from 1. 

Now, we derive the optimal measurement scheme to saturate the QFI. To do so, we compute the effective SLD in the presence of nuisance parameters:
\begin{equation}
\begin{aligned}
L^\mathrm{eff}_{\phi_\mathrm{cl}}
 &=(J^{-1})_{\phi_{AB},\phi_\mathrm{cl}}L_{\phi_{AB}}\\
 &\quad+(J^{-1})_{\phi_{BC},\phi_\mathrm{cl}}L_{\phi_{BC}}\\
 &\quad+(J^{-1})_{\phi_\mathrm{cl},\phi_\mathrm{cl}}L_{\phi_\mathrm{cl}}\\
 &=U^\dagger_\mathrm{gauge}\Bigl[
 (J^{-1})_{\phi_{AB},\phi_\mathrm{cl}}\tilde L_{\phi_{AB}}\\
 &\qquad+(J^{-1})_{\phi_{BC},\phi_\mathrm{cl}}\tilde L_{\phi_{BC}}\\
 &\qquad+(J^{-1})_{\phi_\mathrm{cl},\phi_\mathrm{cl}}\tilde L_{\phi_\mathrm{cl}}
 \Bigr]U_\mathrm{gauge}.
\end{aligned}
\end{equation}
Therefore, the optimal measurement basis is given by the eigenstates of $L^\mathrm{eff}_{\phi_\mathrm{cl}}$, which are
\begin{equation}
    \begin{aligned}
        \ket{u_1} &=\frac{1}{\sqrt{3}} U^\dagger_\mathrm{gauge} (e^{2\pi i/3}\ket{100}+e^{4\pi i/3}\ket{010}+\ket{001})\\
        \ket{u_2} &= \frac{1}{\sqrt{3}} U^\dagger_\mathrm{gauge} (e^{4\pi i/3}\ket{100}+e^{2\pi i/3}\ket{010}+\ket{001})\\
        \ket{u_3} &= \frac{1}{\sqrt{3}} U^\dagger_\mathrm{gauge} (\ket{100}+\ket{010}+\ket{001})\\
    \end{aligned}
\end{equation}
with eigenvalues $\lambda_1 = -\frac{3\sqrt{3}}{2v}, \lambda_2 = \frac{3\sqrt{3}}{2v}, \lambda_3 = 0$. This is essentially a rotated \emph{Tritter} measurement, or Fourier transform of dimension $3$.

We can see that while these measurements are optimal with $O(v^2)$ effective QFI, one needs to first estimate the two nuisance parameters $\phi_{AB}$ and $\phi_{BC}$ in order to find the optimal measurement basis. What is more, the nuisance parameters are affected by the local phase errors, which may only remain stable during the atmospheric coherence time. This implies that we must measure the nuisance parameters for each window. In the limit that the number of photons per window is less than $1$, we will probably never be able to get even a rough estimation. This motivates us to study more robust measurement schemes in the following sections.

\section{Individual-photon measurement Fisher information for $m=3$}
\label{appendix:single_photon_limit_for_m=3}

We calculate the Fisher information per three-photon ACT when the photons are measured sequentially. For this we consider the $n=3$ case of Eq.~\eqref{eq:dm_noisy_atmosphere}. We consider single-photon POVMs that act on all three spatial modes, with each POVM allowed to depend on earlier measurement outcomes. The protocol retains only classical records between photons. The amplitudes $|V_{AB}|$, $|V_{BC}|$, and $|V_{CA}|$ are held fixed while estimating $\phi=\phicl$. Write
\begin{equation}
 \begin{aligned}
 B_{ABC}&=V_{AB}V_{BC}V_{CA}=|V_{AB} V_{BC} V_{CA}| e^{\imag\phicl}
 ,\\ 
 s_2&=\lvert V_{AB}\rvert^2+\lvert V_{BC} \rvert ^2+\lvert V_{CA}\rvert^2.
 \end{aligned}
\end{equation}
We also use $\Gamma  \coloneqq |V_{AB} V_{BC} V_{CA}|$ for simplicity.
All expansions below are uniform in $\phi$ and in the choice of single-photon POVMs as $s_2\to0$. We prove the following theorem.

\begin{theorem}[Optimal individual-photon Fisher information]
\label{thm:seq_fisher}
The maximum classical Fisher information in one three-photon ACT is
\begin{equation}
\label{eq:seq_fisher_optimum}
 \mathcal I_{\mathrm{single}}^{\max}
 =\frac8{81}|V_{AB} V_{BC} V_{CA}|^2+O(|V_{AB} V_{BC} V_{CA}|^2s_2).
\end{equation}
Measuring each photon in the same three-dimensional Fourier basis attains this leading term.
\end{theorem}

The proof has two parts. We first bound the information available after arbitrary measurements of the first two photons. We then compute the outcome distribution of the Fourier measurement and show that it attains the bound.

\subsection{Information left after the first two measurements}
Let $\Sigma_n(V)=\E_{\vec\delta}[(D_{\vec\delta}VD_{\vec\delta}^\dagger/3)^{\otimes n}]$ for $n=2,3$, where $D_{\vec\delta}=\operatorname{diag}(e^{\imag\delta_1},e^{\imag\delta_2},e^{\imag\delta_3})$. As the piston phases are independent uniform random variables, averaging over them preserves only those matrix elements whose row and column strings have the same site occupations. In $\Sigma_2$, these entries depend only on the diagonal of $V$ and on $|V_{ij}|^2$. Hence $\Sigma_2$ is independent of $\phi$.

Now consider sequentially measuring three photons, to get outcomes $x$, $y$, and $z$, with each measurement possibly dependent on previous measurement outcomes. For such a sequential protocol, an outcome branch has effect
\begin{equation}
 E_x\otimes F_{y|x}\otimes G_{z|xy},
\end{equation}
with $\sum_xE_x=\sum_yF_{y|x}=\sum_zG_{z|xy}=I_3$ for the appropriate fixed histories. Since the probability of the first two measurement outcomes is
\begin{equation}
 p_{xy}=\Tr[(E_x\otimes F_{y|x})\Sigma_2]
\end{equation}
which does not depend on $\phi$, all the Fisher information is in the conditional distribution of the third outcome:
\begin{equation}
\label{eq:seq_fisher_chain}
 \mathcal I_{\mathrm{single}}
 =\sum_{x,y}p_{xy}\mathcal I(Z\mid x,y).
\end{equation}
The first two outcomes determine a conditional state of the third photon and the measurement to perform on it.

To upper-bound all sequential protocols, we may enlarge the class of protocols by allowing later measurements to use all previous outcomes and by splitting each of the first two POVM effects into its rank-one spectral components. These changes cannot reduce the maximum Fisher information. We may therefore take
\begin{equation}
 E_x=e_x\ket{u_x}\bra{u_x},\qquad
 F_{y|x}=f_{y|x}\ket{v_{y|x}}\bra{v_{y|x}},
\end{equation}
where $|u_x|=|v_{y|x}|=1$ and
\begin{equation}
 \sum_xe_x=3,\qquad \sum_yf_{y|x}=3.
\end{equation}
Based on this we have the following lemma.

\begin{lemma}[Conditional closure-phase signal]
\label{lem:seq_conditional}
Fix first-two-photon effects $E=e\ket u\bra u$ and $F=f\ket v\bra v$. Define the unnormalized conditional state and its probability by
\begin{equation}
 \tau=\Tr_{12}[(E\otimes F\otimes I_3)\Sigma_3],
 \qquad p=\Tr\tau.
\end{equation}
Let $x,y,z\in\mathbb C^3$ have components
\begin{equation}
\begin{aligned}
 x&=(u_1^*u_2,u_2^*u_3,u_3^*u_1),\\
 y&=(v_1^*v_2,v_2^*v_3,v_3^*v_1),\\
 z_1&=x_2y_3+x_3y_2,\qquad z_2=x_3y_1+x_1y_3,\\
 z_3&=x_1y_2+x_2y_1.
\end{aligned}
\end{equation}
Then $\partial_\phi p=0$,
\begin{equation}
\label{eq:prho3}
 p=\frac{ef}{9}[1+O(s_2)],\qquad
 \rho_{3|u,v}=\frac{\tau}{p}=\frac{I_3}{3}+O(s_2),
\end{equation}
and
\begin{equation}
\label{eq:seq_conditional_derivative}
 \Tr[(\partial_\phi\tau)^2]
 =\frac{2e^2f^2\Gamma^2}{27^2}|z|^2.
\end{equation}
\end{lemma}
\begin{proof}
Let $C$ denote the permutation operation that cycles site labels $1\mapsto2\mapsto3\mapsto1$. Only the block of $\Sigma_3$ spanned by the six permutations $\pi$ of $(1,2,3)$ depends on $\phi$. Its phase-dependent entries give
\begin{equation}
\begin{aligned}
 \partial_\phi\Sigma_3
 =\frac{\imag}{27}\sum_{\pi\in S_3}
 \bigl(B\ket\pi\bra{C\pi}-B^*\ket{C\pi}\bra\pi\bigr).
\end{aligned}
\end{equation}
All other off-diagonal entries involve squared amplitudes, so
$\Sigma_3=I_{27}/27+O(s_2)$ in operator norm. Here the terms of size $\Gamma$ are also $O(s_2)$, since $\Gamma\le(s_2/3)^{3/2}$. Contracting with unit vectors $u,v$ gives
$\tau=ef I_3/27+O(ef s_2)$, which proves the expansions for $p$ and $\rho_{3|u,v}$ (Eq.~\eqref{eq:prho3}). They are uniform over $u,v$.

For the derivative, define the $3\times3$ matrix
\begin{equation}
 M=z_1\ket1\bra2+z_2\ket2\bra3+z_3\ket3\bra1.
\end{equation}
Contracting the first two registers yields
\begin{equation}
 \partial_\phi\tau
 =\frac{ef}{27}\imag(BM-B^*M^\dagger).
\end{equation}
For example, the coefficient of $\ket1\bra2$ comes from the strings $(2,3,1)$ and $(3,2,1)$; their contractions are $x_2y_3$ and $x_3y_2$. The other coefficients follow by cycling the site labels. Finally,
$\Tr M^2=0$ and $\Tr(MM^\dagger)=|z|^2$. Squaring the derivative proves Eq.~\eqref{eq:seq_conditional_derivative}.
\end{proof}
Based on this we now bound the Fisher information of all sequential measurements.

\begin{lemma}[Bound for all sequential measurements]
\label{lem:seq_fisher_bound}
Every adaptive single-photon protocol satisfies
\begin{equation}
 \mathcal I_{\mathrm{single}}
 \le\frac8{81}\Gamma^2+O(\Gamma^2s_2).
\end{equation}
\end{lemma}
\begin{proof}
We first bound the information in any measurement $\{G_z\}$ of a conditional state $\rho=I_3/3+O(s_2)$. Write $\dot\rho=\partial_\phi\rho$. Since
$\Tr(G_z\rho)\ge[1-O(s_2)]\Tr G_z/3$, Cauchy--Schwarz gives
\begin{equation}
\begin{aligned}
 \mathcal I(\{G_z\},\rho)
 &=\sum_z\frac{[\Tr(G_z\dot\rho)]^2}{\Tr(G_z\rho)}\\
 &\le3[1+O(s_2)]\sum_z
       \frac{|\Tr(G_z\dot\rho)|^2}{\Tr G_z}\\
 &\le3[1+O(s_2)]\sum_z\Tr(G_z\dot\rho^2)\\
 &=3[1+O(s_2)]\Tr\dot\rho^2.
\end{aligned}
\end{equation}
The second inequality uses
$|\Tr(G_z\dot\rho)|^2\le\Tr G_z\,\Tr(G_z\dot\rho^2)$.
Zero effects may be omitted.

By Lemma~\ref{lem:seq_conditional}, $\dot\rho=\dot\tau/p$. Thus, each branch of the first two measurements contributes at most
\begin{equation}
\label{eq:seq_branch_information}
 p\,\mathcal I(Z\mid u,v)
 \le\frac{2ef\Gamma^2}{27}|z|^2+O(ef\Gamma^2s_2).
\end{equation}
We now bound $|z|^2$. For arbitrary $x,y\in\mathbb C^3$, define
$\mathcal B(x,y)=(x_2y_3+x_3y_2,x_3y_1+x_1y_3,x_1y_2+x_2y_1)$.
The following elementary inequality is useful:
\begin{equation}
\label{eq:seq_bilinear}
 |\mathcal B(x,y)|^2\le\frac43|x|^2|y|^2.
\end{equation}
To prove it, normalize nonzero $x,y$ to unit length and set
$D=\sum_i|x_i|^2|y_i|^2$. Expanding the left-hand side gives
\begin{equation}
 |\mathcal B(x,y)|^2=1+|\langle x,y\rangle|^2-2D.
\end{equation}
Cauchy--Schwarz gives both $|\langle x,y\rangle|^2\le1$ and
$|\langle x,y\rangle|^2\le3D$. Since
$\min\{1,3D\}\le1/3+2D$, the last display is at most $4/3$. Restoring the norms proves Eq.~\eqref{eq:seq_bilinear}.

For the vectors arising from $u,v$, put $p_i=|u_i|^2$. Then
\begin{equation}
 |x|^2=\sum_{i<j}p_ip_j
 =\frac{1-\sum_ip_i^2}{2}\le\frac13,
\end{equation}
because $\sum_i p_i=1$. Likewise $|y|^2\le1/3$, and hence $|z|^2\le4/27$. Substituting into Eq.~\eqref{eq:seq_branch_information} and summing the branches gives
\begin{equation}
\begin{aligned}
 \mathcal I_{\mathrm{single}}
 &\le\frac{8\Gamma^2}{729}\sum_{x,y}e_xf_{y|x}
            +O(\Gamma^2s_2)\\
 &=\frac8{81}\Gamma^2+O(\Gamma^2s_2),
\end{aligned}
\end{equation}
where $\sum_{x,y}e_xf_{y|x}=3\cdot3=9$. This includes arbitrary choices of the second and third measurements based on the preceding outcomes.
\end{proof}

\subsection{A Fourier measurement attaining the bound}
\begin{proof}[Proof of Theorem~\ref{thm:seq_fisher}]
The upper bound is Lemma~\ref{lem:seq_fisher_bound}. For attainability, let $\omega=e^{2\pi\imag/3}$ and measure each photon in the fixed basis
\begin{equation}
 \ket{f_r}=\frac1{\sqrt3}
       (\ket1+\omega^r\ket2+\omega^{2r}\ket3),
 \qquad r=0,1,2.
\end{equation}
This is a balanced three-port interferometer followed by photodetection. Record the three outcomes $(r,s,t)$.

For any $\pi\in S_3$,
\begin{equation}
 \langle f_r,f_s,f_t|\pi\rangle
 \langle C\pi|f_r,f_s,f_t\rangle
 =\frac1{27}\omega^{r+s+t}.
\end{equation}
Summing the six permutations and adding the conjugate terms gives
\begin{equation}
\begin{aligned}
 p_{rst}
 &=\frac1{27}+R_{rst}
   +\frac{4\Gamma}{243}
       \cos\!\left(\phi+\frac{2\pi(r+s+t)}3\right),\\
 \partial_\phi p_{rst}
 &=-\frac{4\Gamma}{243}
       \sin\!\left(\phi+\frac{2\pi(r+s+t)}3\right),
\end{aligned}
\end{equation}
where $R_{rst}=O(s_2)$ is independent of $\phi$. Thus $p_{rst}^{-1}=27[1+O(s_2)]$. Each residue $q=r+s+t\bmod3$ occurs for nine outcome triples, so
\begin{equation}
\begin{aligned}
 &\sum_{r,s,t}\sin^2\!\left(\phi+\frac{2\pi(r+s+t)}3\right)\\
 &\quad=9\sum_{q=0}^2\sin^2\!\left(\phi+\frac{2\pi q}3\right)
 =\frac{27}{2}.
\end{aligned}
\end{equation}
Consequently,
\begin{equation}
\begin{aligned}
 \mathcal I_{\mathrm{Fourier}}
 &=\sum_{r,s,t}\frac{(\partial_\phi p_{rst})^2}{p_{rst}}\\
 &=27\left(\frac{4\Gamma}{243}\right)^2\frac{27}{2}
       +O(\Gamma^2s_2)\\
 &=\frac8{81}\Gamma^2+O(\Gamma^2s_2).
\end{aligned}
\end{equation}
This attains the sequential upper bound to leading order for every closure phase.
\end{proof}

\section{SLD and QFI computation of the $3$-photon $3$-site case}
\label{sec:details}
In this appendix, we present the exact computation of $L_\mathrm{cl}$ and $F_\mathrm{cl}$ in the $3$-photon $3$-site case. To simplify the notation, we use
\begin{equation}
    a\coloneqq |V_{AB}|, \, b\coloneqq |V_{BC}|, \, c\coloneqq |V_{AC}|.
\end{equation}
Recall that
\begin{equation}
    M_{ABC} = p_\mathrm{trivial} \oplus p_\mathrm{sign} \oplus (M_\mathrm{std} \otimes I^{2\times 2}).
\end{equation}
We first present the exact expression of $M_{ABC}$, using  $\omega = e^{2\pi \imag /3}$:
\begin{align}
    p_\mathrm{trivial} &= \frac{1}{27} \left( 1+a^2+b^2+c^2+2abc\cos(\phicl)\right),\\
    p_\mathrm{sign} &= \frac{1}{27} \left( 1 - a^2 - b^2 - c^2 +2abc\cos(\phicl) \right), \\
M_\mathrm{std}&=\frac1{27}
 \begin{pmatrix}
 \substack{\displaystyle 1-abc\cos\phicl\\\displaystyle +\sqrt{3}abc\sin\phicl}
 &b^2+c^2\omega+a^2\omega^2\\
 b^2+c^2\omega^2+a^2\omega
 &\substack{\displaystyle 1-abc\cos\phicl\\\displaystyle -\sqrt{3}abc\sin\phicl}
 \end{pmatrix}.
\end{align}
For the two $1\times 1$ blocks,
\begin{equation}
\begin{aligned}
    \partial p_\mathrm{trivial} / \partial \phicl &= -\frac{2}{27} abc \sin\phicl \\
    \partial p_\mathrm{sign} / \partial \phicl &= -\frac{2}{27} abc \sin\phicl
\end{aligned}
\end{equation}
Their overall contribution to the QFI is
\begin{equation}
\begin{aligned}
 &\frac{(\partial p_\mathrm{trivial}/\partial\phicl)^2}{p_\mathrm{trivial}}
 +\frac{(\partial p_\mathrm{sign}/\partial\phicl)^2}{p_\mathrm{sign}}\\
 &=\frac{8a^2b^2c^2\sin^2\phicl}{27}\\
 &\quad\times\left[\frac{1+2abc\cos\phicl}
 {(1+2abc\cos\phicl)^2-(a^2+b^2+c^2)}\right],
\end{aligned}
\end{equation}
and their individual SLDs are simply real numbers:
\begin{equation}
\begin{aligned}
    L_\mathrm{cl,trivial} &= \frac{-2abc\sin\phicl}{1+a^2+b^2+c^2+2abc\cos\phicl},\\
    L_\mathrm{cl,sign} &= \frac{-2abc\sin\phicl}{1-a^2-b^2-c^2+2abc\cos\phicl}.
\end{aligned}
\end{equation}

Note that we can transform the $4\times 4$ block of standard irrep to two $2\times 2$ blocks by simply swapping two basis states. This gives
\begin{equation}
\begin{aligned}
 &U_\mathrm{basis}M_{ABC}U_\mathrm{basis}\\
 &\quad=p_\mathrm{trivial}\oplus p_\mathrm{sign}\oplus(I^{2\times2}\otimes M_\mathrm{std})\\
 &\quad=p_\mathrm{trivial}\oplus p_\mathrm{sign}\oplus M_\mathrm{std}\oplus M_\mathrm{std}.
\end{aligned}
\end{equation}
Note that $M_\mathrm{std}$ is essentially a single-qubit Hamiltonian, which means that it can be decomposed into the linear combination of four Pauli matrices. We let $u\coloneqq b^2+c^2\omega + a^2\omega^2$. To simplify the problem a little bit (to avoid Pauli $Y$), we can first apply a diagonal unitary $V M_\mathrm{std} V^\dagger$ where $V$ is defined as
\begin{equation}
    V = \begin{pmatrix}
        e^{-\imag \chi /2} & 0\\
        0 & e^{\imag \chi / 2} 
    \end{pmatrix} = e^{-\imag \chi Z/2},\quad \chi = \arg(u)
\end{equation}
When $u=0$, take $\chi=0$. Observe that
\begin{equation}
\begin{aligned}
 M'_\mathrm{std}&=VM_\mathrm{std}V^\dagger\\
 &=\frac1{27}\Bigl[(1-abc\cos\phicl)I\\
 &\qquad+|u|X-\sqrt3abc\sin\phicl Z\Bigr].
\end{aligned}
\end{equation}
Now, we compute the derivative:
\begin{equation}
\begin{aligned}
 &\partial M'_\mathrm{std}/\partial\phicl\\
 &\quad=\frac1{27}\left[abc\sin\phicl I-\sqrt3abc\cos\phicl Z\right]\\
 &\quad=\frac12(L_\mathrm{cl,std}M'_\mathrm{std}+M'_\mathrm{std}L_\mathrm{cl,std}).
\end{aligned}
\end{equation}
We can assume
\begin{equation}
    L_\mathrm{cl,std} = l_I I + l_X X + l_Z Z.
\end{equation}
Finally, we can solve the SLD equation $\partial M'_\mathrm{std} / \partial \phicl = \frac{1}{2} (L_\mathrm{cl,std} M'_\mathrm{std} + M'_\mathrm{std} L_\mathrm{cl,std} )$:
\begin{align}
l_I &= \frac{
abc\sin\phicl\left(1-4abc\cos\phicl\right)}{(1-abc\cos\phicl)^2
-|u|^2
-3a^2b^2c^2\sin^2\phicl},\\
l_X &= -\frac{|u| l_I }{1-abc\cos\phicl} ,
\\
l_Z &= \frac{ \sqrt{3}abc( \cos\phicl -l_I \sin\phicl)}{1-abc\cos\phicl}. 
\end{align}

Now, the quantum Fisher information contribution of this unnormalized
$2\times2$ block is
\begin{equation}
\begin{aligned}
F_{H,\mathrm{std}}&=\Tr\left(M'_\mathrm{std}L_\mathrm{cl,std}^2\right)\\
 &=\frac1{27}\Bigl[2abc\sin\phicl l_I\\
 &\qquad-2\sqrt3abc\cos\phicl l_Z\Bigr]\\
 &=\frac{2a^2b^2c^2}{27}\Bigg[\frac{3\cos^2\phicl}{1-abc\cos\phicl}\\
 &\quad+\frac{\sin^2\phicl(1-4abc\cos\phicl)^2}
 {\begin{aligned}
 &(1-abc\cos\phicl)\\[-2pt]
 &\quad\times\bigl[(1-abc\cos\phicl)^2\\[-2pt]
 &\qquad-|u|^2-3a^2b^2c^2\sin^2\phicl\bigr]
 \end{aligned}}\Bigg].
\end{aligned}
\end{equation}

Finally, taking the sum of all QFI contributions, we obtain the overall QFI of $M_{ABC}$:
\begin{align}
&\begin{aligned}
 F_\mathrm{cl}&=\frac{4a^2b^2c^2}{27}\Bigg[\\
 &\quad\frac{2\sin^2\phicl(1+2abc\cos\phicl)}
 {(1+2abc\cos\phicl)^2-(a^2+b^2+c^2)^2}\\
 &\quad+\frac{3\cos^2\phicl}{1-abc\cos\phicl}
 \end{aligned}\\
 &\begin{aligned}
 &\quad+\frac{\sin^2\phicl(1-4abc\cos\phicl)^2}
 {\begin{aligned}
 &(1-abc\cos\phicl)\\
 &\quad\times\bigl[(1-abc\cos\phicl)^2\\
 &\qquad-|u|^2-3a^2b^2c^2\sin^2\phicl\bigr]
 \end{aligned}}\Bigg],
 \end{aligned}
\end{align}
In the weak-visibility limit ($a,b,c\ll 1$), we can see that
\begin{equation}
    F_\mathrm{cl} = \frac{4a^2 b^2 c^2}{9} [1+O(a^2+b^2+c^2)].
\end{equation}
The overall SLD is
\begin{equation}
    L_\mathrm{cl} = L_\mathrm{cl,trivial} \oplus L_\mathrm{cl,sign} \oplus L_\mathrm{cl,std} \oplus L_\mathrm{cl,std}.
\end{equation}

Observe that the SLD is also block-diagonal in the Fourier basis of $S_3$. 
This implies that, to achieve the optimal efficiency in closure phase estimation, we can transform the $M_{ABC}$ block of $\rho_{3}^\mathrm{noisy}$ to the Fourier basis of $S_3$. This operation is essentially quantum Fourier transform on the $S_3$ finite group (denoted by $\mathrm{QFT}_{S_3}$).

\section{Generalized cycle measurement} \label{sec:TorusQFT}

Recall that there is a convenient \emph{cycle basis} for the linearly independent closure phases (Eq.~\eqref{eq:cycleBasis}), which suggests that we can put the telescope sites on a high-dimensional torus $\mathbb{T}^M$, where the dimension is the number of closure phases: $M=(m-1)(m-2)/2$. Thus, perform a quantum Fourier transform on $\rho^\text{noisy}_3$ where the boundaries are set by the manifold $\mathbb{T}^{M}$. To implement it, introduce a classical register that keeps track of the coordinates. We can also view this procedure as phase estimation.
\begin{equation}
\begin{aligned}
 &\frac{1}{3^M}\sum_{\vec x\in\{0,1,2\}^M}\ket{\vec x}\bra{\vec x}
 \otimes\rho^\text{noisy}_3\left(\vec\theta+\frac{2\pi\vec x}{3}\right)\\
 &\xrightarrow{\text{QFT}_{\mathbb T^{M}}}
 \frac{1}{3^M}\Bigg[I\otimes\rho_\perp\\
 &\qquad\qquad+\frac{1}{m^3}\sum_t(B_t\cdot S_{\vec c_t}\otimes C_t+h.c.)\Bigg]\,,
\end{aligned}
\end{equation}
where 
\begin{equation}
    B_t = abc \, e^{i \phi_{\text{cl},t}} \,,
\end{equation}
and
\begin{equation}
    S_{\vec c} = \sum_{\vec i} \ket{\vec i }\bra{\vec i + \vec c} \,,
\end{equation}
and
\begin{equation}
    C_t = \sum_{\vec s \in S_3^+ (t)} \ket{\vec s}\bra{\text{cyc}(\vec s)} + \sum_{\vec s \in S_3^- (t)} \ket{\vec s}\bra{\text{cyc}^{-1}(\vec s)}
\end{equation}
where $\text{cyc}$ shifts all elements by one. We want to extract $B_t$, which is now a coherence in a block for each cycle. Measuring with the projectors
\begin{equation}
    P_{t,q}=\frac{1}{3}(I+ \omega^{-q} S_{\vec c_t}\otimes C_t + \omega^{-2q} S_{\vec c_t}^2\otimes C_t^2) \,,
\end{equation}
where $\omega = e^{2\pi i / 3}$ and $q \in \{0,1,2\}$, 
yields measurement probabilities
\begin{equation}
    p_{t,q}= \frac{2}{m^3}(1+2 abc \, \cos(\phi_\text{cl,t} + \frac{2\pi q}{3}) )\,.
\end{equation}
The Fisher information of this classical measurement is (for $\phi_\text{cl}=0$)
\begin{equation}
    \frac{12 (abc)^2}{m^3 (1-v^3)} \,,
\end{equation}
which is optimal. Note that we can do a classical Fourier transform as postprocessing.

The above approach generalizes to $n$ photons in an atmospheric coherence window. We use the same triangular cycle basis. Each spatial coordinate is discretized into $L$ values, where $L>2n$ in order to prevent Fourier aliasing. The transformation using the quantum Fourier transform is
\begin{equation}
\begin{aligned}
 &\frac{1}{L^M}\sum_{\vec x\in\mathbb Z_L^M}\ket{\vec x}\bra{\vec x}
 \otimes\rho^\text{noisy}_n\left(\vec\theta+\frac{2\pi\vec x}{L}\right)\\
 &\xrightarrow{\text{QFT}_{\mathbb T^M}}
 \frac{1}{L^M}\sum_{\vec\sigma\in\Lambda_n}
 e^{i\vec\sigma\cdot\vec\theta}S_{\vec\sigma}\otimes A_{\vec\sigma}^{(n)}\,,
\end{aligned}
\end{equation}
where 
\begin{equation}
    \vec \sigma \cdot \vec \theta \,
\end{equation}
is a closure phase for cycle $\vec \sigma$, and
\begin{equation}
    S_{\vec \sigma} = \sum_{\vec p} \ket{\vec p }\bra{\vec p + \vec \sigma}
\end{equation}
where $\vec p$ is the Fourier dual coordinate vector to $\vec x$, and
\begin{equation}
    A_{\vec \sigma}^{(n)} = \frac{v^n}{m^n}  \sum_{\vec a, \vec b\text{ related by }\vec \sigma} \ket{\vec a}\bra{\vec b} 
\end{equation}
where $v$ denotes the visibility amplitude (taken to be the same for all baselines) and $\vec a,\vec b$ are length-$n$ vectors of telescope site indices. A dimension-$l$ QFT learns closure phases of length $l$, similar to before.

\section{Technical details for the three-photon upper bound}
\label{sec:three_photon_information}
This appendix proves the technical steps invoked, but not proved, between
\cref{thm:joint_3_m_scaling} and Eq.~(62). The low-dimensional
bispectrum representation and the regression bound are already proved
in the main text and are therefore not repeated here.

\subsection{Dimension of the dominant eigenspace}
\label{app:dominant_eigenspace_dimension}
We prove the spectral fact used in
\cref{lem:dominant_eigenspace_dimension}. Since
$H=V\circ\overline V$, the Schur product theorem and $V\succeq0$ imply
$H\succeq0$. Moreover, $H_{ii}=|V_{ii}|^2=1$, so $\Tr H=m$. If
$\lambda_1,\ldots,\lambda_r$ are the eigenvalues of $H$ that are at
least $\tau m$, then
\begin{equation}
    r\tau m
    \leq
    \sum_{\alpha=1}^r\lambda_\alpha
    \leq
    \Tr H
    =m.
\end{equation}
Hence $r\leq1/\tau$.

\subsection{Matrix concentration for learning the dominant eigenspace}
\label{app:matrix_bernstein}
We supply the concentration calculation used in
\cref{lem:learn_H}. Let $q=\binom m2$ and, for an accepted pair record
$(\{i,j\},s)$, define
\begin{equation}
    A_{ij}=\ket{i}\!\bra{j}+\ket{j}\!\bra{i},
    \qquad
    Y=I_m+qsA_{ij}.
\end{equation}
Since every unordered pair is uniformly randomly sampled and
$\E[s|\{i,j\}]=H_{ij}$, we have
\begin{equation}
    \E Y
    =
    I_m+\sum_{i<j}H_{ij}A_{ij}
    =H.
\end{equation}
Let $Z=Y-H$. Since $\|H\|_{\rm op}\leq\Tr H=m$ and
$\|A_{ij}\|_{\rm op}=1$, for $m\geq3$ we have
$\|Z\|_{\rm op}\leq1+q+m\leq m^2$.

For the variance, $s^2=1$ and
$A_{ij}^2=\ket{i}\!\bra{i}+\ket{j}\!\bra{j}$. Each site occurs in
$m-1$ of the $q$ pairs, and hence
$\E A_{ij}^2=(2/m)I_m$. Expanding $Y^2$ gives
\begin{equation}
    \E Y^2
    =
    2H-I_m+\frac{2q^2}{m}I_m.
\end{equation}
Therefore
\begin{equation}
\begin{aligned}
    \E Z^2
    &=\E Y^2-H^2 \\
    &=\frac{2q^2}{m}I_m-(H-I_m)^2
    \preceq\frac{m^3}{2}I_m.
\end{aligned}
\end{equation}
For $M_H$ independent accepted records, the matrix variance parameter
is thus at most $M_Hm^3/2$. The two-sided matrix Bernstein inequality,
applied at threshold $M_H\tau m$, yields
\begin{equation}
\begin{aligned}
    &\Pr\!\left[
        \left\|\frac1{M_H}\sum_{\ell=1}^{M_H}Y_\ell-H\right\|_{\rm op}
        \geq\tau m
    \right] \\
    &\quad\leq
    2m\exp\!\left[-
        \frac{M_H\tau^2}{m(1+2\tau/3)}
    \right] \\
    &\quad\leq
    2m\exp\!\left[-\frac{3M_H\tau^2}{5m}\right],
\end{aligned}
\end{equation}
where the last inequality uses $0<\tau\leq1$. In particular,
$M_H\geq(5m/3\tau^2)\log(2m/\eta)$ makes the failure probability at
most $\eta$.

\subsection{A trace bound from the Hilbert--Schmidt Gram matrix}
\label{app:gram_trace_bound}

The following lemma converts a bound on the Gram matrix of a
collection of operators into a bound on their trace pairings with
an arbitrary matrix.

\begin{lemma}[Trace bound from a Gram matrix]
\label{lem:gram_trace_bound}
Let $R_1,\ldots,R_m$ be Hermitian matrices of the same size, and
let $\Gamma$ be their Hilbert--Schmidt Gram matrix,
$\Gamma_{ij}=\Tr(R_i^\dagger R_j)$.
If $\Gamma\preceq\kappa I$ for some $\kappa\geq0$, then every
complex matrix $A$ of the same size satisfies
\begin{equation}
    \sum_i|\Tr(R_iA)|^2
    \leq\kappa\|A\|_F^2.
    \label{eq:R_trace_bound}
\end{equation}
\end{lemma}

\begin{proof}
Define
\begin{equation}
    W=\sum_i\Tr(R_iA)R_i.
\end{equation}
The Gram-matrix bound gives
\begin{equation}
    \begin{aligned}
        \|W\|_F^2
        &=
        \sum_{i,j}
        \Tr(R_iA)^*
        \Gamma_{ij}
        \Tr(R_jA) \\
        &\leq
        \kappa\sum_i|\Tr(R_iA)|^2.
    \end{aligned}
\end{equation}
Since the $R_i$ are Hermitian, Cauchy--Schwarz yields
\begin{equation}
    \begin{aligned}
        \sum_i|\Tr(R_iA)|^2
        &=
        |\Tr(W^\dagger A)| \\
        &\leq
        \|W\|_F\|A\|_F \\
        &\leq
        \left(
            \kappa\sum_i|\Tr(R_iA)|^2
        \right)^{1/2}
        \|A\|_F.
    \end{aligned}
\end{equation}
If the sum on the left vanishes, the claim is immediate.
Otherwise, squaring and dividing by that sum proves
\cref{eq:R_trace_bound}.
\end{proof}

\subsection{Fixed photon budget and failure-event accounting}
\label{app:failure_accounting}
We now justify the conversion from accepted records to a fixed number
of input photons and close the expected-loss argument. Suppose a stage
requires $M$ accepted records and each attempt is accepted with
probability $p$. Allow
\begin{equation}
    J(M,p,\eta)
    =
    \left\lceil
        \frac{\max\{2M,8\log(1/\eta)\}}{p}
    \right\rceil
\end{equation}
attempts and output zero if fewer than $M$ records are accepted. If
$X$ is the accepted count, then $\E X=Jp\geq2M$, and a binomial
lower-tail bound gives
\begin{equation}
    \Pr[X<M]
    \leq
    \Pr[X<Jp/2]
    \leq
    e^{-Jp/8}
    \leq\eta.
\end{equation}
The values of the accepted records are independent of the acceptance
pattern. Consequently, conditioned on obtaining enough records, the
first $M$ accepted values have exactly the independent uniform-pair or
uniform-triangle distribution used in the main-text lemmas.

Apply this rule to both measurement stages. In addition, after forming
$\widehat H$, output zero whenever the number $r$ of eigenvalues above
$2\tau m$ exceeds $\lceil1/\tau\rceil$. If $r=0$, output zero without
collecting triangle records; on the spectral-accuracy event, the
argument of \cref{lem:bispectrum_low_rank} bounds this zero prediction's
loss by $C_0\tau$ for a universal constant $C_0$. On the same event,
the rank safeguard is never triggered. The two
record-shortage events and the spectral-estimation event have total
failure probability at most $3\eta$; let $\mathcal G$ denote their
complement.

Conditioned on $\mathcal G$, the proofs of
\cref{lem:bispectrum_low_rank,lem:learn_theta} give, for a universal
constant $C_0$,
\begin{equation}
    \E[L(B,\widetilde B)\mid\mathcal G]
    \leq
    C_0\tau+\lambda+\frac{r^3}{M_B}.
\end{equation}
Here the expectation is over the independent triangle records,
conditional on the already fixed pair data. All predictions are
projected onto the unit disk, and the early-termination output is zero.
Since $|B_t|\leq1$, the loss is therefore at most $4$ on every outcome.
It follows that
\begin{equation}
    \E L(B,\widetilde B)
    \leq
    C_0\tau+\lambda+\frac{r^3}{M_B}+12\eta.
\end{equation}
Taking $\tau,\lambda,\eta$ to be sufficiently small constant multiples
of $\epsilon^2$ and
$M_B$ to be a sufficiently large constant multiple of
$r^3/\epsilon^2$ proves the desired expected loss.

Finally, $p_2$ and $p_{\rm dist}$ are bounded below by constants for
$m\geq6$. The fixed-attempt rule therefore changes each accepted-record
budget only by a constant factor, apart from the logarithmic dependence
on $1/\eta$. The rank safeguard gives
$r\leq\lceil1/\tau\rceil$ on every branch on which the regression stage
is run. With $\tau=\Theta(\epsilon^2)$,
\begin{equation}
    M_H
    =
    \widetilde O(m/\epsilon^4),
    \qquad
    M_B
    =
    O(\epsilon^{-8}).
\end{equation}
Each attempt consumes one three-photon ACT, with one photon discarded in
the pair stage. This proves the fixed input-photon bound claimed in
\cref{thm:joint_3_m_scaling}.

\subsection{Polynomial-time classical computation}
\label{app:classical_computation}
It remains to verify the computational claim. Forming $\widehat H$ from
the sparse pair records and diagonalizing it take polynomial time in
$m$. 
Once $U$ is known, forming the features, constructing $G$, accumulating $\widehat c$, solving the $r^3$-dimensional regularized system, and evaluating all $T$ predictions can be done, for example, in
\begin{equation}
    O(T r^6+M_B r^3+ r^9 +T r^3)
\end{equation}
arithmetic operations. Since $r \leq  1/\tau  = O(1/\epsilon^2)$ on
every branch reaching this stage, this is polynomial in $1/\epsilon$.

The regularized system also requires only polynomial precision. Extend
the features in \cref{eq:bispectrum_feature} to all ordered triples.
Orthonormality of the columns of $U$ gives
\begin{equation}
    \frac1{m^3}\sum_{i,j,k}f_{ijk}f_{ijk}^T=I_d.
\end{equation}
Since every summand is positive semidefinite, restricting to
$i<j<k$ yields
\begin{equation}
    0\preceq G\preceq\frac{m^3}{T}I_d
    =\frac6{p_{\rm dist}}I_d.
\end{equation}
Thus the condition number of $G+\lambda I_d$ is at most
$1+6/(p_{\rm dist}\lambda)=O(\epsilon^{-2})$. Standard polynomial-
precision eigensolvers and linear-system algorithms therefore suffice.

\section{Lower bound for single-photon full-bispectral estimation}
\label{sec:single_photon_large_array}

We prove the lower-bound part of
Theorem~\ref{thm:joint_3_m_scaling} by a two-part argument.
The idea is simple.  We construct two visibility matrices that have
exactly the same visibility amplitudes, so that one- and two-photon
statistics cannot distinguish them, but whose bispectra differ by
$\Theta(\epsilon)$ on a constant fraction of all triangles.
We then show that any sequential single-photon measurement obtains only
$O(\epsilon^2/m^2)$ distinguishability between the two candidates from
one three-photon ACT.  Therefore $\Omega(m^2/\epsilon^2)$ photons are
required before the two cases can be reliably distinguished.

Throughout this appendix, each ACT contains exactly three photons in
distinct temporal modes.  For a visibility matrix $V\succeq0$ with
$V_{ii}=1$, let
\begin{equation}
    \rho_\delta(V)
    =
    \frac{1}{m}D_\delta V D_\delta^\dagger,
    \qquad
    \Sigma_n(V)
    =
    \E_\delta[\rho_\delta(V)^{\otimes n}],
    \label{eq:seq_array_state}
\end{equation}
where
$D_\delta=\operatorname{diag}(e^{\imag\delta_1},\ldots,
e^{\imag\delta_m})$ and the piston phases are independently uniform
within each ACT.

A sequential single-photon protocol measures the three photons one at
a time.  Its POVM elements have the form
\[
    E_x\otimes F_{y|x}\otimes M_{z|xy},
\]
where the later measurements may depend arbitrarily on all previous
classical outcomes.  Classical adaptation between different ACTs is
also allowed.

For $T=\binom m3$, define the expected uniform mean-square error
\begin{equation}
    R(\widehat B,V)
    =
    \frac{1}{T}
    \sum_{i<j<k}
    \E_V\left|
        \widehat B_{ijk}-B_{ijk}(V)
    \right|^2 .
    \label{eq:seq_array_risk}
\end{equation}

\begin{theorem}[Single-photon lower bound]
\label{thm:seq_array_lower}
For every $m\ge4$ and $0<\epsilon\le0.1$, there exist two physical
visibility matrices $V_+$ and $V_-$ such that any sequential
single-photon protocol satisfying
\[
    R(\widehat B,V_\pm)\le\epsilon^2
\]
for both candidates requires
\begin{equation}
    N_\gamma
    =
    \Omega\!\left(\frac{m^2}{\epsilon^2}\right)
    \label{eq:seq_array_photon_lower}
\end{equation}
input photons.
\end{theorem}

\subsection{A hard pair of visibility matrices}

Divide the $m$ sites into three groups $A,B,C$, with sizes differing by at most one.  We first make all sites within a group behave identically and specify the correlations between the three groups by the $3\times3$ matrices
\begin{equation}
    C_\pm
    =
    \begin{pmatrix}
        1&r&s_\epsilon\mp5\imag\epsilon\\
        r&1&r\\
        s_\epsilon\pm5\imag\epsilon&r&1
    \end{pmatrix},
    \label{eq:seq_array_core}
\end{equation}
where $r=\frac{1}{\sqrt2}$ and
$s_\epsilon=\sqrt{\frac12-25\epsilon^2}$.
If $a(i)\in\{A,B,C\}$ denotes the group containing site $i$, define
\begin{equation}
    (\widetilde V_\pm)_{ij}
    =
    (C_\pm)_{a(i),a(j)} .
    \label{eq:seq_array_unregularized_candidates}
\end{equation}

One can verify that $C_\pm\succeq0$ for $\epsilon\le0.1$.
Since $\widetilde V_\pm$ are obtained from $C_\pm$ by duplicating its
three indices according to the groups,
$\widetilde V_\pm\succeq0$ as well.

For a technical reason that will become useful below, we add a small
identity component and take the actual hard instances to be
\begin{equation}
    V_\pm
    =
    (1-\eta)I_m+\eta\widetilde V_\pm,
    \qquad
    \eta=\frac{31}{32}.
    \label{eq:seq_array_candidates}
\end{equation}
These matrices still have unit diagonal and are positive semidefinite.
Moreover,
\begin{equation}
    V_\pm\succeq(1-\eta)I_m.
    \label{eq:seq_array_spectral_floor}
\end{equation}
The role of this small identity component is only to guarantee a
uniform lower bound on single-photon measurement probabilities; it
changes all relevant bispectra only by a constant factor.

For every pair of distinct sites,
\[
    (V_\pm)_{ij}
    =
    \eta(\widetilde V_\pm)_{ij},
\]
and therefore
\begin{equation}
    |(V_+)_{ij}|=|(V_-)_{ij}|
    \qquad\text{for every }i,j.
    \label{eq:seq_array_same_amplitude}
\end{equation}
Thus the two candidates have identical visibility amplitudes.

Their bispectra, however, differ on every triangle containing one site
from each group.  For $i\in A$, $j\in B$, and $k\in C$,
\begin{equation}
    B_{ijk}(V_\pm)
    =
    b_\pm
    :=
    \eta^3
    \left(
        \frac{s_\epsilon}{2}
        \pm\frac{5\imag\epsilon}{2}
    \right),
    \label{eq:seq_array_bispectra}
\end{equation}
where $\Delta:=b_+-b_-=5\imag\eta^3\epsilon$.
All other triangle bispectra are identical for $V_+$ and $V_-$.
Moreover, the fraction of triangles containing one site from each
group obeys
\begin{equation}
    p_c
    \coloneqq
    \Pr[\text{cross-group triangle}]
    =
    \frac{|A||B||C|}{\binom m3}
    \ge\frac29.
    \label{eq:seq_array_fraction}
\end{equation}
Hence a constant fraction of all bispectra change by
$|\Delta|=5\eta^3\epsilon=\Theta(\epsilon)$.

To connect this two-point separation to the global error, average
the estimates over these triangles:
\begin{equation}
    \widehat b
    =
    \frac{1}{|A||B||C|}
    \sum_{\substack{i\in A,\ j\in B\\k\in C}}
    \widehat B_{ijk}.
    \label{eq:seq_array_scalar_estimator}
\end{equation}
By Jensen's inequality,
\begin{equation}
    R(\widehat B,V_\pm)
    \ge
    p_c\,
    \E_\pm|\widehat b-b_\pm|^2.
    \label{eq:seq_array_scalar_risk}
\end{equation}
Thus any estimator achieving small global mean-square error must also
estimate the scalar $b_\pm$ accurately.  We next show that sequential
single-photon measurements require $\Omega(m^2/\epsilon^2)$ photons
to distinguish these two possibilities.

\subsection{One ACT contains only \(O(\epsilon^2/m^2)\) information}
Before giving the bound, it is useful to see why a strong suppression
should occur.  Since the two candidates have identical visibility
amplitudes, their piston-averaged one- and two-photon states are
identical.  In particular,
\[
    \Sigma_2(V_+)=\Sigma_2(V_-).
\]
Thus the first two outcomes of any sequential protocol contain no
information by themselves about whether the state is $V_+$ or $V_-$.
They can only determine how the third photon is subsequently measured.
The difference between the two candidates appears only through
three-photon cyclic coherences.

The following lemma quantifies how much of this information can survive
sequential single-photon measurements.

\begin{lemma}[One-ACT distinguishability]
\label{lem:seq_array_one_act}
Let $P_\pm$ denote the complete three-outcome distributions produced by
an arbitrary sequential single-photon measurement on
$\Sigma_3(V_\pm)$.  Then
\begin{equation}
    \chi^2(P_+\|P_-)
    \le
    \frac{8|\Delta|^2}
         {3(1-\eta)^3m^2}
    =
    C\,\frac{\epsilon^2}{m^2},
    \label{eq:seq_array_one_act_chi}
\end{equation}
where
\[
    C
    =
    \frac{200\eta^6}{3(1-\eta)^3}
\]
is an absolute constant.
\end{lemma}

\begin{proof}
We may assume the first two POVM elements to be rank-one,
\[
    E_x=\alpha_x|u_x\rangle\langle u_x|,
    \qquad
    F_{y|x}
    =
    \beta_{y|x}|v_{y|x}\rangle\langle v_{y|x}|,
\]
since retaining the outcome can only increase the
distinguishability.

We first lower bound the probability of every product-measurement
outcome.  Equation~\eqref{eq:seq_array_spectral_floor} implies that,
for every realization of the piston phases,
\begin{equation}
    \rho_\delta(V_-)
    =
    \frac1mD_\delta V_-D_\delta^\dagger
    \succeq
    \frac{1-\eta}{m}I_m.
    \label{eq:seq_array_single_photon_floor}
\end{equation}
Therefore, for arbitrary $E,F,M$,
\begin{align}
    &\quad \, \Tr[
        \Sigma_3(V_-)
        (E\otimes F\otimes M)
    ]
    \nonumber\\
    &=
    \E_\delta\!\left[
        \Tr(E\rho_\delta)
        \Tr(F\rho_\delta)
        \Tr(M\rho_\delta)
    \right]
    \nonumber\\
    &\geq
    \frac{(1-\eta)^3}{m^3}
    \Tr E\,\Tr F\,\Tr M.
    \label{eq:seq_array_product_probability}
\end{align}

Now fix the first two outcomes and write
$E=\alpha|u\rangle\langle u|$ and
$F=\beta|v\rangle\langle v|$.
After averaging over the piston phases, the only terms that distinguish
the two candidates are the cyclic coherences associated with triples
containing one site from each group.  Let $u_a$ and $v_a$ denote the
restrictions of $u$ and $v$ to group $a$.  Contracting these
candidate-dependent terms with the first two POVM elements gives
\begin{equation}
    P_+(x,y,z)-P_-(x,y,z)
    =
    \frac{\alpha\beta}{m^3}
    \Tr[M_{z|xy}D_{u,v}],
    \label{eq:seq_array_probability_difference}
\end{equation}
where $D_{u,v}=L_{u,v}+L_{u,v}^\dagger$ and
\begin{equation}
    L_{u,v}
    =
    \sum_{\substack{a,b,c\in\{A,B,C\}\\a,b,c\ {\rm distinct}}}
    \Delta_{abc}
    \langle v_b,u_b\rangle
    |v_c\rangle\langle u_a|.
    \label{eq:seq_array_contracted_difference}
\end{equation}
Here
$\Delta_{ABC}=\Delta_{BCA}=\Delta_{CAB}=\Delta$, while reversing the
group order replaces $\Delta$ by $\Delta^*$.

Equation~\eqref{eq:seq_array_product_probability} gives
\begin{equation}
    P_-(x,y,z)
    \ge
    \frac{(1-\eta)^3\alpha\beta}{m^3}
    \Tr M_{z|xy}.
    \label{eq:seq_array_denominator}
\end{equation}
For any Hermitian $D$ and positive $M$,
\[
    |\Tr(MD)|^2
    \le
    \Tr M\,\Tr(MD^2).
\]
Using $\sum_zM_{z|xy}=I_m$ therefore gives
\begin{align}
    \chi^2(P_+\|P_-)&= \sum_{x,y,z}\frac{(P_+-P_-)^2}{P_-}\\
    &\le
    \frac{1}{(1-\eta)^3m^3}
    \sum_{x,y}
    \alpha_x\beta_{y|x}
    \|D_{u_x,v_{y|x}}\|_{\mathrm F}^2.
    \label{eq:seq_array_chi_contraction}
\end{align}

The six terms in $L_{u,v}$ act between different ordered pairs of group
subspaces, so they are orthogonal in Frobenius inner product.  Hence
\begin{equation}
    \|D_{u,v}\|_{\mathrm F}^2
    \le
    4|\Delta|^2
    \sum_{\substack{a,b,c\\{\rm distinct}}}
    |\langle v_b,u_b\rangle|^2
    |u_a|^2|v_c|^2.
    \label{eq:seq_array_block_norm}
\end{equation}
For fixed $x$, completeness of the second POVM implies
\[
    \sum_y
    \beta_{y|x}
    |\langle(v_{y|x})_b,(u_x)_b\rangle|^2
    =
    |(u_x)_b|^2.
\]
Writing $q_a=|(u_x)_a|^2$, with $\sum_aq_a=1$, we obtain
\begin{align}
    \sum_y
    \beta_{y|x}
    \|D_{u_x,v_{y|x}}\|_{\mathrm F}^2
    &\le
    4|\Delta|^2
    \sum_{a\ne b}q_aq_b
    \nonumber\\
    &\le
    \frac83|\Delta|^2.
    \label{eq:seq_array_second_sum}
\end{align}
Finally,
$\sum_x\alpha_x=\Tr I_m=m$, so
Eq.~\eqref{eq:seq_array_chi_contraction} gives
\begin{equation}
    \chi^2(P_+\|P_-)
    \le
    \frac{8|\Delta|^2}
         {3(1-\eta)^3m^2},
\end{equation}
which proves the lemma.
\end{proof}

The scaling in Lemma~\ref{lem:seq_array_one_act} is the main point of
the proof.  The candidate-dependent three-photon coherence carries an
$m^{-3}$ normalization.  After the first two photons have been
converted to classical outcomes, POVM completeness recovers only one
factor of $m$ when their outcomes are summed.  The spectral floor in
Eq.~\eqref{eq:seq_array_spectral_floor} ensures that the denominator in
the statistical distance contributes only a constant factor.  The
result is therefore an overall $m^{-2}$ distinguishability per ACT.

\subsection{Accumulating information over many ACTs}

Let $P_\pm^{(K)}$ denote the complete classical transcript after $K$
ACTs.  The protocol may adapt arbitrarily to all previous outcomes.
Conditioned on any previous history, however, the next ACT is fresh and
Lemma~\ref{lem:seq_array_one_act} applies to the measurement selected
for that history.  Using
$D_{\rm KL}(P\|Q)\le\chi^2(P\|Q)$ and the chain rule for relative entropy,
\begin{equation}
    D_{\rm KL}
    \bigl(P_+^{(K)}\|P_-^{(K)}\bigr)
    \le
    C\frac{K\epsilon^2}{m^2}.
    \label{eq:seq_array_many_acts_kl}
\end{equation}
Pinsker's inequality therefore gives
\begin{equation} 
    \operatorname{TV}
    \bigl(P_+^{(K)},P_-^{(K)}\bigr)
    \le
    \sqrt{
        \frac{CK\epsilon^2}{2m^2}
    }.
    \label{eq:seq_array_many_acts_tv}
\end{equation}

We can now finish the two-part argument.  For every complex number
$z$,
\[
    |z-b_+|^2+|z-b_-|^2
    \ge
    \frac{|\Delta|^2}{2}.
\]
Consequently,
\begin{equation}
    \max_{\sigma\in\{+,-\}}
    \E_\sigma|\widehat b-b_\sigma|^2
    \ge
    \frac{|\Delta|^2}{4}
    \left[
        1-\operatorname{TV}
        \bigl(P_+^{(K)},P_-^{(K)}\bigr)
    \right].
    \label{eq:seq_array_two_point}
\end{equation}
Combining this with
Eqs.~\eqref{eq:seq_array_many_acts_tv}, \eqref{eq:seq_array_scalar_risk},
\eqref{eq:seq_array_fraction}, and
$|\Delta|^2=25\eta^6\epsilon^2$ gives
\begin{equation}
    \max_{\sigma\in\{+,-\}}
    R(\widehat B,V_\sigma)
    \ge
    \frac{25\eta^6}{18}\epsilon^2
    \left(
        1-
        \sqrt{
            \frac{CK\epsilon^2}{2m^2}
        }
    \right).
    \label{eq:seq_array_risk_lower}
\end{equation}
Our choice $\eta=31/32$ satisfies
\[
    \frac{25\eta^6}{18}>1.
\]
Therefore, requiring the error to be at most $\epsilon^2$ for both
candidates forces
\[
    \sqrt{
        \frac{CK\epsilon^2}{2m^2}
    }
    \ge
    1-\frac{18}{25\eta^6},
\]
where the right-hand side is a positive constant independent of
$m$ and $\epsilon$.  Hence
\begin{equation}
    K
    =
    \Omega\!\left(
        \frac{m^2}{\epsilon^2}
    \right).
\end{equation}
Since every ACT contains three photons,
\[
    N_\gamma=3K
    =
    \Omega\!\left(
        \frac{m^2}{\epsilon^2}
    \right),
\]
which proves Theorem~\ref{thm:seq_array_lower}.

\section{Multiparameter Cramer-Rao Bounds} \label{sec:multiparameterCramerRao}

Learning the vector $\vec\phi_\text{cl}$ is a \emph{multi-parameter estimation problem}.
The optimal measurements for individual parameters may not commute, and finding the optimal general measurement can be challenging. The Nagaoka-Hayashi ($C_{NH})$ and Holevo ($C_H$) Cramer-Rao bounds constrain the performance of individual and collective measurements, respectively, and can be efficiently evaluated~\cite{conlon2021efficient}. The average variance is set as the cost function to be minimized:
\begin{equation}
    \frac{1}{\binom{m-1}{2}}\sum_{2\leq a < b \leq m} \text{Var}[\phi_{\text{cl},ab}] \,,
\end{equation}
and we treat the remaining pairwise phases as nuisance parameters. We find that $C_{NH}=C_H$ for $n=3$ and general $m$: there is no advantage for collective measurements over three-photon packets. Furthermore, the Cramer-Rao bound given by $1/F_\text{cl}$ is the same as $C_{NH}$; there is no measurement incompatibility and the previous approaches described are optimal. On the other hand, $C_{NH}/C_H = O(m)$ for estimating the vector of closure phases in the $m$-site generalization of the noiseless state (Eq.~\eqref{eq:dm_noisy}). That is, in the limit of infinite photons, a general measurement on all the photons gives an extensive reduction in the estimation variance as a function of array size.

With the multiple-site generalization of the noiseless state (Eq.~\eqref{eq:dm_noisy}), we calculate:
\begin{align}
    C_H &= \frac{3m(1-v)}{2v^2} \,, \\
C_{NH}&=\frac{m(1-v)}{2v^2}
 \begin{aligned}[t]
 \big[&m+\frac{(m-3)^2}{m-1}\\
 &+\frac{2(m-3)\sqrt m}{m-1}\big]\,,
 \end{aligned}
\end{align}
where $v$ is the same visibility amplitude for all baselines,
and thus
\begin{equation} \label{eq:gap}
    \frac{C_{NH}}{C_H}=1+\frac{2(m-3)(m-2+\sqrt{m})}{3(m-1)} \,.
\end{equation}
Thus, the gap is 0 for $m=3$ and grows linearly in $m$ for larger $m$. The presence of a gap means that it is advantageous to do multiple-photon measurements rather than one at a time. We can interpret this result as corresponding to an infinitely long atmospheric coherence time, or taking the number of photons per such coherence time to infinity.

Similarly we compute that the gap between the Holevo and Nagaoka-Hayashi Cramer-Rao bounds using the $\rho^\mathrm{noisy}_3$ is 0:
\begin{equation}
    C_{NH}=C_H=\frac{m^2(1-v^3)}{4v^6} \,.
\end{equation}
Thus, there is no advantage to measuring multiple three-photon packets together. The result of Eq.~\eqref{eq:gap} suggests that there will be an advantage for more than three photons.

\newpage

\end{document}